\documentclass[a4paper]{article}
\pdfoutput=1
\usepackage{jheppub}

\usepackage{amsmath}

\usepackage{amssymb}
\usepackage{graphicx,color}

\usepackage{amsmath,amsthm}
\usepackage{txfonts}

\usepackage{mathrsfs}

\usepackage{bm}

\newcommand{\Ref}[1]{(\ref{#1})}

\newtheorem{Theorem}{Theorem}[section]

\newtheorem{Lemma}[Theorem]{Lemma}

\newcommand{\Z}{\mathbb{Z}}
\newcommand{\R}{\mathbb{R}}
\newcommand{\C}{\mathbb{C}}

\newcommand{\half}{\frac{1}{2}}

\newcommand{\Slc}{\mathrm{SL}(2,\mathbb{C})}
\newcommand{\PSlc}{\mathrm{PSL}(2,\mathbb{C})}
\newcommand{\slc}{\fs\fl_2\mathbb{C}}

\def\be{\begin{eqnarray}}
\def\ee{\end{eqnarray}}

\newcommand{\cc}{\mathcal C}

\newcommand{\ce}{\mathcal E}

\newcommand{\ch}{\mathcal H}
\newcommand{\ci}{\mathcal I}
\newcommand{\cj}{\mathcal J}
\newcommand{\ck}{\mathcal K}
\newcommand{\cl}{\mathcal L}
\newcommand{\cm}{\mathcal M}
\newcommand{\cn}{\mathcal N}

\newcommand{\cp}{\mathcal P}

\newcommand{\cs}{\mathcal S}
\newcommand{\ct}{\mathcal T}

\newcommand{\cw}{\mathcal W}

\newcommand{\sa}{\mathscr{A}}
\newcommand{\sm}{\mathscr{M}}

\newcommand{\fa}{\mathfrak{a}}  \newcommand{\Fa}{\mathfrak{A}}
  
  \newcommand{\Fc}{\mathfrak{C}}
  
\newcommand{\fe}{\mathfrak{e}}  
\newcommand{\ff}{\mathfrak{f}}

\newcommand{\fl}{\mathfrak{l}}  \newcommand{\Fl}{\mathfrak{L}}
  
\newcommand{\fn}{\mathfrak{n}}  
  
\newcommand{\fp}{\mathfrak{p}}  \newcommand{\Fp}{\mathfrak{P}}

\newcommand{\fs}{\mathfrak{s}}  \newcommand{\Fs}{\mathfrak{S}}
\newcommand{\ft}{\mathfrak{t}}  
\newcommand{\fu}{\mathfrak{u}}

  \newcommand{\Fx}{\mathfrak{X}}

\renewcommand{\a}{\alpha}
\renewcommand{\b}{\beta}
\newcommand{\g}{\gamma}
\newcommand{\G}{\Gamma}

\newcommand{\eps}{\varepsilon}

\newcommand{\sig}{\sigma}
\newcommand{\Sig}{\Sigma}
\renewcommand{\l}{\lambda}
\renewcommand{\L }{\Lambda}
\renewcommand{\o}{\omega}
\renewcommand{\O}{\Omega}
\renewcommand{\t}{\tau}

\newcommand{\rmd}{\mathrm d}

\newcommand{\lt}{\left}
\newcommand{\rt}{\right}

\newcommand{\lag}{\left\langle}
\newcommand{\rag}{\right\rangle}

\newcommand{\tr}{\mathrm{tr}}

\newcommand{\act}{\rhd}

\newcommand{\sgn}{\mathrm{sgn}}
\newcommand{\vth}{\vartheta}

\title{4d Quantum Geometry from 3d Supersymmetric Gauge Theory and Holomorphic Block}

\author[]{\ Muxin Han}

\affiliation[1]{Institut f\"ur Quantengravitation, Universit\"at Erlangen-N\"urnberg, Staudtstr. 7/B2, 91058 Erlangen, Germany}

\affiliation[2]{Department of Physics, Florida Atlantic University, 777 Glades Road, Boca Raton, FL 33431-0991, USA}

\emailAdd{hanm(AT)fau.edu} %

\abstract{A class of 3d $\cn=2$ supersymmetric gauge theories are constructed and shown to encode the simplicial geometries in 4-dimensions. The gauge theories are defined by applying the Dimofte-Gaiotto-Gukov construction \cite{DGG11} in 3d-3d correspondence to certain graph complement 3-manifolds. Given a gauge theory in this class, the massive supersymmetric vacua of the theory contain the classical geometries on a 4d simplicial complex. The corresponding 4d simplicial geometries are locally constant curvature (either dS or AdS), in the sense that they are made by gluing geometrical 4-simplices of the same constant curvature. When the simplicial complex is sufficiently refined, the simplicial geometries can approximate all possible smooth geometries on 4-manifold. At the quantum level, we propose that a class of holomorphic blocks defined in \cite{3dblock} from the 3d $\cn=2$ gauge theories are wave functions of quantum 4d simplicial geometries. In the semiclassical limit, the asymptotic behavior of holomorphic block reproduces the classical action of 4d Einstein-Hilbert gravity in the simplicial context.
}


\arxivnumber{}

\begin{document}

\maketitle

\section{Introduction}

3d-3d correspondence, proposed by Dimofte, Gaiotto, and Gukov in \cite{DGG11} (see also \cite{33revisit,3d/3drev}), constructs a class of 3d $\cn=2$ supersymmetric gauge theories $T_{M_3}$ labelled by 3-manifolds $M_3$ \footnote{$T_{M_3}$ is essentially a superconformal field theory living at the IR fix point of the gauge theory. }. In this correspondence, the partition function of the supersymmetric gauge theory $T_{M_3}$ is equivalent to Chern-Simons partition function of the corresponding 3-manifold $M_3$ \cite{3dindice,3dblock}, and the massive supersymmetric vacua of $T_{M_3}$ relate to the flat connections on $M_3$ \cite{CSSduality}. 3d-3d correspondence is a generalization of Alday-Gaiotto-Tachikawa (AGT) correspondence \cite{AGT,Gadde:2011ik}, which proposes a class of 4d $\cn=2$ supersymmetric gauge theories labelled by 2-manifolds. There is also a further generalization by \cite{GGP13}, which proposes a class of 2d $\cn=(0,2)$ supersymmetric gauge theories labelled by 4-manifolds. It has been argued that these correspondences come from the different schemes of reductions from the 6d (0,2) superconformal field theory (SCFT) \cite{Junya,CJ,LY,5braneknots,Gaiotto:2011xs,LTYZ,Tan:2013tq,Tan:2013xba}.

In this paper, we propose that there are a class of 3d $\cn=2$ supersymmetric gauge theories, which turn out to encode the simplicial geometries in 4-dimensions. Given a gauge theory in this class, the massive supersymmetric vacua of the theory contains the classical geometries on a 4d simplicial complex. The resulting 4d simplicial geometries are locally constant curvature (either dS or AdS), in the sense that they are made by gluing geometrical 4-simplices of the same constant curvature. When the simplicial complex is sufficiently refined, the simplicial geometries can approximate all possible smooth geometries on 4-manifold. At the quantum level, we propose that a class of holomorphic blocks from the 3d $\cn=2$ gauge theories are wave functions of quantum 4d simplicial geometries. Holomorphic block is proposed in \cite{3dblock} as the supersymmetric BPS index of 3d $\cn=2$ theory. We find that the holomorphic blocks, defined from the class of 3d theories constructed here, know about the dynamics of 4d geometries. In certain semiclassical limit, the asymptotic behavior of holomorphic block reproduces the classical action of 4d Einstein-Hilbert gravity in the simplicial context.

\begin{figure}[h]
\begin{center}
\includegraphics[width=5cm]{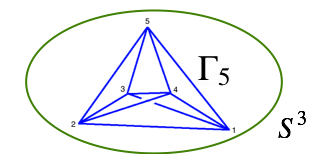}
\caption{The graph complement 3-manifold $S^3\setminus\G_5$ is obtained by firstly embedding the $\G_5$ graph in $S^3$, then removing the graph and its tubular neighborhood from $S^3$. }
\label{gamma5complement}
\end{center}
\end{figure}

The class of 3d $\cn=2$ supersymmetric gauge theories constructed here is a subclass contained in the theories from Dimofte-Gaiotto-Gukov (DGG) construction in \cite{DGG11} for 3d-3d correspondence. The class of 3d $\cn=2$ supersymmetric gauge theories $\{T_{\sm_3}\}$ studied here asociate to a class of 3-manifold $\{\sm_3\}$. The 3-manifolds $\sm_3$ are made by gluing copies of the graph complement 3-manifolds $S^3\setminus\G_5$ (FIG.\ref{gamma5complement}), through the 4-holed spheres associated to the vertices of $\G_5$ graph. The class of $\sm_3$ relate to the class of 4d simplicial complexes (simplicial manifold) $\sm_4$. Namely the fundamental group of the 3-manifold $\pi_1(\sm_3)$ is isomorphic to the fundamental group of the 1-skeleton of the simplicial complex $\pi_1({\rm sk}(\sm_4))$. Moreover a class of (framed) $\Slc$ flat connections on $\sm_3$ are equivalent to the locally constant curvature 4d simplicial geometries on the corresponding $\sm_4$ in Lorentzian signature. As an basic and crucial example, a class of $\Slc$ flat connections on $S^3\setminus\G_5$ are equivalent to the constant curvature geometries on a single (convex) 4-simplex. This example is also an important ingredient in understanding the general relation between $\sm_3$ and $\sm_4$

The massive supersymmetric vacua of the theory $T_{\sm_3}$ on $\R^2\times S^1$ relates to $\Slc$ flat connections on $\sm_3$ by 3d-3d correspondence. Because of the above relation between the flat connections on $\sm_3$ and simplicial geometries on the corresponding $\sm_4$, a class of massive supersymmetric vacua of the theory $T_{\sm_3}$ gives all the 4d locally constant curvature simplicial geometries on $\sm_4$. For any 3d $\cn=2$ supersymmetric gauge theory on $\R^2\times S^1$, there are 2 natural parameters parametrizing the moduli space of supersymmetric vacua: the complex mass parameters $\vec{x}$ from the reduction on $S^1$ and effective background Fayet-Iliopoulos (FI) parameters $\vec{y}$ preserving supersymmetry. The correspondence of $T_{\sm_3}$ between supersymmetric vacua and 4d simplicial geometries relates the susy parameters $\vec{x},\vec{y}$ to the geometric quantities in 4d. Namely, for those vacua satisfying the 4-geometry correspondence, the complex masses $\vec{x}$ relate to the triangle areas $\mathbf{a}_\ff$ of the simplicial geometry on $\sm_4$, and the effective FI parameters $\vec{y}$ relate to the deficit angles $\eps_{\ff}$ (in the bulk) and the dihedral angles $\Theta_{\ff}$ (on the boundary). That supersymmetry is preserved on those vacua in $T_{\sm_3}$ is equivalent to the existence of locally constant curvature simplicial geometry on $\sm_4$. 

It has been argued in \cite{3dblock,Pasquetti:2011fj} that for a generic 3d $\cn=2$ gauge theory, its ellipsoid partition function $Z_{S^3_b}$ and spherical index $Z_{S^2\times_q S^1}$ (partition function on $S^2\times_q S^1$) admit the holomorphic factorizations. They are both factorized into a class of universal holomorphic building blocks $\{B^\a\}_\a$, known as holomorphic blocks in 3-dimensions \footnote{The recent work \cite{Nieri:2015yia} shows that the holomorphic blocks also exist in 4 dimensions for 4d gauge theories.}. This result has been shown to be valid for all gauge theories $T_{M_3}$ in 3d-3d correspondence. In general the holomorphic block $B^\a$ is a supersymmetric BPS index of 3d $\cn=2$ gauge theory, which can also be understood as the partition function on $D^2\times_q S^1$ with a topological twist \cite{3dblock}. The label $\a$ of holomorphic block $B^\a$ 1-to-1 corresponds to the branches of massive supersymmetric vacua in the theory on $\R^2\times S^1$ (the asymptotic regime of $D^2\times_q S^1$). Given the class of theories $T_{\sm_3}$, we pick out the supersymmetric vacua $\a_{\rm 4d}$ satisfying the 4-geometry correspondence, and construct the corresponding holomorphic block $B_{\sm_3}^{\a_{\rm 4d}}$ \footnote{A key reason of using holomorphic block here is that not all the SUSY vacua of $T_{\sm_3}$ satisfying the correspondence to 4d simplicial geometry. The holomorphic blocks under consideration here are the ones labelled by $\a_{4d}$. The 4d geometrical meaning of other SUSY vacua is not completely clear at the moment, and is a research undergoing.}. We propose that the resulting holomorphic block is a wave function quantizing the locally constant curvature simplicial geometries on $\sm_4$, which encodes the dynamics of 4d geometry. Indeed, in a certain semiclassical limit, the asymptotic behavior of the resulting holomorphic block reproduces the classical action of 4d Einstein-Hilbert gravity (with cosmological constant) on the simplicial complex $\sm_4$. The Einstein-Hilbert action in the simplicial context is also known as \emph{Einstein-Regge action} \cite{regge,BD}. The classical action recovered here is the Einstein-Regge action evaluated at the simplicial geometries made by gluing constant curvature 4-simplices. Therefore the class of holomorphic blocks $B_{\sm_3}^{\a_{\rm 4d}}$ from $T_{\sm_3}$ may be viewed as certain quantization of simplicial gravity in 4 dimensions.

The following table summarizes the relation between 3d $\cn=2$ supersymmetric gauge theory $T_{\sm_3}$ and the simplicial geometry on $\sm_4$

\begin{table}[h]
\begin{center}
\begin{tabular}{|c|c|}
\hline
$T_{\sm_3}$ & $\sm_4$ \\
\hline
A class of massive supersymmetric vacua on $\R^2\times S^1$ & Locally constant curvature 4d simplicial geometries\\
\hline
Preserving supersymmetry on the vacua & The existence of 4d simplicial geometries\\
\hline
Complex mass parameters $x_i$ & Triangle areas $\mathbf{a}_{\ff}$\\
\hline
Effective FI parameters ${y}_i$ preserving supersymmetry & Bulk deficit angles $\eps_{\ff}$ and boundary dihedral angles $\Theta_{\ff}$\\
\hline
Holomorphic block (partition function on $D^2\times_q S^1$) & Semiclassical wave function of 4d simplicial geometries\\
\hline
Effective twisted superpotential & Einstein-Regge action of 4d geometry\\
\hline
Deformation parameter $\hbar=\ln q$ & Cosmological constant $\kappa$ in Planck unit\\
\hline
\end{tabular}
\end{center}
\label{default}
\end{table}

\noindent
Note that the 1st rows in the above table of correspondences can be formulated in the language of flat connections on 3-manifold because of the 3d-3d correspondence. The correspondence between flat connections on $S^3\setminus\G_5$ and constant curvature 4-simplex geometries has been proposed in the author's recent work \cite{HHKR,curvedMink,3dblockHHKR}. In this paper, the correspondence is generalized to the general situation of simplicial complex with arbitrarily many 4-simplices.

The holomorphic blocks from supersymmetric gauge theories $T_{\sm_3}$ can be understood in the full framework of M-theory. The relation between 4d simplicial geometry and supersymmetric gauge theory proposed in this paper relates M-theory to simplicial geometry in 4d. The 3d-3d correspondence used in the construction of $T_{\sm_3}$ can be resulting from certain reduction of M5-brane IR dynamics, i.e. 6d (0,2) SCFT. The holomorphic blocks $B^{\a}_{\sm_3}$ playing central role here is interpreted as the partition function of two M5-branes embedded in an 11d M-theory background \cite{3dblock}
\be
M_3\times D^2\times_q S^1\subset T^*M_3\times TN\times_q S^1
\ee
where $D^2\subset TN$ is a cigar inside Taub-NUT space. $M_3$ is $S^3$ or $\Fx_3$ depending on the number of geometrical 4-simplices. The codimension-2 graph defect is given by a stack of additional intersecting M5-branes, which may be formulated field-theoretically as surface operator with junctions \cite{5braneknots,Chun:2015gda}. It is interesting to re-understand and re-interpret our correspondence in the framework of full M-theory with branes. The detailed discussion in this perspective will appear elsewhere \cite{future}.    

On the other hand, 6d (0,2) SCFT has a holographic dual to M-theory on $\text{AdS}_7\times S^4$ \cite{Witten:1996hc,Witten:1998wy,D'Hoker:2008qm,Fiorenza:2012tb}. Here we have shown that 4d (locally constant curvature) simplicial gravity emerging from holomorphic blocks $B^{\a_{\rm 4d}}_{\sm_3}$. Given that the holomorphic block is the partition function of 6d (0,2) SCFT on certain background, its relation with 4d gravity might have interesting relation to AdS/CFT \cite{Gang:2014ema,Li:2014uqa}.

It is important to mention that the idea of studying $S^3\setminus\G_5$ and the class of $\sm_3$ comes from the covariant formulation of Loop Quantum Gravity (LQG) \cite{book,book1,CLQG,review,review1,Perez2012,Rovelli:2011eq}. The studies of spinfoam models in LQG \cite{EPRL,FK,HHKR,QSF,QSF1,QSFasymptotics} motivates the relation between Chern-Simons theory on 3-manifolds $\sm_3$ and the geometries on simplicial 4-manifolds $\sm_4$. The holomorphic block studied in this paper defines the \emph{Spinfoam Amplitude} in LQG language, which describes the evolution of quantum gravity on the simplicial complex $\sm_4$. Therefore the present work relates LQG to supersymmetric gauge theory and M5-brane dynamics in String/M-theory, which is another significant physical consequence of the present work. There are many possible future developments from LQG perspectives. For example, it is interesting to further understand the perturbative behavior of the holomorphic block from the point of view of semiclassical low energy approximation in LQG \cite{lowE,lowE1,lowE2}. We should also investigate and understand the behavior of the holomorphic blocks under the refinement of the 4d simplicial complex (suggested by the studies on spinfoam model \cite{Banburski:2014cwa}). It is also interesting to relate the present result to the canonical operator formulation of LQG \cite{QSD,Thiemann2006,Thiemann2006a,master,link,link1,masterPI,Bonzom:2011hm}, since the Ward identity as an operator constraint equation $\hat{\bf A} B^{\a}=0$ from $T_{\sm_3}$ might relate to Hamiltonian constraint equation in LQG.

The paper is organized as follows: In Section \ref{symplectic data}, we give an ideal triangulation of the graph complement 3-manifold $S^3\setminus\G_5$, and analyze the symplectic coordinates for framed flat connections by the ideal triangulation. In Section \ref{3d/3dduality}, we construct the supersymmetric gauge theories labelled by $S^3\setminus\G_5$ and $\sm_3$ by using the ideal triangulation and symplectic data studied in Section \ref{symplectic data}. In Section \ref{HB3d}, we give a brief review of holomorphic blocks for 3d $\cn=2$ supersymmetric gauge theories, and apply the construction to the theories $T_{S^3\setminus\G_5}$ and $T_{\sm_3}$. In Section \ref{4dQG}, we identify the supersymmetric vacua in $T_{\sm_3}$ (including $T_{S^3\setminus\G_5}$), which correspond to the locally constant curvature simplicial geometries on 4d simplicial complex $\sm_4$. We also relate the susy parameters to the geometrical quantities in 4d simplicial geometry. In Section \ref{HBQG}, we study the class of holomorphic blocks as the quantum states of 4d simplicial geometry. We show that in the semiclassical limit, the asymptotics of holomorphic block give 4d Einstein-Regge action with cosmological constant.

\section{3-manifold, Ideal Triangulation, and Symplectic Data}\label{symplectic data}

\subsection{Ideal Triangulation of $\G_5$-Graph Complement 3-Manifold}\label{idealtriangulation}

\begin{figure}[h]
\begin{center}
\includegraphics[width=5cm]{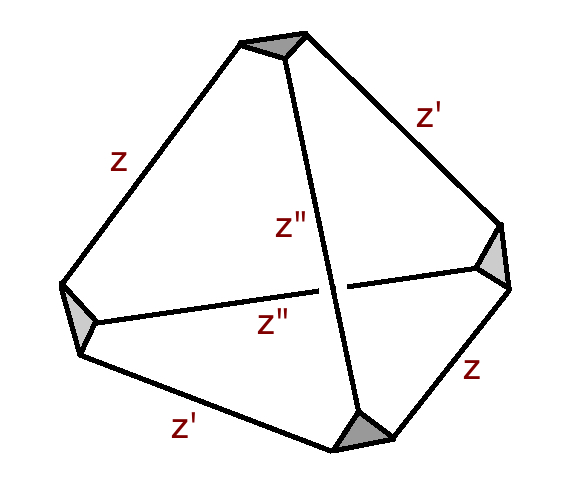}
\caption{An ideal tetrahedron.}
\label{tetrahedron}
\end{center}
\end{figure}

The 3-manifold $M_3=S^3\setminus\G_5$, being the complement of $\G_5$-graph in $S^3$, can be triangulated by a set of (topological) ideal tetrahedra. An ideal tetrahedron can be understood as a tetrahedron whose vertices are located at ``infinity''. It is convenient to truncate the vertices to define the ideal tetrahedron as a ``truncated tetrahedron'' as in FIG.\ref{tetrahedron}. There are 2 types of boundary components for the ideal tetrahedron: (a) the original boundary of the tetrahedron, and (b) the new boundary components created by truncating the tetrahedron vertices. Following e.g. \cite{Dimofte2011,DGG11,DGV}, the type-(a) boundary is referred to as \emph{geodesic boundary}, and the type-(b) boundary is referred to as \emph{cusp boundary}. 

In general, a graph complement 3-manifold $M_3$ also has 2 types of boundary components: (A) the boundary components created by removing the neighborhood of vertices of the graph, and (B) the boundary components created by removing the neighborhood of edges. Each type-(A) boundary component is a $n$-holed sphere, where the number of holes is the same as the vertex valence. Each type-(B) boundary component is an annulus, which begins and ends at holes of the type-(A) boundary. For a graph complement 3-manifold $M_3$, the type-(A) boundary is referred to as \emph{geodesic boundary}, and the type-(B) boundary is referred to as \emph{cusp boundary} of $M_3$. An ideal triangulation of $M_3$ decomposes $M_3$ into a set of ideal tetrahedra, such that the geodesic boundary of $M_3$ is triangulated by the geodesic boundary of the ideal tetrahedra, while the cusp boundary of $M_3$ is triangulated by the cusp boundary of the ideal tetrahedra.

\begin{figure}[h]
\begin{center}
\includegraphics[width=15cm]{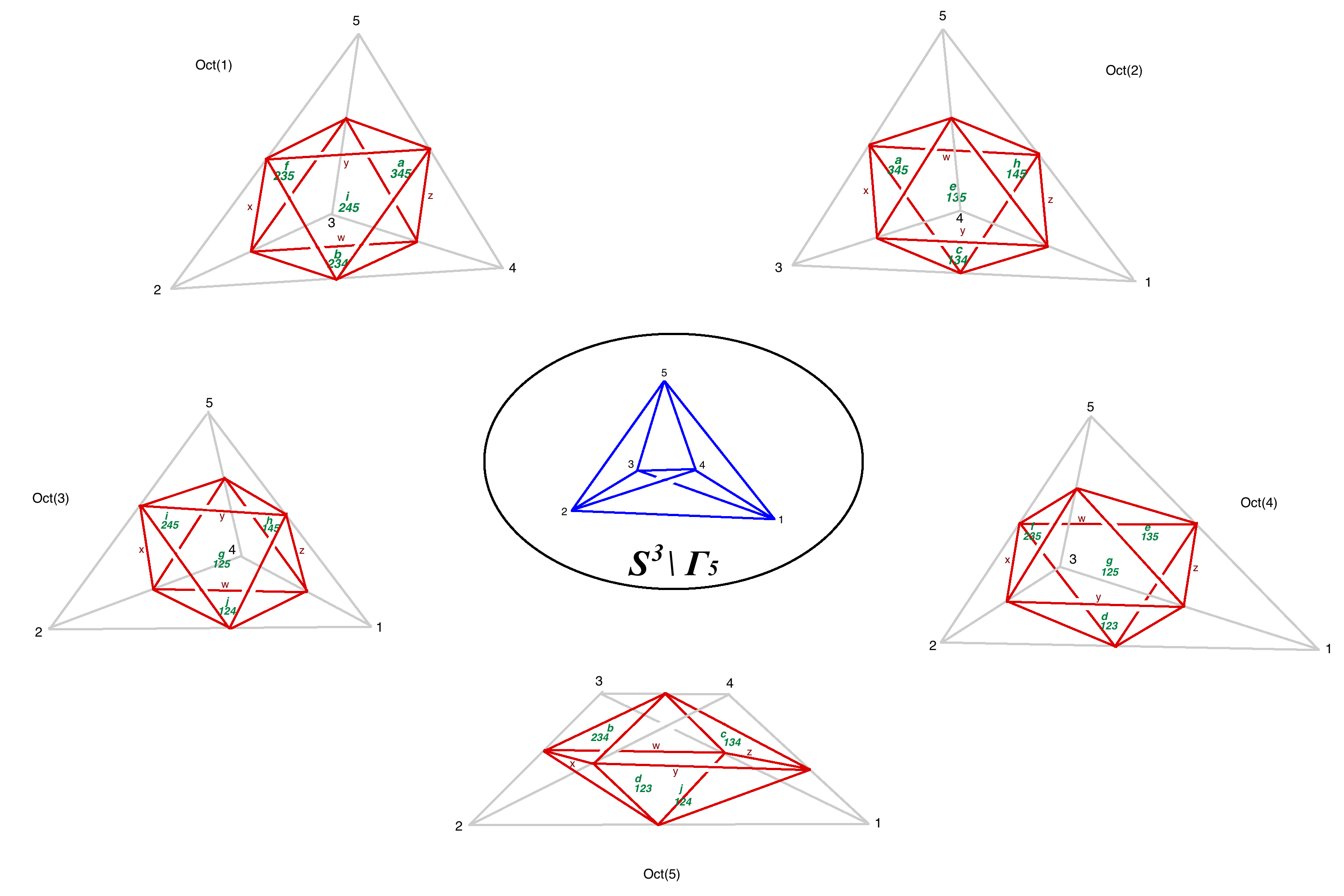}
\caption{Ideal triangulation of $S^3\setminus\G_5$ by using 5 ideal octahedra (red), which correspond to 20 ideal tetrahedra. The cusp boundaries of the ideal octahedra are not drown in the figure. The faces with green label $a,b,c,d,e,f,g,h,i,j$ are the faces where a pair of octahedra are glued. The labels show how the 5 ideal octahedra are glued together. In each ideal octahedron, we have chosen the edges with red label $x,y,z,w$ to form the equator of the octahedron.}
\label{5oct}
\end{center}
\end{figure}

\begin{figure}[h]
\begin{center}
\includegraphics[width=7cm]{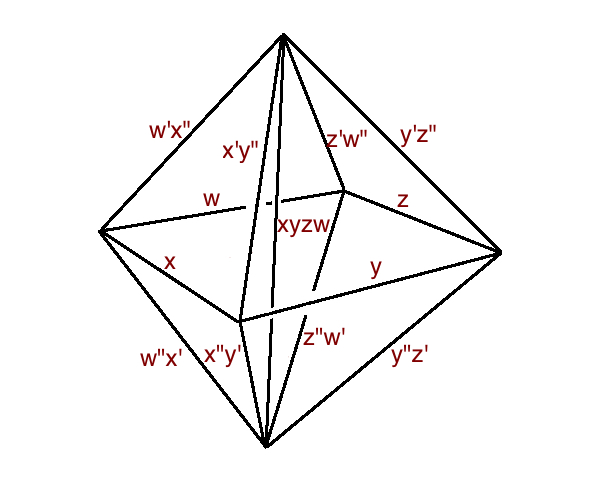}
\caption{Chosen the equator edges with labels $x,y,z,w$, an ideal octahedron can be subdivided into 4 ideal tetrahedra by drawing a vertical line connecting the remaining 2 vertices which doesn't belong to the equator.}
\label{oct_coordinate}
\end{center}
\end{figure}

For the 3-manifold $S^3\setminus\G_5$ that we are interested in, the geodesic boundary is made of 5 four-holed spheres, while the cusp boundary is made of 10 annuli connecting the four-holed spheres. The $\G_5$-graph drawn in the middle of FIG.\ref{5oct} naturally subdivides $S^3\setminus\G_5$ into 5 tetrahedron-like region (5 grey tetrahedra in FIG.\ref{5oct}, whose vertices coincide with the vertices of the graph). Each tetrahedron-like region should actually be understood as an ideal octahedron (with vertices truncated), so that the octahedron faces contribute the geodesic boundary (4-holed spheres) of $S^3\setminus\G_5$, while the octahedron cusp boundaries contribute the cusp boundary (annuli) of $S^3\setminus\G_5$. The way to glue 5 ideal octahedra to form $S^3\setminus\G_5$ is shown in FIG.\ref{5oct}. Each ideal octahedron can be subdivided into 4 idea tetrahedra as shown in FIG.\ref{oct_coordinate}. A specific way of subdividing an octahedron into 4 tetrahedra is specified by a choice of octahedron equator. As a result, the $\G_5$-graph complement $S^3\setminus\G_5$ can be triangulated by 20 ideal tetrahedra.

\subsection{Phase Space Coordinates of $\Slc$ Flat Connections}\label{coordinates}

Given a 3-manifold $M_3$ with both geodesic and cusp boundaries, a \emph{framed} $\Slc$ flat connection on $M_3$ is an $\Slc$ flat connection $A$ on $M_3$ with a choice of flat section $s$ (called the \emph{framing flag}) in an associated flag bundle over every cusp boundary \cite{DGV,GMN09,FG03}. The flat section $s$ may be viewed as a $\C^2$ vector field on a cusp boundary, defined up a complex rescaling and satisfying the flatness equation $(\rmd-A)s=0$. Consequently the vector $s(\fp)$ at a point $\fp$ of the cusp boundary is an eigenvector of the monodromy of $A$ around the cusp based at $\fp$. Similarly, a framed flat connection on $\partial M_3$ is a flat connection $\Fa$ on $\partial M_3$ with the same choice of {framing flag} on every cusp boundary. Moreover if a cusp boundary component in $\partial M_3$ is a small disc, the monodromy of a framed flat connection $\Fa$ around the disc is unipotent. Here the moduli space of framed $\Slc$ flat connections on $\partial M_3$ is denoted by $\cp_{\partial M_3}$, which has a phase space structure. The moduli space of framed $\Slc$ flat connections on $M_3$ is denoted by $\cl_{M_3}$, which isomorphic to a Lagrangian submanifold in $\cp_{\partial M_3}$ \cite{DGV}. In this paper when we talk about the framed flat connections on $M_3$ and $\partial M_3$, we assume the framing are generic, so that the reducible flat connections are excluded \footnote{The holonomies of reducible flat connections only rotate a proper subspace of $\C^2$, e.g. the abelian flat connection.}. 
 
The boundary of an ideal tetrahedron is a sphere with 4 cusp discs (in the truncated tetrahedron picture). The framed flat connections on the boundary can be understood as the flat connections on a 4-holed sphere (the geodesic boundary). The moduli space of $\Slc$ flat connection on a sphere with a number of holes can be parametrized by Fock-Goncharov edge coordinates, which is a complex number $x_E\in \C^{\times}$ associated with each edge $E$ of an ideal triangulation of $n$-holed sphere \cite{FG03} (also see \cite{DGV} or \cite{GMN09} for a nice summary). The boundary of the ideal tetrahedron provides an ideal triangulation of the boundary. Moreover in this case of an ideal tetrahedron, the monodromy around each hole/cusp on the boundary is unipotent, i.e. the product of edge coordinates around each hole equals $-1$. Therefore it is standard to call the six edge coordinates on the boundary $z,z',z''$, equal on opposite edges (as shown in FIG.\ref{tetrahedron}), and satisfying $zz'z''=-1$. Thus
\be
\cp_{\partial\Delta}=\{z,z',z''\in\C^{\times} \,|\,zz'z''=-1\}\simeq(\C^{\times})^2.
\ee
The Atiayh-Bott-Goldmann form $\O=\int_{\partial M_3}\tr\lt(\delta \Fa\wedge\delta \Fa\rt)$ endows $\cp_{\partial\Delta}$ a holomorphic symplectic structure $\O=\rmd Z''\wedge\rmd Z$. $Z=\ln z,\,Z'=\ln z',\,Z''=\ln z''$ define the logarithmic lifts of the phase space coordinates, satisfying
\be
Z+Z'+Z''=i\pi,\ \ \ \text{and}\ \ \ \lt\{Z,Z''\rt\}=\{Z',Z\}=\{Z'',Z'\}=1. 
\ee
If we extend the flat connection into the bulk, the moduli space of $\Slc$ flat connection on an ideal tetrahedron, denoted by $\cl_{\Delta}$, is isomorphic to a holomorphic Lagrangian submanifold in the phase space $\cp_{\partial\Delta}$. The Lagangian submanfold is defined by a holomorphic algebraic curve (see e.g. \cite{DGV}):
\be
\cl_{\Delta}\simeq\{z^{-1}+z''-1=0\}\subset \cp_{\partial\Delta}.
\ee

The graph complement 3-manifold $S^3\setminus\G_5$ can be decomposed into ideal octahedra, as it is discussed above. The geodesic boundary of an ideal octahedron is a sphere with 6 holes. The ideal triangulation of octahedron boundary provided by the octahedron consists of 8 octahedron edges. Thus The moduli space of $\Slc$ flat connection on a 6-holed sphere is of $\dim_\C=12$. By the same reason as the ideal tetrahedron, the monodromy around each hole on the boundary is unipotent, which gives 6 constraints. The phase space of an ideal octahedron is 
\be
\cp_{\partial\mathrm{oct}}=\lt\{\chi_{E=1,\cdots,8}\,\Big|\,\prod_{E\ \text{at hole}}(-\chi_E)=1\rt\}
\ee
which is of $\dim_{\C}=6$. As it is discussed, an ideal octahedron can be decomposed into 4 ideal tetrahedra, the phase space $\cp_{\partial\mathrm{oct}}$ can be obtained via a symplectic reduction from 4 copies of $\cp_{\partial\Delta}$. The edge coordinates of $\cp_{\partial\mathrm{oct}}$ can be expressed as a linear combination of the tetrahedron edge coordinates. In general for any edge on the boundary or in the bulk, it associates 
\be
x_E=\prod(z,z',z''\ \text{incident at}\ E)\ \ \ \text{or}\ \ \ X_E=\sum(Z,Z',Z''\ \text{incident at}\ E),
\ee 
being a product or sum over all the tetrahedron edge coordinates incident at the edge $E$. For a boundary edge, $X_E$ is the edge coordinates of the phase space. For a bulk edge, $x_E$ or $X_E$ is rather a constraint which is often denoted by $c_E$ or $C_E$, satisfying
\be
c_E=1\ \ \ \text{or}\ \ \ C_E=2\pi i, 
\ee
because the monodromy around a bulk edge is trivial \cite{DGG11,DGV}. We denotes the edge coordinates in 4 copies of $\cp_{\partial\Delta}$ by $X,Y,Z,W$ and their prime and double prime. All the edge coordinates are expressed in FIG.\ref{oct_coordinate}, where we have a single constraint
\be
C=X+Y+Z+W
\ee
We make a symplectic transformation in 4 copies of $\cp_{\partial\Delta}$ from the coordinates $(X,X")$,$(Y,Y'')$,$(Z,Z'')$,$(W,W'')$ to a set of new symplectic coordinates $(X,P_X),(Y,P_Y),(Z,P_Z),(C,\G)$, where
\be
P_X=X''-W'',\ \ P_Y=Y''-W'',\ \ P_Z=Z''-W'',\ \ \G=W''
\ee
and each pair are canonical conjugate variables, Poisson commutative with other pairs. The octahedron phase space $\cp_{\partial\mathrm{oct}}$ is a symplectic reduction by imposing the constraint $C=2\pi i$ and removing the ``gauge orbit'' variable $\G$. A set of symplectic coordinates of $\cp_{\partial\mathrm{oct}}$ are given by $(X,P_X),(Y,P_Y),(Z,P_Z)$.

Now we glue 5 ideal octahedra into $S^3\setminus\G_5$ as in FIG.\ref{5oct}. The phase space $\cp_{\partial(S^3\setminus\G_5)}$ can be obtained from the product of phase spaces $\cp_\times=(\cp_{\partial\Delta})^{\times 20}$ of 20 ideal tetrahedra, followed by a symplectic reduction with the 5 constraints $C_\a=2\pi i$ ($\a=1,\cdots,5$) in the 5 octahedra.

The geodesic boundary of $S^3\setminus\G_5$ consists of five 4-holed spheres, which are denoted by $\cs_{a=1,\cdots,5}$. In FIG.\ref{5oct}, each $\cs_a$ are made by the triangles from the geodesic boundaries of the octahedra. We compute all the edge coordinates on the geodesic boundary of $S^3\setminus\G_5$ in Table \ref{edges}
\begin{table}[h]
\caption{Edge coordinates of 4-holed spheres. Recall in FIG.\ref{5oct} that the octahedra are glued through the triangles labelled by $a,b,c,d,e,f,g,h,i,j$. Here e.g. $a_2'$ labels the triangles symmetric to the triangle $a$ with respect to the equator of Oct(2). The ``primed triangles'' with the primed labels triangulate the geodesic boundary of $S^3\setminus\G_5$. Here $X_{\a},Y_\a,Z_\a,W_\a$ ($\a=1,\cdots,5$) are the tetrahedron edge coordinates from the 4 tetrahedra triangulating Oct($a$).}
\begin{center}
\begin{tabular}{|c|c|c|}
\hline
$\cs_1$: & $h_2'\cap h_3':\ \ Z_2+Z_3$        & $h_3'\cap e_4':\ \ Y_3''+Z_3'+Z_4''+W_4'$\\
         & $h_2'\cap e_4':\ \ Z_2''+W_2'+Z_4$ & $h_3'\cap c_5':\ \ Z_3''+W_3'+Y_5''+Z_5'$\\
         & $h_2'\cap h_3':\ \ Y_2''+Z_2'+Z_5$ & $e_4'\cap c_5':\ \ Y''_4+Z_4'+Z_5''+W_5'$\\
\hline
$\cs_2$: & $i_1'\cap i_3':\ \ X_1''+Y_1'+X_3$ & $i_3'\cap f_4':\ \ X_3''+Y_3'+W_4''+X_4'$\\
         & $i_1'\cap f_4':\ \ X_1+X_4$        & $i_3'\cap b_5':\ \ W_3''+X_3'+X_5''+Y_5'$\\
         & $i_1'\cap b_5':\ \ W_1''+X_1'+X_5$ & $f_4'\cap b_5':\ \ X''_4+Y_4'+W_5''+X_5'$\\
\hline
$\cs_3$: & $b_1'\cap a_2':\ \ Z_1'+W_1''+X_2$        & $a_2'\cap d_4':\ \ W_2''+X_2'+Y_4'+Z_4''$\\
         & $b_1'\cap d_4':\ \ W'_1+X_1''+X_4'+Y_4''$ & $a_2'\cap d_5':\ \ X_2''+Y_2'+Z_5'+W_5''$\\
         & $b_1'\cap d_5':\ \ W_1+W_5'+X_5''$        & $d_4'\cap d_5':\ \ Y_4+W_5$\\
\hline
$\cs_4$: & $a_1'\cap c_2':\ \ Z_1+X_2'+Y_2''$        & $c_2'\cap j_3':\ \ Y_2'+Z_2''+Z_3'+W_3''$\\
         & $a_1'\cap j_3':\ \ Y''_1+Z_1'+W_3'+X_3''$ & $c_2'\cap j_5':\ \ Y_2+Y_5'+Z_5''$\\
         & $a_1'\cap j_5':\ \ Z_1''+W_1'+X_5'+Y_5''$ & $j_3'\cap j_5':\ \ W_3+Y_5$\\
\hline
$\cs_5$: & $f_1'\cap e_2':\ \ Y'_1+Z_2''+W_2'+X_2''$ & $e_2'\cap g_3':\ \ Z_2'+W_2''+Y_3'+Z_3''$\\
         & $f_1'\cap g_3':\ \ Y_1+X_3'+Y_3''       $ & $e_2'\cap g_4':\ \ W_2+Z_4'+W_4''$\\
         & $f_1'\cap g_4':\ \ X_1'+Y_1''+W_4'+X_4''$ & $g_3'\cap g_4':\ \ Y_3+W_4$\\
\hline
\end{tabular}
\end{center}
\label{edges}
\end{table}%

In our discussion, it turns out to be convenient to use complex Fenchel-Nielsen (FN) coordinates \cite{kabaya,DGV} for the boundary phase space $\cp_{\partial(S^3\setminus\G_5)}$. The complex FN length variables $\l_{ab}=e^{\L_{ab}}$ are simply the eigenvalues of monodromies meridian to the 10 annuli (cusp boundaries) $\ell_{ab}$ connecting 4-holed spheres $\cs_a$ and $\cs_b$. They relate the above edge coordinates by $2\L=\sum_{E\text{ around hole}}(X_E-i\pi)$ \cite{FG03,DGV}. The resulting 10 complex FN length variables are listed in the following, expressed in terms of $(X_a,P_{X_a}),(Y_a,P_{Y_a}),(Z_a,P_{Z_a}),(C_a,\G_a)$ from Oct($a$). 
\be
\L_{12}&=&\frac{1}{2}\lt(-C_3-C_4-C_5+P_{Y_3}+P_{Y_4}+P_{Y_5}+X_3+X_4+X_5+Y_3+Y_4+Y_5+3 i \pi\rt)\\
\L_{13}&=&\frac{1}{2}\lt(-{C_2}-{C_5}+P_{Y_2}+P_{Y_4}-P_{Z_4}+P_{Z_5}+X_2+X_5+Y_2+Y_5+2 Z_5+i \pi \rt) \\
\L_{14}&=&\frac{1}{2} \lt(-{C_3}+P_{Y_2}+P_{Y_5}-P_{Z_2}+P_{Z_3}-P_{Z_5}+X_3+Y_3+2 Z_3\rt) \\
\L_{15}&=&\frac{1}{2}\lt(-C_2-C_4+P_{Y_3}+P_{Z_2}-P_{Z_3}+P_{Z_4}+X_2+X_4+Y_2+Y_4+2 Z_2+2 Z_4\rt) \\
\L_{23}&=&\frac{1}{2} \lt(-P_{X_1}+P_{X_4}-P_{X_5}-P_{Y_4}+X_4-Y_4\rt) \\
\L_{24}&=&\frac{1}{2} \lt(-P_{X_3}+P_{X_5}-P_{Y_1}-P_{Y_5}-X_1+X_5-Y_1-Y_5+i \pi \rt) \\
\L_{25}&=&\frac{1}{2} \lt(P_{X_1}+P_{X_3}-P_{X_4}-P_{Y_1}-P_{Y_3}+X_1+X_3-Y_1-Y_3\rt)\\
\L_{34}&=&\frac{1}{2}\lt(C_1-C_5+P_{X_2}+P_{X_5}-P_{Y_2}-P_{Z_1}-P_{Z_5}-X_1+X_2+X_5-Y_1-Y_2+Y_5-2 Z_1+i \pi \rt) \\
\L_{35}&=&\frac{1}{2}\lt(-{C_1}+P_{X_1}-P_{X_2}-P_{X_4}-P_{Z_1}+P_{Z_4}+X_1-X_4+Y_1-Y_4+ 2 i \pi \rt) \\
\L_{45}&=&\frac{1}{2}\lt(-C_3-P_{X_2}+P_{X3}+P_{Y_1}-P_{Z_1}+P_{Z_2}-P_{Z_3}-X_2+X_3-Y_2+Y_3+2 i \pi \rt).
\ee
The above $\L_{ab}$ are mutually Poisson commutative and commuting with all the edge coordinates in Table \ref{edges}. 

The definition of complex FN twist variable $\t_{ab}=e^{\ct_{ab}}$ depends on a choice of longitude path along each annulus, traveling from $\cs_a$ to $\cs_b$ (see \cite{DGV} for details). Here we make a simple choice by drawing each path in a cusp boundary component of a \emph{single} octahedron, since there is always a piece of the annulus being a cusp boundary component of a octahedron, which connects a pair of triangles respectively in $\cs_a$ and $\cs_b$. See Table \ref{twists}. 

\begin{table}[h]
\caption{The twist variables and the corresponding octahedron whose cusp boundary component contains the chosen path}
\begin{center}
\begin{tabular}{|c|c|c|c|c|c|c|c|c|c|c|}
\hline
FN twist & $\t_{12}$ & $\t_{13}$ & $\t_{14}$ & $\t_{15}$ & $\t_{23}$ & $\t_{24}$ & $\t_{25}$ & $\t_{35}$ & $\t_{45}$\\
\hline
Octahedron & Oct(3) & Oct(5) & Oct(2) & Oct(4) & Oct(4) & Oct(5) & Oct(1) & Oct(2) & Oct(3) \\
\hline
\end{tabular}
\end{center}
\label{twists}
\end{table}%
 
The resulting 10 FN complex twists are listed in the following:
\be
\ct_{12}&=&P_{X_3}-P_{Z_3}-Y_3-Z_3+i \pi  \\
\ct_{13}&=& P_{Z_5} \\
\ct_{14}&=& -P_{Y_2}+P_{Z_2}-Y_2+Z_2 \\
\ct_{15}&=& P_{Z_4} \\
\ct_{23}&=& P_{X_4}-P_{Y_4} \\
\ct_{24}&=& P_{X_5}-P_{Y_5} \\
\ct_{25}&=& P_{X_1}-P_{Y_1} \\
\ct_{34}&=& -P_{Z_1} \\
\ct_{35}&=& -{C_2}+P_{X_2}+2 X_2+Y_2+Z_2 \\
\ct_{45}&=& P_{Y_3}+Y_3+Z_3-i \pi.  
\ee 
$\ct_{ab}$ are mutually Poisson commutative, and satisfying $\{\L_{ab},\ct_{cd}\}=\delta_{ab,cd}$.

The twist variable $\ct_{ab}$ commutes with the edge coordinates in Table \ref{edges}, except for the edges of the pair of triangles connected by the path defining $\ct_{ab}$ \cite{DV14}. Therefore for any 4-holed sphere $\cs_a$, the 6 edge coordinates $X_E$ are possibly noncommutative with 4 twists $\ct_{ab}$ ($b\neq a$). We consider a linear combination anzatz $F=\sum_E c_E X_E$ and assume $F$ is commutative with the 4 twists, i.e.
\be
\sum_E c_E\{X_E,\ct_{ab}\}=0\ \ \ \ b\neq a
\ee
which gives 4 linear equations for 6 unknown $c_E$ of $\cs_a$. There are 2 linear independent solutions $\{c_E^{(1)}\}$ and $\{c_E^{(2)}\}$, we define 
\be
M_a=\sum_E c_E^{(1)} X_E,\ \ \ \ P_a=\sum_E c_E^{(2)} X_E.\label{MP}
\ee
which turn out to satisfy $\{M_a,P_{b}\}=\delta_{ab}, \{M_a,M_{b}\}=\{P_a,P_{b}\}=0$, and Poisson commute with all $\L_{ab}$ and $\ct_{ab}$. It turns out that $c_E=\pm\half$. Explicitly $M_{a=1,\cdots,5}$ are given by
\be
M_1&=& \frac{1}{2} \big(-{C_2}-{C_3}+C_4+P_{Y_2}-P_{Y_3}+P_{Y_5}+2P_{Z_3}-P_{Z_4}-P_{Z_5}+X_2\nonumber\\
&&\ +X_3-X_4+Y_2+Y_3-Y_4+2Z_3+2 i \pi \big) \\
M_2&=&\frac{1}{2} \lt(P_{X_1}-2 P_{X_3}+P_{X_4}+P_{X_5}+P_{Y_3}-P_{Y_5}-X_3-X_5+Y_3-Y_5-i \pi \rt)   \\
M_3&=& \frac{1}{2} \lt(-{C_1}+P_{X_2}-P_{X_5}+P_{Y_4}-P_{Z_1}-P_{Z_4}+X_1+2 X_2+Y_1-2 i \pi \rt) \\
M_4&=& \frac{1}{2}\big(-C_1-C_3+P_{X_3}-P_{X_5}+P_{Y_1}+P_{Y_2}-P_{Z_2}+P_{Z_3}+P_{Z_5}+X_1 \nonumber\\
&&\ +X_3 - X_5+ Y_1+2Y_2+Y_3-Y_5+2 Z_3+3 i \pi \big) \\
M_5&=& \frac{1}{2} \big(C_2-C_4-P_{X_1}+P_{X_3}+P_{X_4}+P_{Y_1}-2P_{Y_3}-P_{Z_2}+P_{Z_3}-P_{Z_4}\nonumber\\
&&\ -X_1-X_2+X_3+X_4-Y_1-Y_2-Y_3+Y_4-2Z_2+4 i \pi \big), 
\ee
and $P_{a=1,\cdots,5}$ are given by
\be
P_1&=& \frac{1}{2}\big({C_4}+{C_5}+P_{Y_2}-P_{Y_3}-P_{Y_4}-P_{Z_2}+P_{Z_3}-P_{Z_5}-X_4-X_5-Y_4\nonumber\\
&&\ -Y_5-2 Z_2-3 i \pi \big) \\
P_2&=& \frac{1}{2}\lt(P_{X_3}-P_{X_5}-P_{Y_1}-P_{Y_3}-P_{Y_4}-X_1+X_3-X_4-Y_1-Y_3-Y_4+6 i \pi \rt) \\
P_3&=& \frac{1}{2}\big(-C_5+P_{X_1}-P_{X_4}+P_{X_5}-P_{Y_2}+P_{Z_4}-P_{Z_5}-X_2-X_4+X_5\nonumber\\
&&\ -Y_2-Y_4+Y_5+7 i \pi \big) \\
P_4&=& \frac{1}{2}\big(C_1+C_3-P_{X_2}+P_{X_5}-P_{Y_5}-P_{Z_1}+P_{Z_2}-P_{Z_3}-X_1-X_2-X_3\nonumber\\
&&\ +X_5-Y_1-Y_2-Y_3+Y_5-2 Z_3+i \pi \big) \\
P_5&=& \frac{1}{2}\big(-C_2-P_{X_1}+P_{X_2}+P_{X_4}-P_{Y_3}+P_{Z_1}-P_{Z_2}+P_{Z_3}-X_1+X_2\nonumber\\
&&\ -Y_1+Y_2+6 i \pi \big) 
\ee
$(M_a,P_a)$ are the symplectic variables parametrizing the $\Slc$ flat connections on 4-holed sphere $\cs_a$ with fixed conjugacy classes $\l_{ab}$. 

In the following we simply set $C_\a=2\pi i$ in the definition of $\L_{ab},\ct_{ab},M_a,P_a$ to remove their $C_\a$ dependence. Here we consider $\L_{ab}$ and $M_a$ to be the position variables, while $\ct_{ab}$ and $P_a$ are considered to be the momentum variables, i.e.
\be
m_I=(\L_{ab},M_a),\ \ \ \ p_I=(\ct_{ab},P_a).
\ee 
satisfying the standard Poisson brackets $\{m_I,p_J\}=\delta_{IJ}$, $\{m_I,m_J\}=\{p_I,p_J\}=0$.

We obtain the $\mathbf{Sp}(30,\mathbb{Q})$ matrix transforming from $\vec{\varphi}\equiv(X_\a,Y_\a,Z_\a)_{\a=1}^5$ and $\vec{\chi}\equiv(P_{X_\a},P_{Y_\a},P_{Z_\a})_{\a=1}^5$ to $\vec{m}$ and $\vec{p}$
\be
\left(
\begin{array}{c}
\vec{m}\\ \vec{p}
\end{array}
\right)=\left(
\begin{array}{cc}
\mathbf{A} & \mathbf{B}\\ \mathbf{C} & \mathbf{D}
\end{array}
\right)\left(
\begin{array}{c}
\vec{\varphi}\\ \vec{\chi}
\end{array}\right)+i\pi\left(
\begin{array}{c}
\vec{\nu}_1\\ \vec{\nu}_2
\end{array}
\right)\label{ABCD}
\ee
where $\vec{\nu}_1,\vec{\nu}_2$ are 15-dimensional vectors with $\Z/2$ entries. 
\be
\vec{\nu}_1&=&(-\frac{3}{2}, -\frac{3}{2}, -1, -2, 0, \frac{1}{2}, 0, \frac{1}{2}, 0, 0, 0, -\frac{1}{2}, -2 , -\frac{1}{2}, 2 )^T\nonumber\\
\vec{\nu}_2&=&(1, 0, 0, 0, 0, 0, 0, 0, -2, -1, \frac{1}{2}, 3 , \frac{5}{2}, \frac{5}{2}, 2 )^T
\ee
The $15\times 15$ matrix blocks $\mathbf{A},\mathbf{B},\mathbf{C},\mathbf{D}$ are given by 
\be
\mathbf{A}=\left(
\begin{array}{ccccccccccccccc}
 0 & 0 & 0 & 0 & 0 & 0 & \frac{1}{2} & \frac{1}{2} & 0 & \frac{1}{2} & \frac{1}{2} & 0 & \frac{1}{2} &
   \frac{1}{2} & 0 \\
 0 & 0 & 0 & \frac{1}{2} & \frac{1}{2} & 0 & 0 & 0 & 0 & 0 & 0 & 0 & \frac{1}{2} & \frac{1}{2} & 1 \\
 0 & 0 & 0 & 0 & 0 & 0 & \frac{1}{2} & \frac{1}{2} & 1 & 0 & 0 & 0 & 0 & 0 & 0 \\
 0 & 0 & 0 & \frac{1}{2} & \frac{1}{2} & 1 & 0 & 0 & 0 & \frac{1}{2} & \frac{1}{2} & 1 & 0 & 0 & 0 \\
 0 & 0 & 0 & 0 & 0 & 0 & 0 & 0 & 0 & \frac{1}{2} & -\frac{1}{2} & 0 & 0 & 0 & 0 \\
 -\frac{1}{2} & -\frac{1}{2} & 0 & 0 & 0 & 0 & 0 & 0 & 0 & 0 & 0 & 0 & \frac{1}{2} & -\frac{1}{2} & 0 \\
 \frac{1}{2} & -\frac{1}{2} & 0 & 0 & 0 & 0 & \frac{1}{2} & -\frac{1}{2} & 0 & 0 & 0 & 0 & 0 & 0 & 0 \\
 -\frac{1}{2} & -\frac{1}{2} & -1 & \frac{1}{2} & -\frac{1}{2} & 0 & 0 & 0 & 0 & 0 & 0 & 0 & \frac{1}{2} &
   \frac{1}{2} & 0 \\
 \frac{1}{2} & \frac{1}{2} & 0 & 0 & 0 & 0 & 0 & 0 & 0 & -\frac{1}{2} & -\frac{1}{2} & 0 & 0 & 0 & 0 \\
 0 & 0 & 0 & -\frac{1}{2} & -\frac{1}{2} & 0 & \frac{1}{2} & \frac{1}{2} & 0 & 0 & 0 & 0 & 0 & 0 & 0 \\
 0 & 0 & 0 & \frac{1}{2} & \frac{1}{2} & 0 & \frac{1}{2} & \frac{1}{2} & 1 & -\frac{1}{2} & -\frac{1}{2} & 0 &
   0 & 0 & 0 \\
 0 & 0 & 0 & 0 & 0 & 0 & -\frac{1}{2} & \frac{1}{2} & 0 & 0 & 0 & 0 & -\frac{1}{2} & -\frac{1}{2} & 0 \\
 \frac{1}{2} & \frac{1}{2} & 0 & 1 & 0 & 0 & 0 & 0 & 0 & 0 & 0 & 0 & 0 & 0 & 0 \\
 \frac{1}{2} & \frac{1}{2} & 0 & 0 & 1 & 0 & \frac{1}{2} & \frac{1}{2} & 1 & 0 & 0 & 0 & -\frac{1}{2} &
   -\frac{1}{2} & 0 \\
 -\frac{1}{2} & -\frac{1}{2} & 0 & -\frac{1}{2} & -\frac{1}{2} & -1 & \frac{1}{2} & -\frac{1}{2} & 0 &
   \frac{1}{2} & \frac{1}{2} & 0 & 0 & 0 & 0 \\
\end{array},
\right)\ee

\be
\mathbf{B}=\left(
\begin{array}{ccccccccccccccc}
 0 & 0 & 0 & 0 & 0 & 0 & 0 & \frac{1}{2} & 0 & 0 & \frac{1}{2} & 0 & 0 & \frac{1}{2} & 0 \\
 0 & 0 & 0 & 0 & \frac{1}{2} & 0 & 0 & 0 & 0 & 0 & \frac{1}{2} & -\frac{1}{2} & 0 & 0 & \frac{1}{2} \\
 0 & 0 & 0 & 0 & \frac{1}{2} & -\frac{1}{2} & 0 & 0 & \frac{1}{2} & 0 & 0 & 0 & 0 & \frac{1}{2} & -\frac{1}{2}
   \\
 0 & 0 & 0 & 0 & 0 & \frac{1}{2} & 0 & \frac{1}{2} & -\frac{1}{2} & 0 & 0 & \frac{1}{2} & 0 & 0 & 0 \\
 -\frac{1}{2} & 0 & 0 & 0 & 0 & 0 & 0 & 0 & 0 & \frac{1}{2} & -\frac{1}{2} & 0 & -\frac{1}{2} & 0 & 0 \\
 0 & -\frac{1}{2} & 0 & 0 & 0 & 0 & -\frac{1}{2} & 0 & 0 & 0 & 0 & 0 & \frac{1}{2} & -\frac{1}{2} & 0 \\
 \frac{1}{2} & -\frac{1}{2} & 0 & 0 & 0 & 0 & \frac{1}{2} & -\frac{1}{2} & 0 & -\frac{1}{2} & 0 & 0 & 0 & 0 &
   0 \\
 0 & 0 & -\frac{1}{2} & \frac{1}{2} & -\frac{1}{2} & 0 & 0 & 0 & 0 & 0 & 0 & 0 & \frac{1}{2} & 0 &
   -\frac{1}{2} \\
 \frac{1}{2} & 0 & -\frac{1}{2} & -\frac{1}{2} & 0 & 0 & 0 & 0 & 0 & -\frac{1}{2} & 0 & \frac{1}{2} & 0 & 0 &
   0 \\
 0 & \frac{1}{2} & -\frac{1}{2} & -\frac{1}{2} & 0 & \frac{1}{2} & \frac{1}{2} & 0 & -\frac{1}{2} & 0 & 0 & 0
   & 0 & 0 & 0 \\
 0 & 0 & 0 & 0 & \frac{1}{2} & 0 & 0 & -\frac{1}{2} & 1 & 0 & 0 & -\frac{1}{2} & 0 & \frac{1}{2} &
   -\frac{1}{2} \\
 \frac{1}{2} & 0 & 0 & 0 & 0 & 0 & -1 & \frac{1}{2} & 0 & \frac{1}{2} & 0 & 0 & \frac{1}{2} & -\frac{1}{2} & 0
   \\
 0 & 0 & -\frac{1}{2} & \frac{1}{2} & 0 & 0 & 0 & 0 & 0 & 0 & \frac{1}{2} & -\frac{1}{2} & -\frac{1}{2} & 0 &
   0 \\
 0 & \frac{1}{2} & 0 & 0 & \frac{1}{2} & -\frac{1}{2} & \frac{1}{2} & 0 & \frac{1}{2} & 0 & 0 & 0 &
   -\frac{1}{2} & 0 & \frac{1}{2} \\
 -\frac{1}{2} & \frac{1}{2} & 0 & 0 & 0 & -\frac{1}{2} & \frac{1}{2} & -1 & \frac{1}{2} & \frac{1}{2} & 0 &
   -\frac{1}{2} & 0 & 0 & 0 \\
\end{array}
\right),
\ee

\be
\mathbf{D}=\left(
\begin{array}{ccccccccccccccc}
 0 & 0 & 0 & 0 & 0 & 0 & 1 & 0 & -1 & 0 & 0 & 0 & 0 & 0 & 0 \\
 0 & 0 & 0 & 0 & 0 & 0 & 0 & 0 & 0 & 0 & 0 & 0 & 0 & 0 & 1 \\
 0 & 0 & 0 & 0 & -1 & 1 & 0 & 0 & 0 & 0 & 0 & 0 & 0 & 0 & 0 \\
 0 & 0 & 0 & 0 & 0 & 0 & 0 & 0 & 0 & 0 & 0 & 1 & 0 & 0 & 0 \\
 0 & 0 & 0 & 0 & 0 & 0 & 0 & 0 & 0 & 1 & -1 & 0 & 0 & 0 & 0 \\
 0 & 0 & 0 & 0 & 0 & 0 & 0 & 0 & 0 & 0 & 0 & 0 & 1 & -1 & 0 \\
 1 & -1 & 0 & 0 & 0 & 0 & 0 & 0 & 0 & 0 & 0 & 0 & 0 & 0 & 0 \\
 0 & 0 & -1 & 0 & 0 & 0 & 0 & 0 & 0 & 0 & 0 & 0 & 0 & 0 & 0 \\
 0 & 0 & 0 & 1 & 0 & 0 & 0 & 0 & 0 & 0 & 0 & 0 & 0 & 0 & 0 \\
 0 & 0 & 0 & 0 & 0 & 0 & 0 & 1 & 0 & 0 & 0 & 0 & 0 & 0 & 0 \\
 0 & 0 & 0 & 0 & \frac{1}{2} & -\frac{1}{2} & 0 & -\frac{1}{2} & \frac{1}{2} & 0 & -\frac{1}{2} & 0 & 0 & 0 &
   -\frac{1}{2} \\
 0 & -\frac{1}{2} & 0 & 0 & 0 & 0 & \frac{1}{2} & -\frac{1}{2} & 0 & 0 & -\frac{1}{2} & 0 & -\frac{1}{2} & 0 &
   0 \\
 \frac{1}{2} & 0 & 0 & 0 & -\frac{1}{2} & 0 & 0 & 0 & 0 & -\frac{1}{2} & 0 & \frac{1}{2} & \frac{1}{2} & 0 &
   -\frac{1}{2} \\
 0 & 0 & -\frac{1}{2} & -\frac{1}{2} & 0 & \frac{1}{2} & 0 & 0 & -\frac{1}{2} & 0 & 0 & 0 & \frac{1}{2} &
   -\frac{1}{2} & 0 \\
 -\frac{1}{2} & 0 & \frac{1}{2} & \frac{1}{2} & 0 & -\frac{1}{2} & 0 & -\frac{1}{2} & \frac{1}{2} &
   \frac{1}{2} & 0 & 0 & 0 & 0 & 0 \\
\end{array}
\right).
\ee
$\mathbf{C}$ is determined by $\mathbf{D}^T\mathbf{A}=\mathbf{I}+\mathbf{B}^T\mathbf{C}$. Here the block matrix $\mathbf{B}$ is invertible. Both $\mathbf{A}\mathbf{B}^T$ and $\mathbf{D}\mathbf{B}^{-1}$ are symmetric matrics.

In addition, the expressions of $\L_{ab},M_a;\ct_{ab},P_a$ don't involve $\G_\a$ (the conjugate variable to the constraint $C_\a$ of each octahedron). $C_\a$ has been set to be $2\pi i$ in the definitions of $\L_{ab},M_a;\ct_{ab},P_a$. So the symplectic transformation in the $(C_\a,\G_a)$-subspace is trivial. $C_\a,\G_\a$ still survive as the symplectic coordinates of $\cp_{\times}$. The collection of $(\L_{ab},M_a,C_\a;\ct_{ab},P_a,\G_\a)$ is a complete set of symplectic coordinates of $\cp_{\times}$. In the coordinate system, the symplectic reduction from $\cp_\times$ to $\cp_{\partial(S^3\setminus\G_5)}$ is simply the removal of the coordinates $C_\a,\G_\a$. The Atiayh-Bott-Goldmann symplectic form on $\cp_{\partial(S^3\setminus\G_5)}$ then reduces to
\be
\O=\sum_{a<b}\rmd \ct_{ab}\wedge \rmd \L_{ab}+\sum_{a=1}^5\rmd P_a\wedge \rmd M_a.
\ee

\section{3-dimensional Supersymmetric Gauge Theories Labelled by 3-Manifolds}\label{3d/3dduality}

\subsection{Supersymmetric Gauge Theory Corresponding to $S^3\setminus\G_5$}\label{SUSYgraphcomplement}

Here we apply the construction by Dimofte, Gaiotto, and Gukov in \cite{DGG11} to define a 3-dimensional $\cn=2$ supersymmetric gauge theory $T_{S^3\setminus\G_5}$ labelled by the graph complement 3-manifold $S^3\setminus\G_5$. 

3-dimensional $\cn=2$ chiral multiplet and vector multiplet can be obtained by the dimensional reduction from 4-dimensional $\cn=1$ chiral and vector. The SUSY algebra in 3d from the dimensional reduction gives ($\mu=0,1,2$)
\be
\{Q_\a,\bar{Q}_{\dot{\a}}\}=2\sig^\mu_{\a\dot{\a}}P_\mu+2\sig^3_{\a\dot{\a}}Z\label{susyalg}
\ee
where the central charge $Z=P_3$, and the reduction is along $x^3$-direction. In superspace language and in WZ-gauge
\be
\Phi(y,\theta)&=&\phi(y)+\sqrt{2}\theta^\a\psi_\a(y)+\theta^\a\theta_\a F(y),\ \ \ \ y^\mu=x^\mu+i\theta^\a\sig^\mu_{\a\dot{\a}}\bar{\theta}^{\dot{\a}}\nonumber\\
V(x,\theta,\bar{\theta})&=&-\theta^\a\sig^\mu_{\a\dot{\a}}\bar{\theta}^{\dot{\a}}A_\mu(x)-\sig^{3}_{\a\dot{\a}}\theta^\a\bar{\theta}^{\dot{\a}}\sig^{\mathrm{3d}}(x)+i\theta^\a\theta_\a\bar{\theta}_{\dot{\a}}\bar{\l}^{\dot{\a}}(x)-i\bar{\theta}_{\dot{\a}}\bar{\theta}^{\dot{\a}}{\theta}^\a{\l}_\a(x)+\half\theta^\a\theta_\a\bar{\theta}_{\dot{\a}}\bar{\theta}^{\dot{\a}}D(x).
\ee	
The vector multiplet contains a real scalar $\sig^{\mathrm{3d}}(x)$ in addition to the gauge field $A_\mu(x)$ and two Majorana fermion. By dimensional reduction from 4-dimensions, $\sig^{\mathrm{3d}}(x)$ comes from the component of 4d gauge field along the direction of reduction ($x^3$-direction). Here we only consider the vector multiplet with abelian gauge group U(1). In 3-dimensions, $\cn=2$ vector multiplet can be dualized to a linear multiplet \cite{3dSUSY}
\be
\Sig=\bar{D}^{\a} D_\a V=-2\sig^{\mathrm{3d}} -2i\bar{\theta}^\a\l_\a+2i\l^\a\theta_\a-2\theta^\a\sig^\mu_{\a\dot{\a}}\bar{\theta}^{\dot{\a}}\eps_{\mu\nu\rho}\partial^\nu A^\rho+2\bar{\theta}^\a\theta_\a D+ o(\theta^2\bar{\theta},\bar{\theta}^2\theta).
\ee
where e.g. $\bar{\theta}^\a\l_\a\equiv\sig^3_{\dot{\a}\a}\bar{\theta}^{\dot{\a}}\l^\a$.

In Dimofte-Gaiotto-Gukov (DGG) construction, the field theory $T_\Delta$ corresponding to an ideal tetrahedron is a single chiral multiplet coupled to a background U(1) gauge field, with a level $1/2$ Chern-Simons term:
\be
\Fl_{\Delta}[V_Z]=\int\rmd^4\theta\ \Phi_Z^\dagger e^{V_Z}\Phi_Z+\frac{(-1/2)}{4\pi}\int\rmd^4\theta\ \Sig_Z V_Z,\label{chiral}
\ee
where only the chiral multiplet $\Phi_Z$ is dynamical. The level $1/2$ Chern-Simons term cancels the anomaly generated by the gauge coupling of $\Phi_Z$ \cite{3dSUSY}. In defining $T_\Delta$, a canonical conjugate pair, or namely a polarization, has been chosen to be $(Z,Z'')$ in the tetrahedron phase space $\cp_{\partial\Delta}$. The chiral multiplet $\Phi_Z$ is associated with the chosen polarization. The R-charge of $\Phi_Z$ is assigned to be $\mathrm{Im}(Z)/\pi$. The 3d supersymmetric gauge theories considered here always preserve the U(1) R-symmetry. 

Given a 3-manifold obtained by gluing $N$ ideal tetrahedron, as a intermediate step we consider $N$ copies of tetrahedron theory $T_\Delta$:
\be
\Fl_{\{\Delta_i\}_{i=1}^N}[\vec{V}_{Z}]=\sum_{i=1}^N\lt[\int\rmd^4\theta\ \Phi_{Z_i}^\dagger e^{V_{Z_i}}\Phi_{Z_i}+\frac{(-1/2)}{4\pi}\int\rmd^4\theta\ \Sig_{Z_i} V_{Z_i}\rt]. 
\ee
For an ideal octahedron with $N=4$, we need a symplectic transformation which changes the tetrahedron polarizations $(X,X")$,$(Y,Y'')$,$(Z,Z'')$,$(W,W'')$ to the polarization of $\cp_{\partial\mathrm{Oct}}$ (plus the constraint and gauge orbit) $(X,P_X),(Y,P_Y),(Z,P_Z),(C,\G)$. The symplectic transformation is of ``type-GL'' in \cite{DGG11} 
\be
\left(
\begin{array}{cc}
U & 0 \\
0 & (U^{-1})^T
\end{array}
\right),\ \ \ \ U=\left(
\begin{array}{cccc}
1 & 0 & 0 & 0 \\
0 & 1 & 0 & 0 \\
0 & 0 & 1 & 0 \\
1 & 1 & 1 & 1
\end{array}
\right).
\ee
Type-GL symplectic transformation corresponds to the following operation on the supersymmetric field theory by DGG-construction:
\be
\Fl_{\{\Delta_i\}_{i=1}^N}[\vec{V}_{Z}]\mapsto \Fl'_{\{\Delta_i\}_{i=1}^N}[\vec{V}^{\mathrm{new}}_{Z}]:=\Fl_{\{\Delta_i\}_{i=1}^N}[U^{-1}\vec{V}^{\mathrm{new}}_{Z}]
\ee
Imposing constraint $C=2\pi i$ corresponds to adding a superpotential $W$ to the field theory. In the case of octahedron, $\cw=\Phi_X\Phi_Y\Phi_Z\Phi_W$ so that $\cw$ has R-charge 2 by $C=X+Y+Z+W=2\pi i$. As a result the supersymmetic field theory corresponding to an ideal octahedron is 
\be
&&\Fl_{\mathrm{Oct}}\lt[V_X,V_Y,V_Z,V_C\rt]\nonumber\\
&=&\sum_{n=X,Y,Z}\lt[\int\rmd^4\theta\ \Phi_{n}^\dagger e^{V_{n}}\Phi_{n}+\frac{(-1/2)}{4\pi}\int\rmd^4\theta\ \Sig_{n} V_{n}\rt]+ \int \rmd^2\theta\ \Phi_X\Phi_Y\Phi_Z\Phi_W+\nonumber\\
&&+ \int\rmd^4\theta\ \Phi_{W}^\dagger e^{V_{C}-V_X-V_Y-V_Z}\Phi_{W}+\frac{(-1/2)}{4\pi}\int\rmd^4\theta \lt(\Sig_{C}-\Sig_X-\Sig_Y-\Sig_Z\rt) \lt(V_{C}-V_X-V_Y-V_Z\rt)
\ee
The superpotential breaks the U(1) symmetry coupled to $V_C$, which forces $V_C=0$. The resulting theory $T_{\mathrm{Oct}}$ has the flavor symmetry $U(1)_X\times U(1)_Y\times U(1)_Z$ coupled to external gauge fields $V_X,V_Y,V_Z$. As a general results from DGG-construction, the flavor symmetry of the resulting 3d supersymmetric gauge theory labelled by a 3-manifold $M_3$ is $U(1)^{\half\dim\cp_{\partial M_3}}$, where $\cp_{\partial M_3}$ is the moduli space of $\Slc$ flat connection on the boundary of $M_3$.

The graph complement $S^3\setminus\G_5$ is made by 5 ideal octahedra. The corresponding supersymmetric field theory is given by 5 copies of $T_{\mathrm{Oct}}$ followed by a sequence of operations, which corresponds to the symplectic transformations in $\cp_{\partial(S^3\setminus\G_5)}$. We denote the sequence of external gauge fields by ${V}_I=\{V_{X_\a},V_{Y_\a},V_{Z_\a}\}_{\a=1}^5$, which is a 15-dimensional vector. 5 copies of $T_{\mathrm{Oct}}$ can be written as
\be
\Fl_{\mathrm{Oct}^5}\lt[{V}_I\rt]&=&\sum_{M=1}^{20}\lt[\int\rmd^4\theta\ \Phi_{M}^\dagger e^{\mathbf{K}_{MI}V_I}\Phi_{M}+\frac{(-1/2)}{4\pi}\int\rmd^4\theta\lt(\mathbf{K}_{MI'}\Sig_{I'} \rt)\lt(\mathbf{K}_{MI}V_I\rt)\rt]+\nonumber\\
&&\ + \sum_{\a=1}^5\int \rmd^2\theta\ \Phi_{X_\a}\Phi_{Y_\a}\Phi_{Z_\a}\Phi_{W_\a}
\ee
where the indices $I,I'=1,\cdots,15$, and repeating the indices $I,I'$ means the summation over $I$ and $I'$. Here $\mathbf{K}$ is a direct sum
\be
\mathbf{K}=\oplus_{\a=1}^5 K_{(\a)}, \ \ \ \ 
K_{(\a)}=\left(
\begin{array}{cccc}
1 & 0 & 0  \\
0 & 1 & 0  \\
0 & 0 & 1  \\
-1 & -1 & -1 
\end{array}
\right)
\ee
where the matrix $K_{(\a)}$ acts on $(V_{X_\a},V_{Y_\a},V_{Z_\a})$ in the subspace labelled by $\a$. $\Fl_{\mathrm{Oct}^5}\lt[{V}_I\rt]$ is simply 5 copies of $\Fl_{\mathrm{Oct}}$ with $V_{C_\a}$ set to be zero.

The $\mathbf{Sp}(30,\Z/2)$ matrix in Eq.\Ref{ABCD} can be decomposed into a sequence of elementary symplectic transformations \cite{hua,DG12}:
\be
\lt(\begin{array}{cc}
\mathbf{A} & \mathbf{B}\\ \mathbf{C} & \mathbf{D}
\end{array}
\right)=
\lt(\begin{array}{cc}
\mathbf{I} & \mathbf{0}\\ \mathbf{D}\mathbf{B}^{-1} & \mathbf{I}
\end{array}
\right)
\lt(\begin{array}{cc}
\mathbf{0} & \mathbf{I}\\ -\mathbf{I} & \mathbf{0}
\end{array}\rt)
\lt(\begin{array}{cc}
\mathbf{I} & \mathbf{0}\\ \mathbf{A}\mathbf{B}^T & \mathbf{I}
\end{array}
\right)
\lt(\begin{array}{cc}
(\mathbf{B}^{-1})^T & \mathbf{0}\\ \mathbf{0} & \mathbf{B}
\end{array}
\right).\label{4steps}
\ee
Each step in the above corresponds to a symplectic operation on the supersymmetric field theory \cite{DGG11,Witten:2003ya}. 

\begin{enumerate}
\item ``{GL-type}''
\be
\lt(\begin{array}{cc}
(\mathbf{B}^{-1})^T & \mathbf{0}\\ \mathbf{0} & \mathbf{B}
\end{array}
\right):\ \ \ \ \ \Fl_{\mathrm{Oct}^5}\lt[\vec{V}\rt]\mapsto \Fl_1\lt[\vec{V}^{\text{new}}\rt]=\Fl_{\mathrm{Oct}^5}\lt[B^T\vec{V}^{\text{new}}\rt].\label{GLtypetrans}
\ee 
$\vec{V}^{\text{new}}$ are the background gauge fields associated with the polarization after the symplectic transformation. The Lagrangian $\Fl_1\lt[\vec{V}\rt]$ reads
\be
\Fl_1\lt[\vec{V}\rt]&=&\sum_{M=1}^{20}\lt[\int\rmd^4\theta\ \Phi_{M}^\dagger e^{\mathbf{K}_{MI}\mathbf{B}^T_{IJ}V_J}\Phi_{M}+\frac{(-1/2)}{4\pi}\int\rmd^4\theta\lt(\mathbf{K}_{MI'}\mathbf{B}^T_{I'J'}\Sig_{J'} \rt)\lt(\mathbf{K}_{MI}\mathbf{B}^T_{IJ}V_J\rt)\rt]+\nonumber\\
&&\ + \sum_{\a=1}^5\int \rmd^2\theta\ \Phi_{X_\a}\Phi_{Y_\a}\Phi_{Z_\a}\Phi_{W_\a}.
\ee

\item ``T-type''
\be
\lt(\begin{array}{cc}
\mathbf{I} & \mathbf{0}\\ \mathbf{A}\mathbf{B}^T & \mathbf{I}
\end{array}
\right):\ \ \ \ \ 
\Fl_1\lt[\vec{V}\rt]\mapsto \Fl_2[\vec{V}^{\text{new}}]=\Fl_1\lt[\vec{V}^{\text{new}}\rt]+\frac{1}{4\pi}\int\rmd^4\theta \ \vec{\Sig}^{\text{new}}\cdot\mathbf{A}\mathbf{B}^T\cdot\vec{V}^{\text{new}}\label{TAB}
\ee
The Lagrangian $\Fl_2[\vec{V}]$ reads
\be
\Fl_2[\vec{V}]&=&\sum_{M=1}^{20}\int\rmd^4\theta\ \Phi_{M}^\dagger e^{\mathbf{K}_{MI}\mathbf{B}^T_{IJ}V_J}\Phi_{M}+\frac{1}{4\pi}(\mathbf{A}\mathbf{B}^T-\half\mathbf{B}\mathbf{K}^{T}\mathbf{K}\mathbf{B}^T)_{IJ}\int\rmd^4\theta\ \Sig_{I}V_J\nonumber\\
&&\ + \sum_{\a=1}^5\int \rmd^2\theta\ \Phi_{X_\a}\Phi_{Y_\a}\Phi_{Z_\a}\Phi_{W_\a}
\ee
Here $\mathbf{K}^{T}\mathbf{K}$ is a direct sum of square symmetric matrices
\be
\mathbf{K}^{T}\mathbf{K}=\oplus_{\a=1}^5 K_{(\a)}^T K_{(\a)},\ \ \ \ K_{(\a)}^T K_{(\a)}=\left(
\begin{array}{cccc}
2 & 1 & 1  \\
1 & 2 & 1  \\
1 & 1 & 2   
\end{array}
\right)
\ee

\item ``{S-type}''
\be
\lt(\begin{array}{cc}
\mathbf{0} & \mathbf{I}\\ -\mathbf{I} & \mathbf{0}
\end{array}\rt):\ \ \ \ \ 
\Fl_2\lt[\vec{V}\rt]\mapsto \Fl_3[\vec{V}^{\text{new}}]=\Fl_2\lt[\vec{\tilde{V}}\rt]-\frac{1}{2\pi}\int\rmd^4\theta\ \Sig_I^{\text{new}}\tilde{V}_I \label{Strans}
\ee
Here all the background gauge fields become dynamical gauge fields $\vec{\tilde{V}}$. The last term describes the new background gauge fields coupled with the monopole currents $J_{\mathrm{topological}}=\rmd \tilde{A}$, which are charged under the topological U(1) symmetries.
\be
\Fl_3[\vec{V}]&=&\sum_{M=1}^{20}\int\rmd^4\theta\ \Phi_{M}^\dagger e^{\mathbf{K}_{MI}\tilde{V}_I}\Phi_{M}+\frac{1}{4\pi}(\mathbf{B}^{-1}\mathbf{A}-\half\mathbf{K}^{T}\mathbf{K})_{IJ}\int\rmd^4\theta\ \tilde{\Sig}_{I}\tilde{V}_J -\frac{1}{2\pi}(\mathbf{B}^T)^{-1}_{IJ}\int\rmd^4\theta\ \Sig_I\tilde{V}_J\nonumber\\
&&\ + \sum_{\a=1}^5\int \rmd^2\theta\ \Phi_{X_\a}\Phi_{Y_\a}\Phi_{Z_\a}\Phi_{W_\a}
\ee  
where we have made a field redefinition $\vec{\tilde{V}}\mapsto (\mathbf{B}^T)^{-1}\vec{\tilde{V}}$ for all the 15 dynamical gauge fields. Here we have ignore the Yang-Mills terms of the dynamical gauge fields, since they are exact by SUSY transformation, and the partition function is independent of YM coupling $g^2_{YM}$ \cite{Kapustin09,Hama11}. 

\item ``T-type''
\be
\lt(\begin{array}{cc}
\mathbf{I} & \mathbf{0}\\ \mathbf{D}\mathbf{B}^{-1} & \mathbf{I}
\end{array}
\right):\ \ \ \ \ 
\Fl_3\lt[\vec{V}\rt]\mapsto \Fl_{S^3\setminus\G_5}[\vec{V}^{\text{new}}]=\Fl_3\lt[\vec{V}^{\text{new}}\rt]+\frac{1}{4\pi}\int\rmd^4\theta \ \vec{\Sig}^{\text{new}}\cdot\mathbf{D}\mathbf{B}^{-1}\cdot\vec{V}^{\text{new}}.\label{Ttypetrans}
\ee

\end{enumerate}

\noindent
The final Lagrangian can be written as follows
\be
\Fl_{S^3\setminus\G_5}[\vec{V}]&=&\sum_{M=1}^{20}\int\rmd^4\theta\ \Phi_{M}^\dagger e^{\mathbf{K}_{MI}\tilde{V}_I}\Phi_{M}+ \sum_{\a=1}^5\int \rmd^2\theta\ \Phi_{X_\a}\Phi_{Y_\a}\Phi_{Z_\a}\Phi_{W_\a}+\nonumber\\
&&\ +\frac{\tilde{\ft}_{IJ}}{4\pi}\int\rmd^4\theta\ \tilde{\Sig}_{I}\tilde{V}_J +\frac{\fs_{IJ}}{2\pi}\int\rmd^4\theta\ \Sig_I\tilde{V}_J+ \frac{\ft_{IJ}}{4\pi}\int\rmd^4\theta \ \Sig_I{V}_J\label{GCtheory}
\ee
where $\tilde{\ft}$ is the dynamical Chern-Simons level matrix, ${\ft}$ is the background Chern-Simons level matrix, and $\fs$ is the Chern-Simons level matrix for monopole currents coupled with background gauge fields
\be
\tilde{\ft}_{IJ}=(\mathbf{B}^{-1}\mathbf{A}-\half\mathbf{K}^{T}\mathbf{K})_{IJ},\ \ \ \ft_{IJ}=(\mathbf{D}\mathbf{B}^{-1})_{IJ},\ \ \ \fs_{IJ}=-(\mathbf{B}^T)^{-1}_{IJ}.
\ee
In the supersymmetric gauge theory $T_{S^3\setminus\G_5}$, there are 15 dynamical vector multiplet (gauge fields) $\tilde{V}_{I=1,\cdots,15}$ carrying the gauge group $\mathrm{U}(1)^{15}$. There are 15 background vector multiple $\vec{V}=(V_{\L_{ab}},V_{M_a})$ labelled by the position variables $(\L_{ab},M_a)$ in the final polarization of $\cp_{\partial(S^3\setminus\G_5)}$. The 15 background vector multiple $\vec{V}$ are all coupled with the monopole currents charged under topological U(1). The topological U(1) symmetries give the total flavor symmetry $\mathrm{U}(1)^{15}$ for $T_{S^3\setminus\G_5}$, where 15 equals $\half\dim_\C\cp_{\partial(S^3\setminus\G_5)}$. 

Both the Chern-Simons level matrices $\ft,\fs$ are of integer entries:
\be
\ft&=&\left(
\begin{array}{ccccccccccccccc}
 0 & 0 & 1 & 0 & 0 & 0 & 1 & 0 & -1 & 1 & -1 & 0 & 0 & 0 & 0 \\
 0 & 2 & -1 & 0 & 1 & -1 & 0 & 1 & 1 & -1 & 0 & 0 & -1 & 0 & 0 \\
 1 & -1 & -2 & 0 & 0 & 0 & 0 & 0 & 0 & 0 & 1 & 0 & 0 & 0 & 0 \\
 0 & 0 & 0 & 2 & 0 & 0 & 0 & 0 & 1 & -1 & 0 & 0 & 0 & 0 & 1 \\
 0 & 1 & 0 & 0 & 2 & -1 & 1 & 1 & 0 & 0 & 0 & 1 & -1 & 0 & 0 \\
 0 & -1 & 0 & 0 & -1 & 2 & 0 & 0 & -1 & 1 & 0 & 0 & 0 & 1 & 0 \\
 1 & 0 & 0 & 0 & 1 & 0 & 2 & 0 & 0 & 0 & 0 & 1 & 0 & 0 & 0 \\
 0 & 1 & 0 & 0 & 1 & 0 & 0 & 1 & 1 & 0 & 0 & 0 & 0 & 0 & 0 \\
 -1 & 1 & 0 & 1 & 0 & -1 & 0 & 1 & 0 & -1 & 0 & 0 & 0 & 0 & 0 \\
 1 & -1 & 0 & -1 & 0 & 1 & 0 & 0 & -1 & 1 & 0 & 0 & 0 & 1 & -1 \\
 -1 & 0 & 1 & 0 & 0 & 0 & 0 & 0 & 0 & 0 & 0 & 0 & 0 & 0 & 0 \\
 0 & 0 & 0 & 0 & 1 & 0 & 1 & 0 & 0 & 0 & 0 & 0 & 0 & 0 & 0 \\
 0 & -1 & 0 & 0 & -1 & 0 & 0 & 0 & 0 & 0 & 0 & 0 & 0 & 0 & 0 \\
 0 & 0 & 0 & 0 & 0 & 1 & 0 & 0 & 0 & 1 & 0 & 0 & 0 & 0 & 0 \\
 0 & 0 & 0 & 1 & 0 & 0 & 0 & 0 & 0 & -1 & 0 & 0 & 0 & 0 & 1 \\
\end{array}
\right),\\
\fs&=&\left(
\begin{array}{ccccccccccccccc}
 0 & 1 & 0 & 1 & 1 & 0 & -1 & -1 & -1 & -1 & -1 & 0 & 0 & 0 & 0 \\
 0 & 0 & 1 & -1 & -1 & 0 & 1 & 1 & 1 & 0 & 1 & 0 & -1 & -2 & -2 \\
 0 & 0 & 0 & 0 & -1 & 1 & 0 & 0 & 1 & 0 & 0 & 0 & 0 & 0 & 1 \\
 0 & 0 & 0 & -1 & -1 & -1 & 0 & 1 & 0 & -1 & -1 & -2 & 0 & 0 & 0 \\
 0 & 1 & 1 & 0 & 0 & 0 & 0 & 0 & 0 & -1 & 1 & 0 & 0 & -1 & -1 \\
 1 & 1 & 0 & 1 & 0 & 0 & -1 & -1 & -1 & 0 & -1 & 0 & 0 & 2 & 1 \\
 -1 & 1 & 0 & 0 & 0 & 0 & -1 & 0 & 0 & -1 & 0 & 0 & 0 & 0 & 0 \\
 0 & 0 & 1 & -1 & 0 & 0 & 0 & 0 & 0 & 0 & 1 & 0 & -1 & -1 & -1 \\
 0 & 0 & 1 & 0 & 0 & 0 & 1 & 1 & 0 & 0 & 0 & -1 & 0 & -1 & -1 \\
 0 & 0 & 0 & 1 & 0 & 0 & -1 & -1 & 0 & 0 & 0 & 1 & 0 & 1 & 1 \\
 0 & 0 & 0 & 0 & 0 & -1 & 0 & 0 & -1 & 0 & 0 & 0 & 0 & 0 & 0 \\
 -1 & 0 & 0 & 0 & 0 & 0 & 0 & 0 & 0 & -1 & 0 & 0 & 0 & 0 & 0 \\
 0 & 0 & 0 & 0 & 0 & 0 & 0 & 0 & 0 & 0 & -1 & 0 & 1 & 1 & 1 \\
 0 & 0 & 0 & 0 & 0 & 0 & -1 & -1 & -1 & 0 & 0 & 0 & 0 & 1 & 0 \\
 0 & 0 & 0 & 0 & 0 & 0 & 0 & 1 & 0 & -1 & -1 & -1 & 0 & 0 & 0 \\
\end{array}
\right)
\ee
The rows and columns of $\ft$ are ordered with respect to 
\be
\lt(V_{\L_{12}},V_{\L_{13}},V_{\L_{14}},V_{\L_{15}},V_{\L_{23}},V_{\L_{24}},V_{\L_{25}},V_{\L_{34}},V_{\L_{35}},V_{\L_{45}},V_{M_1},V_{M_2},V_{M_3},V_{M_4},V_{M_5}\rt)
\ee
The columns of $\fs$ are with respect to the same ordering.

In $\tilde{\ft}$, the term $\mathbf{B}^{-1}\mathbf{A}$ is of integer entries:
\be
\mathbf{B}^{-1}\mathbf{A}=\left(
\begin{array}{ccccccccccccccc}
 1 & 0 & 0 & 0 & 0 & 0 & 0 & 0 & 0 & 0 & 0 & 0 & -1 & 0 & 0 \\
 0 & 1 & 0 & 0 & 0 & 0 & -1 & 0 & 0 & -1 & 0 & 0 & -1 & 0 & 0 \\
 0 & 0 & 1 & -1 & 0 & 0 & 0 & 0 & 0 & 0 & 1 & 0 & -1 & -1 & -1 \\
 0 & 0 & -1 & 2 & 1 & 1 & -1 & -1 & 0 & 0 & 0 & 1 & 0 & 1 & 1 \\
 0 & 0 & 0 & 1 & 1 & 1 & 0 & 0 & 1 & 0 & 0 & 1 & 0 & 0 & 1 \\
 0 & 0 & 0 & 1 & 1 & 1 & 0 & 0 & 0 & 0 & 0 & 1 & 0 & 0 & 0 \\
 0 & -1 & 0 & -1 & 0 & 0 & 2 & 1 & 1 & 1 & 1 & 0 & 0 & -1 & -1 \\
 0 & 0 & 0 & -1 & 0 & 0 & 1 & 2 & 1 & 0 & 0 & -1 & 0 & -1 & -1 \\
 0 & 0 & 0 & 0 & 1 & 0 & 1 & 1 & 1 & 0 & 0 & 0 & 0 & -1 & -1 \\
 0 & -1 & 0 & 0 & 0 & 0 & 1 & 0 & 0 & 2 & 1 & 1 & 0 & 0 & 0 \\
 0 & 0 & 1 & 0 & 0 & 0 & 1 & 0 & 0 & 1 & 2 & 1 & 0 & -1 & -1 \\
 0 & 0 & 0 & 1 & 1 & 1 & 0 & -1 & 0 & 1 & 1 & 2 & 0 & 0 & 0 \\
 -1 & -1 & -1 & 0 & 0 & 0 & 0 & 0 & 0 & 0 & 0 & 0 & 1 & 1 & 1 \\
 0 & 0 & -1 & 1 & 0 & 0 & -1 & -1 & -1 & 0 & -1 & 0 & 1 & 3 & 2 \\
 0 & 0 & -1 & 1 & 1 & 0 & -1 & -1 & -1 & 0 & -1 & 0 & 1 & 2 & 2 \\
\end{array}
\right)
\ee
But $-\half\mathbf{K}^{T}\mathbf{K}$ has the half-integer entries, which cancels the gauge anomalies generated by the chiral multiplets.

A way to understand the idea behind the above manipulation of 3d supersymmetric gauge theories is to consider the partition function $Z_{S^3_b}(T_{M_3})$ of the theory $T_{M_3}$ on a 3d ellipsoid $S_b^3$, which gives a state-integral model of $M_3$ \cite{DGG11,Dimofte2011}. Some details is reviewed in Appendix \ref{ellipsoidpartition}.

\subsection{Gluing Copies of $S^3\setminus\G_5$ and Supersymmetric Gauge Theories } \label{sm3 theories}

\begin{figure}[h]
\begin{center}
\includegraphics[width=15cm]{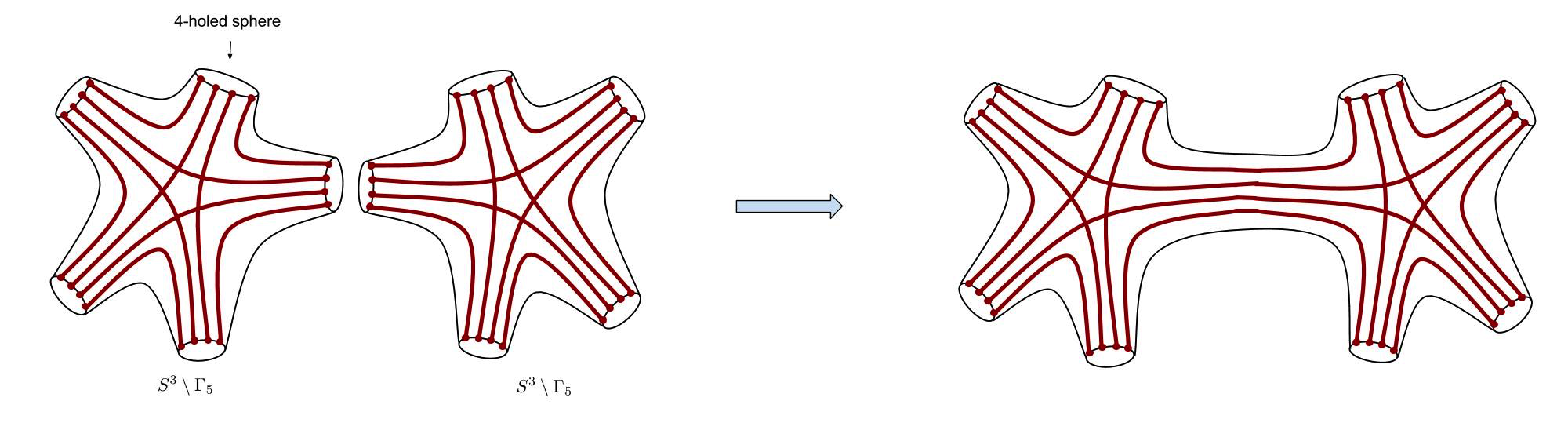}
\caption{Gluing 2 copies of $S^3\setminus\G_5$ through a pair of 4-holed spheres. In the figure we view the 3-manifolds from 4-dimensions (and suppress 1-dimension in drawing the figure). The thick red curves are the annuli cusps in the boundary of the 3-manifolds.}
\label{gluing3-fold}
\end{center}
\end{figure}

Now we consider to glue many copies of $S^3\setminus\G_5$, to construct a class of 3-manifolds, which are generally labeled by $\sm_3$. In $\sm_3$, any pair of $S^3\setminus\G_5$ are glued through a pair of 4-holed spheres, being a component of the geodesic boundary in $\partial(S^3\setminus\G_5)$. See FIG.\ref{gluing3-fold} for a illustration. In general $\sm_3$ is the complement of a certain graph in a 3-manifold $\Fx_3$, where $\Fx_3$ may have nontrivial 1st homology group $H_1(\Fx_3)$. It can be seen by e.g. gluing 3 copies of $S^3\setminus\G_5$ and close an annulus cusp to form a torus cusp. 

By the ideal triangulation of $S^3\setminus\G_5$, the 4-holed spheres are triangulated by the geodesic triangle boundaries of the ideal tetrahedra. 2 copies of $S^3\setminus\G_5$ are glued by identifying the 4 pairs of triangles which triangulate the pair of 4-holed spheres. Therefore gluing a pair of $S^3\setminus\G_5$ change 6 pairs of external edges in the pair of 4-holed spheres into 6 internal edges in the ideal triangulation. The 6 internal edges associate with 6 constraints ($X_E,X_E'$ are the edge coordinates on the pair of 4-holed spheres respectively)\footnote{$\{C_E,C_{E'}\}=0$ since the pair of 4-holed spheres are of opposite orientations $\{X_E,X_{E'}\}=\eps_{EE'}$, $\{X'_E,X'_{E'}\}=\eps'_{EE'}$, with $\eps_{EE'}=-\eps'_{EE'}$.}
\be
C_{E=1,\cdots,6}\equiv X_E+X_E'=2\pi i+\hbar \label{CC0}
\ee
imposed on the product phase space $\cp_{\partial(S^3\setminus\G_5)}\times \cp_{\partial(S^3\setminus\G_5)}$, whose symplectic reduction gives $\cp_{\partial\sm_3}$.

As an example, we identify the 4-holed sphere $\cs_1$ of $S^3\setminus\G_5$ with the 4-holed sphere $\cs_1'$ of another $S^3\setminus\G_5$. The annulus cusp $\ell_{1a}$, which connects $\cs_1,\cs_a$ of $S^3\setminus\G_5$, is continued with the annulus $\ell_{1a}'$, which connects $\cs_1',\cs_a'$ of the second copy of $S^3\setminus\G_5$. Firstly the constraints $C_E=2\pi i+\hbar$ implies that the logarithmic monodromy eigenvalues $\L_{1a}$ is independent of path homotopies \cite{Dimofte2011,DGV}, i.e.
\be
\cc_a\equiv \L_{1a}+\L_{1a}'=0, \label{CC1}
\ee
where $\L_{1a}'$ is the logarithmic eigenvalues of the monodromies meridian to $\ell_{1a}'$\footnote{For the second 4-holed sphere $\cs_1'$ the monodromy $\L'_{1a}$ still relates the edge coordinates $X_E'$ by $2\L'=\sum_{E\text{ around hole}}(X_E'-i\pi)$. But because of the opposite orientation, the Poisson bracket with FN twist is of opposite sign $\{\L'_{1a},\ct'_{1a}\}=-1$.}. Secondly the constraints also implies the relation between the 4-holed sphere coordinates $(M_1,P_1)$ and $(M_1',P_1')$, i.e.\footnote{Again because of the opposite orientation, $\{M'_{1},P'_{1}\}=-1$ where $M_a',P_a'$ are defined by substituting $X_E$ with $X_E'$ in Eq.\Ref{MP}.}
\be
\cc_M=M_1+M_1'=\zeta_1\lt(i\pi+\frac{\hbar}{2}\rt),\ \ \ \ \cc_P=P_1+P_1'=\zeta_2\lt(i\pi+\frac{\hbar}{2}\rt),\label{CC2}
\ee
where $\zeta_1,\zeta_2\in\Z$ in general, and $\zeta_1=\zeta_2=2$ in the example that $\cs_1$ glued with $\cs_1'$. It is clear that Eqs.\Ref{CC1} and \Ref{CC2} are a set of constraints equivalent to Eq.\Ref{CC0}.

We consider the symplectic transformations to:
\be
\lt(\begin{array}{c}
\L_{1a}\\
\cc_a\\
\ct_{1a}+\ct'_{1a}\\
\G_a
\end{array}\rt)=\lt(\begin{array}{cccc}
1 & 0 & 0 & 0\\
1 & 1 & 0 & 0\\
0 & 0 & 1 & -1\\
0 & 0 & 0 & 1
\end{array}\rt)\lt(\begin{array}{c}
\L_{1a}\\
\L_{1a}'\\
\ct_{1a}\\
-\ct'_{1a}
\end{array}\rt),\ \ \ \ \ 
\lt(\begin{array}{c}
\cc_{M}\\
\cc_P\\
\G_M\\
\G_P
\end{array}\rt)=
\lt(\begin{array}{cccc}
1 & 0 & 0 & 0\\
0 & 0 & 0 & -1\\
0 & 0 & 1 & 0\\
0 & 1 & 0 & 0
\end{array}\rt)
\lt(\begin{array}{cccc}
1 & 1 & 0 & 0\\
0 & 1 & 0 & 0\\
0 & 0 & 1 & 0\\
0 & 0 & -1 & 1
\end{array}\rt)\lt(\begin{array}{c}
M_1\\
M_1'\\
P_1\\
-P_1'
\end{array}\rt)\label{CCCCCC6}
\ee
where the first transformation is of GL-type, and the second is a composition of S-type and GL-type. Then we make a further GL-type transformation in the subspace of constraints and their momenta. 
\be
\lt(\begin{array}{c}
C_E\\
\G_E
\end{array}\rt)=\lt(\begin{array}{cc}
U^{-1} & 0 \\
0 & U^T 
\end{array}\rt)\lt(\begin{array}{c}
\cc_a\\
\cc_M\\
\cc_P\\
\G_a\\
\G_M\\
\G_P
\end{array}\rt)+\lt(\begin{array}{c}
\nu_E\\
0
\end{array}\rt)\lt(i\pi+\frac{\hbar}{2}\rt)
\ee
where the $6\times6$ matrix $U$ and 6-dimensional vector $\nu_E$ read
\be
U=\left(
\begin{array}{cccccc}
 0 & 0 & 0 & \frac{1}{2} & \frac{1}{2} & \frac{1}{2} \\
 0 & \frac{1}{2} & \frac{1}{2} & 0 & 0 & \frac{1}{2} \\
 \frac{1}{2} & 0 & \frac{1}{2} & 0 & \frac{1}{2} & 0 \\
 \frac{1}{2} & \frac{1}{2} & 0 & \frac{1}{2} & 0 & 0 \\
 \frac{1}{2} & 0 & -\frac{1}{2} & \frac{1}{2} & 0 & \frac{1}{2} \\
 0 & \frac{1}{2} & \frac{1}{2} & -\frac{1}{2} & \frac{1}{2} & 0 \\
\end{array}
\right),\ \ \ \ \nu_E=\left(
\begin{array}{c}
0 \\
0 \\
6 \\
6 \\
0 \\
0 \\
\end{array}\right).
\ee

It is straight-forward to apply the symplectic transformation to two free copies of theories $T_{S^3\setminus\G_5}$, by the same  procedure as in Eqs.\Ref{GLtypetrans}, \Ref{TAB}, and \Ref{Strans}. The resulting Lagrangian is denoted by $\Fl'_{\sm_3}$. $\Fl'_{\sm_3}$ has $2\times 20$ chiral multiplets and gauge group $\mathrm{U(1)}^{2\times 15+1}$, where an additional U(1) gauge symmetry comes from the S-type symplectic transformation in Eq.\Ref{CCCCCC6}. The 3d $\cn=2$ supersymmetric gauge theory $T_{\sm_3}$ labelled by $\sm_3$ is given by imposing superpotentials to $\Fl'_{\sm_3}$, where each superpotential $\cw_{E}$ is associated to a constraint $C_E$.
\be
\Fl_{\sm_3}=\Fl_{\sm_3}'+\sum_{E=1}^6\int\rmd^2 \theta\, \cw_{E}
\ee 
The constraints $C_E=X_E+X_E'$ where $X_E$ involves all unprimed, primed, and double-primed $\cp_{\partial\Delta}$ coordinates (see Table \ref{edges} the expressions of $X_E$). Recall that our constructions of $\Fl_{S^3\setminus\G_5}$ and $\Fl_{\sm_3}$ use the choice of polarization $(Z,Z'')$ for each ideal tetrahedron. The chiral superfield $\Phi_Z$ in the Lagrangian is associated to the polarization $(Z,Z'')$. Therefore for $C_E$ involving primed or double-primed coordinate, the superpotential $\cw_E$ involves monopole operators, and in general has a complicated expression in $\Fl_{\sm_3}$, in contrast to the superpotential $\cw_\a=\Phi_{X_\a}\Phi_{Y_\a}\Phi_{Z_\a}\Phi_{W_\a}$ in $\Fl_{S^3\setminus\G_5}$. But each $\cw_E$ can be defined as a monomial of chiral superfields in a certain mirror Lagrangian to $\Fl'_{\sm_3}$, which are constructed by a different choices of polarizations in $\cp_{\partial\Delta}$.

For $\sm_3$ obtained by gluing an arbitrary number $N$ copies of $S^3\setminus\G_5$, the corresponding 3d $\cn=2$ supersymmetric theories $T_{\sm_3}$ can be constructed in the same manner. There is a description of $T_{\sm_3}$, containing $20N$ chiral multiplet and having $\mathrm{U}(1)^{15N+\check{N}}$ as gauge group, where $\check{N}$ is the number of 4-holed spheres shared by 2 copies of $S^3\setminus\G_5$. The flavor group of $T_{\sm_3}$ is $\mathrm{U(1)}^{\half\dim\cp_{\partial\sm_3}}$.

\section{Holomorphic Block of 3-dimensional Supersymmetric Gauge Theories}\label{HB3d}

The holomorphic block has been firstly proposed in \cite{3dblock,Pasquetti:2011fj} as a BPS index for 3d supersymmetric gauge theory (it has been generalized to 4d gauge theory \cite{Nieri:2015yia}). It has also been studied recently in \cite{YK} via supersymmetric localization technique. A brief review of the object is provided in Appendix \ref{BPS} for self-containedness.

Let's consruct holomorphic block of the 3d supersymmetric gauge theory $T_{S^3\setminus\G_5}$ labelled by the graph complement 3-manifold $S^3\setminus\G_5$, which is defined in Section \ref{SUSYgraphcomplement}. $T_{S^3\setminus\G_5}$ has the gauge group $\mathrm{U(1)}^{15}$ and the flavor symmetry group $\mathrm{U(1)}^{15}$. It is straight-forward to obtain the perturbative expression of holomorphic block integral
\be
B^\a_{S^3\setminus\G_5}\lt(\vec{x},q\rt)=\int_{\cj_\a}\prod_{I=1}^{15}\frac{\rmd s_I}{s_I}\ 
\exp\lt[\frac{1}{\hbar}\widetilde{\cw}_{S^3\setminus\G_5}(\vec{s},\vec{m},\hbar)\rt].
\ee
Both $\vec{x}$ and $\vec{s}$ are 15-dimensional vectors, with $s_I=e^{\sig_I}$ and $x_I=e^{m_I}$ ($I=1,\cdots,15$). The twisted superpotential has the leading contribution in $\hbar$:
\be
\widetilde{\cw}_{S^3\setminus\G_5}\lt(\vec{s},\vec{m},\hbar\rt)&=&\sum_{\mu=1}^5\lt[\mathrm{Li}_2(e^{-\sig_{X,\,\mu}})+\mathrm{Li}_2(e^{-\sig_{Y,\,\mu}})+\mathrm{Li}_2(e^{-\sig_{Z,\,\mu}})+\mathrm{Li}_2(e^{\sig_{X,\,\mu}+\sig_{Y,\,\mu}+\sig_{Z,\,\mu}-2\pi i})\rt]+\frac{1}{2}\vec{\sig}^T\mathbf{B}^{-1}\mathbf{A}\vec{\sig}\nonumber\\
&&-\vec{\sig}^T\mathbf{B}^{-1}\vec{m}+\frac{1}{2}\vec{m}^T\mathbf{D}\mathbf{B}^{-1}\vec{m}+i\pi\vec{\sig}^T\mathbf{B}^{-1}\vec{\nu}_1-i\pi\vec{m}^T\mathbf{D}\mathbf{B}^{-1}\vec{\nu}_1+i \pi\vec{\nu}_2\cdot\vec{m}+o(\hbar)
\ee
Here each $\sig_I$ stands for a twisted mass (shifted by R-charge) of a chiral multiplet $\sig_I\equiv \{\sig_{X,\,\mu},\sig_{Y,\,\mu},\sig_{Z,\,\mu}\}_{\mu=1}^5$. The matrices $\mathbf{A},\mathbf{B},\mathbf{D}$ and the vector $\vec{\nu}_2$ are given in Eq.\Ref{ABCD}. The parameters $\vec{m}$ in the holomorphic block is also identified to the phase space position coordinates $\vec{m}$ in Eq.\Ref{ABCD}. Note that the leading contribution of the integrand in $B^\a\lt(\vec{x},q\rt)$ is formally the same as the leading contribution of the integrand in the ellipsoid partition function discussed in Section \ref{ellipsoidpartition}. 

It is also straight-forward to construct the nonperturbative block integral of $B^\a_{S^3\setminus\G_5}\lt(\vec{x},q\rt)$:
\be
B^\a_{S^3\setminus\G_5}(\vec{x},q)=\int_{\cj_\a}\prod_{I=1}^{15}\frac{\rmd s_I}{2\pi i s_I} \mathrm{CS}[k,\nu;x,s,q]\prod_{\Phi=1 }^{20}B_\Delta\Big(z_{\Phi}(s;R_{\Phi}),q\Big),
\ee
where the label $I=(X_\mu,Y_\mu,Z_\mu)_{\mu=1}^5$. There are 20 chiral multiplet blocks $B_\Delta$ in the integrand, 
\be
\prod_{\Phi }B_\Delta\Big(z_{\Phi}(s;R_{\Phi}),q\Big)=\prod_{\mu=1}^5\lt[B_\Delta\lt(s_{X,\,\mu},q\rt)\,B_\Delta\lt(s_{Y,\,\mu},q\rt)\,B_\Delta\lt(s_{Z,\,\mu},q\rt)\,B_\Delta\lt(q s_{X,\,\mu}^{-1}s_{Y,\,\mu}^{-1}s_{Z,\,\mu}^{-1};q\rt)\rt]
\ee
The Chern-Simons contribution $\mathrm{CS}[k,\nu;x,s,q]$ is given by 
\be
\mathrm{CS}[k,\nu;x,s,q]=\prod_{t}\theta\lt((-q^{\half})^{b_t}(x,s)^{a_t};q\rt)^{n_t}.
\ee
The exponents $a_t,b_t,n_t$ are given in Eqs.\Ref{exponentrule}, \Ref{nalpha}, and \Ref{nalphabeta} in Appendix \ref{abn}. The integration cycle $\cj_\a$ is the downward gradient flow cycle for $\widetilde{\cw}_{S^3\setminus\G_5}$ (See Eq.\Ref{floweqn} for definition) in the neighborhood of a saddle point $\a$. $\cj_\a$ extends toward $\ln s_I \to\pm\infty$ away from the saddle point (see \cite{3dblock} for details).

The supersymmetric ground states $|\a\rangle$ at the asympetotic boundary of $D^2\times_q S^1$ are given by the solutions to Eq.\Ref{expvac}, which is in this case (from the derivative of $\sig_I\equiv\sig_{X,\,\mu},\sig_{Y,\,\mu},\sig_{Z,\,\mu}$):
\be
\lt(1-s^{-1}_{Z,\,\mu}\rt)\prod_Js_J^{(\mathbf{B}^{-1}\mathbf{A})_{IJ}}=\lt(1-s_{X,\,\mu}s_{Y,\,\mu}s_{Z,\,\mu}\rt)e^{-i\pi\mathbf{B}^{-1}_{IK}(\vec{\nu}_1)_K}\prod_{J}x_J^{(\mathbf{B}^{-1})_{IJ}}.\label{sIxI}
\ee
where $(\mathbf{B}^{-1}\mathbf{A})_{IJ}\in\Z$, $(\mathbf{B}^{-1})_{IJ}\in\Z$, and $\mathbf{B}^{-1}_{IK}\vec{\nu}^K_1\in\Z$. On the other hand, Eq.\Ref{yi} gives $y_I\equiv\exp p_I$
\be
&&\quad\quad\quad\quad\quad\quad\quad y_I=e^{i\pi\lt(\vec{\nu}_2-\mathbf{D}\mathbf{B}^{-1}\vec{\nu}_1\rt)_I}\prod_Js_J^{-(\mathbf{B}^{-1})_{JI}}\prod_K x_K^{(\mathbf{D}\mathbf{B}^{-1})_{IK}},\nonumber\\
&&\!\!\!\!\!\!\!\!\!\!\!\!\!\!\!\!\!\!\!\!\!\!\!\!\!\!
\text{or in terms of logarithmic variables:}\quad 
\vec{p}=-(\mathbf{B}^{-1})^{T}\vec{\sig}+\mathbf{D}\mathbf{B}^{-1}\vec{m}+i\pi\lt(\vec{\nu}_2-\mathbf{D}\mathbf{B}^{-1}\vec{\nu}_1\rt)+2\pi i\vec{N},\label{yIsIxI}
\ee
where ${N}_I\in\Z$, $(\mathbf{D}\mathbf{B}^{-1})_{IK}\in\Z$ and $(\vec{\nu}_2-\mathbf{D}\mathbf{B}^{-1}\vec{\nu}_1)_L\in\Z$. We can find $\vec{s}$ or $\vec{\sig}$ 
\be
&& s_I=\prod_{J}y_J^{\mathbf{B}_{JI}}\prod_{K}x_K^{(\mathbf{B}^T\mathbf{D}\mathbf{B}^{-1})_{IK}}\prod_{L} e^{i\pi\mathbf{B}_{LI}\lt(\vec{\nu}_2-\mathbf{D}\mathbf{B}^{-1}\vec{\nu}_1\rt)_L},\nonumber\\
\text{or}&&
\vec{\sig}=-\mathbf{B}^T\vec{p}+\mathbf{B}^T\mathbf{D}\mathbf{B}^{-1}\vec{m}+i\pi\mathbf{B}^T\lt(\vec{\nu}_2-\mathbf{D}\mathbf{B}^{-1}\vec{\nu}_1\rt)+2\pi i\mathbf{B}^T\vec{N}.\label{sIinyI}
\ee 
$\vec{s}=\vec{s}(\vec{x},\vec{y})$ is multivalued because the entries of $\mathbf{B}$ are half-integers. $\vec{s}(\vec{x},\vec{y})$ involves the square-roots of $\vec{x},\vec{y}$. 

Inserting Eq.\Ref{sIinyI} into Eq.\Ref{sIxI}, we obtain an algebraic equations $\mathbf{A}_I(x,y)=0$ ($I=1,\cdots,15$) locally describing the holomorphic Lagrangian submanifold $\cl_{\mathrm{SUSY}}(T_{S^3\setminus\G_5})$. The resulting $\mathbf{A}_I(x,y)=0$ (considering all possible choices of $\pm_I$) is identical to the set of algebraic equations, which describes the embedding of $\cl_{S^3\setminus\G_5}(\mathbf{t})$ in the phase space $\cp_{\partial( S^3\setminus\G_5)}$. The construction of $\cl_{S^3\setminus\G_5}(\mathbf{t})$ is carried out in Appendix \ref{AppA} by following the procedure in \cite{Dimofte2011}. Note that from the derivation in Appendix \ref{AppA}, Eq.\Ref{sIxI} already equivalently characterizes $\cl_{S^3\setminus\G_5}(\mathbf{t})$, although $\vec{s},\vec{x}$ aren't canonical conjugate. Eq.\Ref{yIsIxI} is simply a change of coordinates.  

A supersymmetric ground state $|\a\rangle\leftrightarrow\vec{s}^{(\a)}(\vec{x})$ is a solution to Eq.\Ref{sIxI}. It determines a unique $\vec{y}^{(\a)}(\vec{x})$ satisfying $\mathbf{A}_I(x,y)=0$ by
\be
y_I^{(\a)}(\vec{x})=e^{i\pi\lt(\vec{\nu}_2-\mathbf{D}\mathbf{B}^{-1}\vec{\nu}_1\rt)_I}\prod_J\lt[s^{(\a)}_J(\vec{x})\rt]^{-(\mathbf{B}^{-1})_{JI}}\prod_K x_K^{(\mathbf{D}\mathbf{B}^{-1})_{IK}}.
\ee


Rigorously speaking, the Lagrangian submanifold $\cl_{M_3}(\mathbf{t})$ for any ideal triangulated $M_3$ might have mild but nontrivial dependence of the ideal triangulation $\mathbf{t}$, because of its construction uses a specific triangulation. $\cl_{M_3}(\mathbf{t})$ captures those flat connections whose framing data is generic with respect to the 3d triangulation $\mathbf{t}$, namely the parallel-transported flags inside each tetrahedron define non-degenerate cross-ratios $z,z',z''\in \C\setminus\{0,1\}$. Thus $\cl_{M_3}(\mathbf{t})$ is an open subset in the moduli space $\cl_{M_3}$ of framed $\Slc$ flat connections. If the triangulation of a 3-manifold $M_3$ is not regular enough, $\cl_{M_3}(\mathbf{t})$ constructed by following the procedure in \cite{Dimofte2011} or Appendix \ref{AppA} might only capture a small part of $\cl_{M_3}$, although generically the closure of $\cl_{M_3}(\mathbf{t})$ is independent of regular $\mathbf{t}$ and usually isomorphic to $\cl_{M_3}$. 

Comparing Eqs.\Ref{sIxI} and \Ref{sIinyI} and the construction in Appendix \ref{AppA} manifests the isomorphism 
\be
\cl_{\mathrm{SUSY}}(T_{S^3\setminus\G_5})\simeq\cl_{S^3\setminus\G_5}(\mathbf{t})
\ee
in the case of the graph complement 3-manifold $M_3=S^3\setminus\G_5$. $\cl_{S^3\setminus\G_5}(\mathbf{t})$ constructed in Appendix \ref{AppA} captures the right part of framed flat connections which is useful in Section \ref{4dQG}. And we do believe that the triangulation in Section \ref{idealtriangulation} is indeed regular enough to make $\cl_{S^3\setminus\G_5}(\mathbf{t})\simeq \cl_{S^3\setminus\G_5}$.

Importantly the proof of the isomorphism in Appendix \ref{AppA} identifies the parameter $p_I$ with the momentum variables in Eq.\Ref{ABCD}. Therefore the symplectic structure $\O_{\mathrm{SUSY}}$ is identified to the Atiyah-Bott-Goldman symplectic form, by the identification of the parameters in gauge theory and the symplectic coordinates of $\cp_{\partial (S^3\setminus\G_5)}$, i.e. $\vec{m}=(\L_{ab},M_a)$ and $\vec{p}=(\ct_{ab},P_a)$. Therefore the leading order contribution to the holomorphic block is given by a contour integral of the Liouville 1-form in terms of the right symplectic coordinates:
\be
B_{S^3\setminus\G_5}^\a\lt(\vec{x},q\rt)=\exp\lt[\frac{1}{\hbar}\int_{\Fc\subset\hat{\cl}_{\mathrm{SUSY}}}^{(x_I,y_I^{(\a)})} \vth+o\lt(\ln\hbar\rt)\rt],\quad\text{where}\quad \vth=\sum_{a<b}\ct_{ab}\rmd \L_{ab}+\sum_{a=1}^5P_a\rmd M_a\ .
\ee

Now we consider the theory $T_{\sm_3}$ defined in Section \ref{sm3 theories}. As an example we consider again $\sm_3$ obtained by gluing 2 copies of $S^3\setminus\G_5$ by identifying $\cs_1$ and $\cs_1'$. The holomorphic block of $T_{\sm_3}$ has the perturbative expression
\be
B^\a_{\sm_3}\lt(\vec{x},q\rt)=\int_{\cj_\a}\frac{\rmd \check{s}}{\check{s}} \prod_{I=1}^{15}\frac{\rmd s_I}{s_I}\prod_{J=1}^{15}\frac{\rmd s'_J}{s'_J}\ 
\exp\lt[\frac{1}{\hbar}\widetilde{\cw}_{\sm_3}(\vec{s},\vec{s}',\check{s},\vec{m},\vec{m}',\hbar)\rt],
\ee
where $\check{s}=\exp(\check{\sig})$ comes from the addition U(1) gauge symmetry as 2 copies of $S^3\setminus\G_5$ are glued through a 4-holed sphere, and the twisted superpotential reads (recall that $\vec{m}=(\L_{ab},M_a)$)
\be
\widetilde{\cw}_{\sm_3}(\vec{s},\vec{s}',\check{s},\vec{m},\vec{m}',\hbar)=\widetilde{\cw}_{S^3\setminus\G_5}\lt(\vec{s},\vec{m},\hbar\rt)\Big|_{M_1=i\pi\zeta_1-\check{\sig}}+\widetilde{\cw}_{S^3\setminus\G_5}\lt(\vec{s}',\vec{m}',\hbar\rt)\Big|_{\L_{1a}'=-\L_{1a},\ M_1'=\check{\sig}}+i\pi\zeta_2\check{\sig}.
\ee

We use again Eq.\Ref{expvac} to obtain the supersymmetric ground state $|\a\rangle$ at the asymptotic boundary of $D^2\times_q S^1$. The derivatives of $\widetilde{\cw}_{\sm_3}$ in $\vec{s}$ and $\vec{s}'$ have been computed above. We only need to insert $M_1=i\pi\zeta_1-\check{\sig}$, $\L_{1a}'=-\L_{1a}$, and $ M_1'=\check{\sig}$ in the results Eq.\Ref{sIxI}. The derivative in $\check{s}$ gives a new equation. Recall the definition of $y_i=\exp (p_i)$ in Eq.\Ref{yi} for $T_{S^3\setminus\G_5}$, where $p_i$ is a derivative of $\widetilde{\cw}_{S^3\setminus\G_5}$ in $m_i$. But the derivative of $\widetilde{\cw}_{\sm_3}$ in $\check{s}$ (or $\check{\sig}$) is computed by the derivative of $M_1$ or $M_1'$ in each $\widetilde{\cw}_{S^3\setminus\G_5}$. The new equation from the derivative in $\check{s}$ is essentially the constraint $\cc_P$ in Eq.\Ref{CC2} \footnote{According to our orientation convention in Section \ref{sm3 theories}, $\exp\lt(\frac{\partial\widetilde{\cw}_{S^3\setminus\G_5}\lt(\vec{s}',\vec{m}',\hbar\rt)}{\partial M_1'} \rt)=\exp(-P_1')$.}
\be
\exp\lt(P_1\Big|_{M_1=i\pi\zeta_1-\check{\sig}}+P_1'\Big|_{\L_{1a}'=-\L_{1a},\ M_1'=\check{\sig}}\rt)=\exp\lt(i\pi\zeta_2\rt).\label{derchecks}
\ee
Or more explicitly in terms of $\vec{\sig},\vec{\sig}'$ \footnote{The matrix element $(\mathbf{D}\mathbf{B}^{-1})_{M_1,M_1}$ is actually zero, see the Chern-Simons level matrix $\ft$. But we keep it in the formula for the generality, because $(\mathbf{D}\mathbf{B}^{-1})_{M_5,M_5}=1$ if the pair of $S^3\setminus\G_5$ are glued by identifying $\cs_5,\cs'_5$.}
\be
\exp\lt[({\sig}_{I}-\sig_I')(\mathbf{B}^{-1})_{I,M_1}-({m}_I-m_I')(\mathbf{D}\mathbf{B}^{-1})_{I,M_1}-(i\pi\zeta_1-2\check{\sig})(\mathbf{D}\mathbf{B}^{-1})_{M_1,M_1}+i\pi\zeta_2\rt]=1\label{derchecks1}
\ee

The effective FI parameters $y_i$ are derived by Eq.\Ref{yi}. Most of $y_i$'s are still given by Eq.\Ref{yIsIxI} (inserting $M_1=i\pi\zeta_1-\check{\sig}$, $\L_{1a}'=-\L_{1a}$, and $ M_1'=\check{\sig}$), except that there is no effective FI parameter for $\L_{1a}'$, while the effective FI parameter $y_{\L_{1a}}\equiv\t_{\ell_a}=\exp(\ct_{\ell_a})$ ($\ell_a$ denotes the annulus extended from $\ell_{1a}$) is given by 
\be
\t_{\ell_a}=\exp\lt(\ct_{1a}\Big|_{M_1=i\pi\zeta_1-\check{\sig}}+\ct'_{1a}\Big|_{\L_{1a}'=-\L_{1a},\ M_1'=\check{\sig}}\rt).\label{tauTT}
\ee

The algebraic equations determining the supersymmetric ground states containing 2 copies of Eq.\Ref{sIxI} for unprimed and primed $s_I$ and $x_I$. We still use Eq.\Ref{yIsIxI} but only understand it as changes of variables from $\sig_I,\sig_I'\mapsto y_I,y_I'$. As before, the changes of variables results in the two sets of algebraic equations $\mathbf{A}_I(x,y)=0$ defining 2 copies $\cl_{\rm SUSY}(T_{S^3\setminus\G_5})$. However the constraints $M_1=i\pi\zeta_1-\check{\sig}$, $\L_{1a}'=-\L_{1a}$, and $ M_1'=\check{\sig}$ are imposed to $\mathbf{A}_I(x,y)=0$, and the new equation Eq.\Ref{derchecks} from the derivative in $\check{s}$ also has to be imposed as well. Moreover Eq.\Ref{tauTT} motivate us to introduce the new variables $\t_{\ell_a}$
\be
\mathbf{A}_I(x,y)\Big|_{M_1=i\pi\zeta_1-\check{\sig},\ct_{1a}=\t_{\ell_a}-\ct_{1a}'}=0,\quad\text{and}\quad\mathbf{A}_J(x',y')\Big|_{\L_{1a}'=-\L_{1a},\ M_1'=\check{\sig},P_1'=i\pi\zeta_2-P_1}=0,\label{A=0andA=0}
\ee
where $I,J=1,\cdots,15$. Now we need to eliminate $\check{\sig}$ because it comes from a dynamical gauge field, and we also need to eliminate $P_1$ and $\ct_{1a}'$ because their conjugate variables $M_1$ and $\L_{1a}'$ have already been eliminated. The elimination uses 6 of the above equations, so it results in 24 equations with 48 variables
\be
\vec{\mathbf{A}}_{\sm_3}\lt(\vec{x},\vec{y}\rt)=0,\quad \text{with}\quad \vec{x}=(e^{\L_\ell},e^{M_\cs}),\quad\vec{y}=(e^{\ct_\ell},e^{P_\cs})\label{Asm3=0}
\ee
where $\ell$ labels the annulus cusps in $\partial\sm_3$ and $\cs$ labels the 4-holed spheres in $\partial\sm_3$.

The algebraic equations $\vec{\mathbf{A}}_{\sm_3}\lt(\vec{x},\vec{y}\rt)=0$ defines the moduli space of supersymmetric vacua $\cl_{\rm SUSY}(T_{\sm_3})$. By construction, we have the isomorphism 
\be
\cl_{\rm SUSY}(T_{\sm_3})\simeq\cl_{\sm_3}(\mathbf{t}).
\ee
It is because Eq.\Ref{A=0andA=0} is essentially the application of the symplectic transformation Eq.\Ref{CCCCCC6} to 2 copies of $\mathbf{A}_I(x,y)=0$, followed by imposing the constraints Eqs.\Ref{CC1} and \Ref{CC2}. The conjugate momenta of the constraints $\cc_a,\cc_M,\cc_P$ are $\G_a=-\ct_{1a}'$, $\G_M=P_1$, and $\G_P=M_1'=\check{\sig}$. These momenta are precisely the variables eliminated in the last step of deriving $\vec{\mathbf{A}}_{\sm_3}\lt(\vec{x},\vec{y}\rt)=0$. Therefore the above procedure to obtain Eq.\Ref{Asm3=0} coincides with the procedure of deriving $\cl_{\sm_3}(\mathbf{t})$ from $\cl_{S^3\setminus\G_5}(\mathbf{t})$ in \cite{Dimofte2011}. As a supersymmetric ground state $|\a\rangle$, a solution $({s}_I,s'_I,\check{s})^{(\a)}(\vec{x})$ to Eq.\Ref{expvac} corresponds a unique solution $\vec{y}^{(\a)}(\vec{x})$ to $\vec{\mathbf{A}}_{\sm_3}\lt(\vec{x},\vec{y}\rt)=0$. 

The moduli space of framed flat connections $\cl_{\sm_3}(\mathbf{t})$ can also be characterized by a pair of Eq.\Ref{sIxI}, imposing the constraints $M_1=i\pi\zeta_1-\check{\sig}$, $\L_{1a}'=-\L_{1a}$, and $ M_1'=\check{\sig}$ to $\vec{x}$, as well as combining Eq.\Ref{derchecks1}. After eliminating 7 of the $s_I,s_I'$ variables (including $\check{s}$) by using 7 equations, we end up with 24 equations with 48 $\vec{s},\vec{x}$ variables, which relate $\vec{\mathbf{A}}_{\sm_3}\lt(\vec{x},\vec{y}\rt)=0$ by a change of coordinates.


The above results can be generalized straightforwardly to arbitrary $\sm_3$ made by gluing many copies of $S^3\setminus\G_5$. Eqs.\Ref{expvac} and \Ref{yi} can be applied to the twisted superpotential of $T_{\sm_3}$, to derive the moduli space of supersymmetric vacua $\cl_{\rm SUSY}(T_{\sm_3}):\ \vec{\mathbf{A}}_{\sm_3}\lt(\vec{x},\vec{y}\rt)=0$. We always have the isomorphism $\cl_{\rm SUSY}(T_{\sm_3})\simeq\cl_{\sm_3}(\mathbf{t})$ with the moduli space of framed flat connections, and have the 1-to-1 correspondence between $\vec{y}^{(\a)}(\vec{x})$ and $\vec{s}^{(\a)}(\vec{x})$. By the isomorphism, $\O_{\rm SUSY}$ is identified to the Atiyah-Bott-Goldman symplectic form on $\cp_{\partial\sm_3}$. The parameters $\vec{x},\vec{y}$ are identified to the symplectic coordinates on $\cp_{\partial\sm_3}$. 
\be
\vec{x}=(\l^2_{T^2},\l_{\ell},{\sf m}_\cs),\quad \vec{y}=(\t_{T^2},\t_{\ell},{\sf p}_\cs)
\ee
Here $\l_{T^2}=\exp(\L_{T^2}), \t_{T^2}=\exp(\ct_{T^2})$ are the eigenvalues of meridian and longitude holonomies of each torus cusp in $\partial\sm_3$. The torus cusp happens e.g. when we glue a pair of $S^3\setminus\G_5$ through 2 pairs of 4-holed spheres, or when we glue 3 copies of $S^3\setminus\G_5$, and each pair of $S^3\setminus\G_5$ share a 4-holed sphere. $\l_\ell=\exp(\L_\ell)$ is the eigenvalue of meridian holonomy to each annulus cusp. $\l_\ell$ is the complex FN length variable constructed in Section \ref{coordinates} in the case of $S^3\setminus\G_5$. $\t_\ell=\exp(\ct_{\ell})$ is the conjugate complex FN twist variable. ${\sf m}_\cs=\exp(M_\cs),{\sf p}_\cs=\exp(P_\cs)$ are the position and momentum coordinates associates to each 4-holed sphere $\cs\subset\partial\sm_3$. 

The leading order contribution to the holomorphic block is given by a contour integral:
\be
B_{\sm_3}^\a\lt(\vec{x},q\rt)=\exp\lt[\frac{1}{\hbar}\int_{\Fc\subset\hat{\cl}_{\rm SUSY}}^{(\vec{x},\vec{y}^{(\a)})} \vth+o\lt(\ln\hbar\rt)\rt],
\ee
where $\vth$ is the Liouville 1-form
\be
\vth=\sum_{T^2}\ln{ \t_{T^2}}\frac{\rmd \l^2_{T^2}}{\l^2_{T^2}}+\sum_{\ell}\ln{ \t_{\ell}}\frac{\rmd \l_\ell}{\l_\ell}+\sum_{\cs}\ln{{\sf p}_\cs}\frac{\rmd {\sf m}_\cs}{{\sf m}_\cs}.
\ee
This result will be important in deriving the relation with 4-dimensional simplicial gravity.

\section{Supersymmetric Vacua and 4-dimensional Simplicial Geometry}\label{4dQG}

\subsection{Supersymmetric Vacua of $T_{\sm_3}$ and 4-dimensional Simplicial Geometry}

Let's consider the class of 3d $\cn=2$ supersymmtric gauge theories labelled by the class of $\sm_3$, being the gluing of many copies of $S^3\setminus\G_5$ in FIG.\ref{gluing3-fold}. The parameter space (moduli space) of massive supersymmetric vacua $\cl_{\mathrm{SUSY}}(T_{\sm_3})$ is isomorphic to the moduli space of framed $\Slc$ flat connections $\cl_{\sm_3}(\mathbf{t})$ which can be captured by the ideal triangulation $\mathbf{t}$ in Section \ref{3d/3dduality}. 

In this section, we would like to show that given a 3-manifold $\sm_3$, it corresponds to a unique simplicial manifold $\sm_4$ in 4-dimensions. There exists a class of supersymmetric vacua in $\cl_{\mathrm{SUSY}}(T_{\sm_3})$, or namely a class of framed flat connections in $\cl_{\sm_3}(\mathbf{t})$, that equivalently describes the simplicial geometry on $\sm_4$. 

\begin{figure}[h]
\begin{center}
\includegraphics[width=5cm]{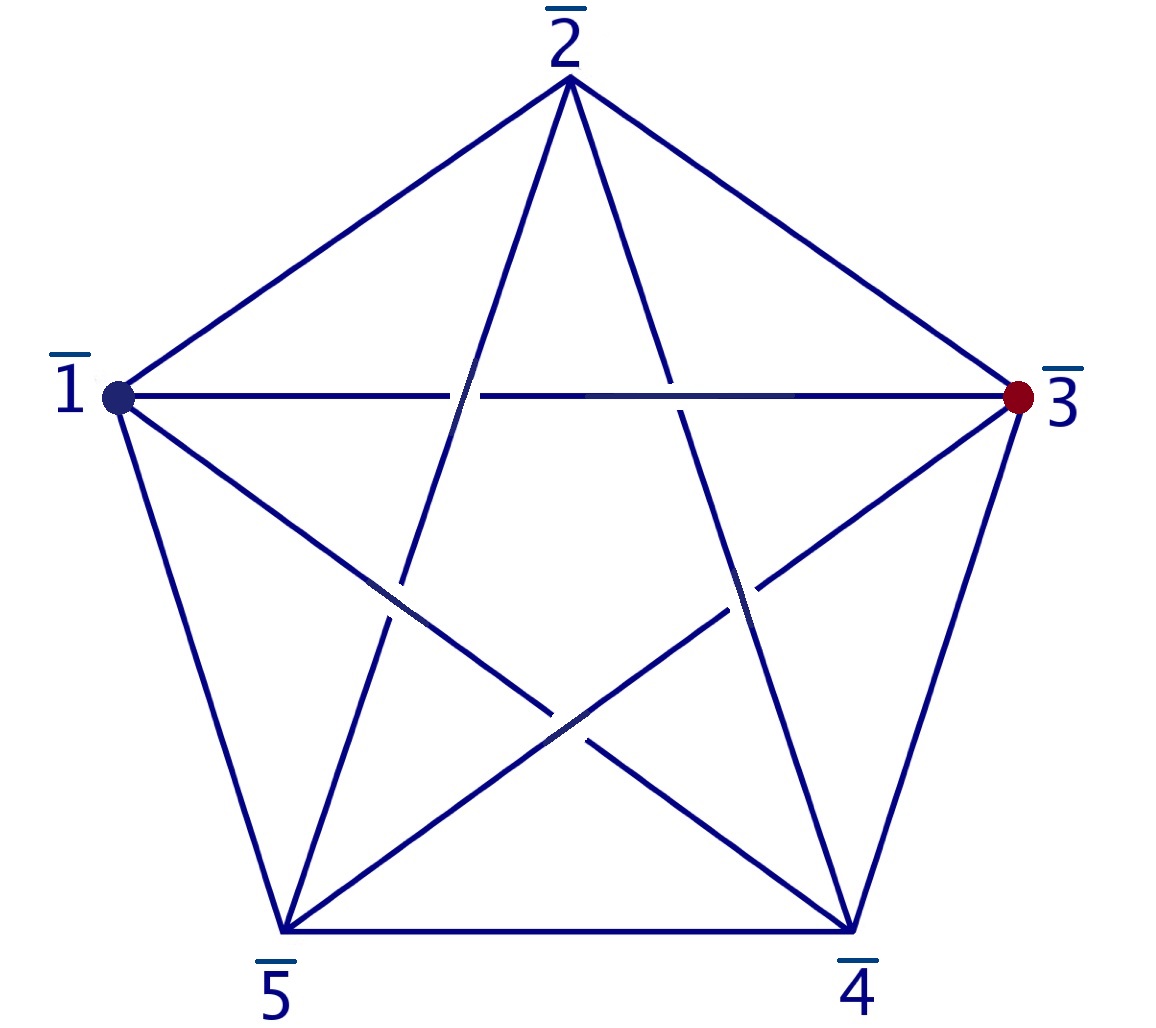}
\caption{A 4-simplex.}
\label{4simplex}
\end{center}
\end{figure}

As the simplest and most important example in the correspondence between the supersymmetric vacua in $\cl_{\mathrm{SUSY}}(T_{\sm_3})$ and the simplicial geometry on $\sm_4$, we consider the supersymmetric gauge theory $T_{S^3\setminus\G_5}$, whose $\cl_{\mathrm{SUSY}}(T_{S^3\setminus\G_5})$ is isomorphic to $\cl_{S^3\setminus\G_5}(\mathbf{t})$. In our correspondence, the 3-manifold $S^3\setminus\G_5$ corresponds to a 4-manifold, which is a 4-simplex FIG.\ref{4simplex}. A 4-simplex may be viewed as the simplest 4-manifold, in the sense that 4-simplex is the building block of the simplicial decomposition of arbitrary 4-manifold. It turns out that there is a class of supersymmetric vacua in $\cl_{\mathrm{SUSY}}(T_{S^3\setminus\G_5})$ which equivalently describes all the geometries of a 4-simplex with a constant curvature $\kappa$.

There is a simple idea behind this correspondence. The relation between the manifolds of different dimensions can be related by considering two types of fundamental groups living in different dimensions. Firstly, let's consider a $(d-1)$-dimensional manifold $M_{d-1}$, which may be taken as a $(d-1)$-sphere with certain codimension-2 defects, and let's consider the fundamental group $\pi_1(M_{d-1})$. On the other hand, we consider a $d$-dimensional polyhedron $M_d$, and the fundamental group of its 1-skeleton, denoted by $\pi_1(\mathrm{sk}(M_d))$. We claim that for any $d$-dimensional polyhedron $M_d$, there exists a $(d-1)$-dimensional manifold $M_{d-1}$, such that we have an isomorphism between the two fundamental groups
\be
\pi_1(M_{d-1})\simeq\pi_1(\mathrm{sk}(M_d)).\label{isofund}
\ee
A simple topological proof is follows\footnote{I thank an anonymous referee for pointing it out.}: Let $M_d$ be a $d$-polyhedron, $\partial M_d$ its topological boundary, which is decomposed into cells by $\mathrm{sk}(M_d)$. Let $\G \in \partial M_d$ be the $(d-3)$-skeleton of the dual cell decomposition of $\partial M_d$. Then $\mathrm{sk}(M_d)$ is homotopic to $\partial M_d\setminus\G$.

Now we assume there are flat connections $\o_{\mathrm{flat}}$ of structure group $G$ on $M_{d-1}$, and there are geometries on $M_d$. Each geometry on $M_d$ gives a spin connection $\o_{\mathrm{Spin}}$. Once a pair $(M_d,M_{d-1})$ satisfying Eq.\Ref{isofund} is found, the spin connection on $M_d$ are related to the flat connection on $M_{d-1}$. Indeed, when we evaluate holomonies of $\o_{\mathrm{Spin}}$ along the 1-skeleton of $M_d$, we obtain a homomorphism (modulo conjugation) from $\pi_1(\mathrm{sk}(M_d))$ to $\mathrm{SO}(d)$ or $\mathrm{SO}(d-1,1)$ for Euclidean or Lorentzian signature. On the other hand, the flat connections $\o_{\mathrm{flat}}$ is a homomorphism (modulo conjugation) from $\pi_1(M_{d-1})$ to the structure group $G$. Therefore we obtain the following commuting triangle if $G$ is $\mathrm{SO}(d)$ or $\mathrm{SO}(d-1,1)$: 
\be
&\pi_1(M_{d-1})\ \ \ \ \ \ \ \ \ \ \ \ \ \ \ \ \ \stackrel{S}{\longleftarrow}\ \ \ \ \ \ \ \ \ \ \ \ \ \ \ \ \ \pi_1(\mathrm{sk}(M_d))&\nonumber\\
&&\nonumber\\
&\o_{\mathrm{flat}}\searrow\ \ \ \ \ \ \ \ \ \ \ \ \ \ \ \ \ \ \ \ \ \ \ \ \ \ \swarrow\o_{\mathrm{spin}}&\nonumber\\
&&\nonumber\\
&\mathrm{SO}(d)\quad\text{or}\quad\mathrm{SO}(d-1,1)&\label{Smap}
\ee 
where $S$ denotes the isomorphism from $\pi_1(\mathrm{sk}(M_d))$ to $\pi_1(M_{d-1})$. As the representations of the two types of fundamental groups, the spin connection on $M_d$ and flat connection on $M_{d-1}$ are related by
\be
\o_{\mathrm{Spin}}=\o_{\mathrm{flat}}\circ S.\label{FlatSpin}
\ee  

\begin{figure}[h]
\begin{center}
\includegraphics[width=13cm]{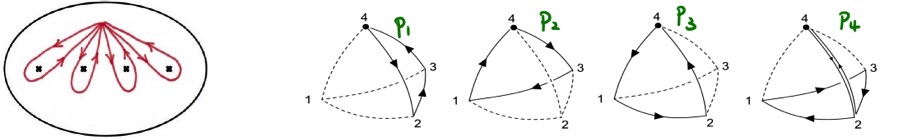}
\caption{A 4-holed sphere and a tetrahedron. }
\label{4holedsphere}
\end{center}
\end{figure}

In the following we give 2 explicit examples of the pair ($M_{d-1},M_d$) which is used in our analysis. The simplest example of the pair ($M_{d-1},M_d$) satisfying the isomorphism Eq.\Ref{isofund} is in the case of $d=3$: Let $M_{d-1}$ be a 4-holed sphere $S^2\setminus\{\mathrm{4\ pts}\}$, and $M_d$ be a tetrahedron. The fundamental group of 4-holed sphere is given by $\pi_1(S^2\setminus\{\mathrm{4\ pts}\})=\langle\fl_{i=1,\cdots,4}|\fl_4\fl_3\fl_2\fl_1=\fe \rangle$ where $\fl_i$ is a noncontractible loop circling around a hole. On the other hand, for the 1-skeleton of a tetrahedron, its fundamental group is generated by 4 closed paths $p_{i=1,\cdots,4}$ along the 1-skeleton, each of which circles around a triangle as in FIG.\ref{4holedsphere}. It is not difficult to see that if one connects all the 4 paths, it actually gives a trivial path. Therefore $\pi_1(\mathrm{sk}(\mathrm{Tetra}))=\langle p_{i=1,\cdots,4}| p_4p_3p_2p_1=1 \rangle$. Obviously 
\be
\pi_1\lt(S^2\setminus\{\mathrm{4\ pts}\}\rt)\simeq\pi_1\Big(\mathrm{sk}(\mathrm{Tetra})\Big).\label{fundtetra}
\ee
We choose the structure group to be SO(3), so that the spin connection on a tetrahedron are related to the flat connection on 4-holed sphere. By using this relation, all possible constant curvature tetrahedron geometries can be reconstructed by the flat connections on 4-holed sphere \cite{curvedMink}. 

Let's go to 1-dimension higher and consider a pair $(M_{d-1},M_d)$ with $d=4$: Let $M_{d-1}$ be the graph complement 3-manifold $S^3\setminus\G_5$, and $M_d$ be a 4-simplex. The fundamental group $\pi_1(S^3\setminus\G_5)$ can be computed in a generalized Wirtinger presentation \cite{brown,HHKR}: $\pi_1(S^3\setminus\G_5)$ is generated by a set of loops $\fl_{ab}$ meridian to the edges $\ell_{ab}$ of the $\G_5$ graph, modulo the following relations:
\be
\text{vertex 1}:&&\fl_{14}\fl_{13}^{(1)}\fl_{12}\fl_{15}=\fe,\nonumber\\
\text{vertex 2}:&&\fl_{12}^{-1}\fl_{24}^{}\fl_{23}\fl_{25}=\fe,\nonumber\\
\text{vertex 3}:&&\fl_{23}^{-1}(\fl_{13}^{(2)})^{-1}\fl_{34}\fl_{35}=\fe,\nonumber\\
\text{vertex 4}:&&\fl_{34}^{-1}\fl_{24}^{-1}\fl_{14}^{-1}\fl_{45}=\fe,\nonumber\\
\text{vertex 5}:&&\fl_{25}^{-1}\fl_{35}^{-1}\fl_{45}^{-1}\fl_{15}^{-1}=\fe,\nonumber\\
\text{crossing}:&&\fl_{13}^{(1)}=\fl_{24}\fl_{13}^{(2)}\fl_{24}^{-1}.\label{pi1M3}
\ee
The fundamental group $\pi_1(\mathrm{sk}(\mathrm{Simplex}))$ can be computed in a similar way as the case of tetrahedron, i.e. stating from a fixed base point and drawing closed paths $p_{ab}$ around each triangle $\ff_{ab}$ (the triangle that doesn't connect to the vertices $\bar{a}$ and $\bar{b}$). $\pi_1(\mathrm{sk}(\mathrm{Simplex}))$ is generated by the closed paths $p_{ab}$ modulo the following relations \cite{HHKR}
\be
\text{Tetra 1}:&&p_{14}p_{13}^{(1)}p_{12}p_{15}=1,\nonumber\\
\text{Tetra 2}:&&p_{12}^{-1}p_{24}^{}p_{23}p_{25}=1,\nonumber\\
\text{Tetra 3}:&&p_{23}^{-1}(p_{13}^{(2)})^{-1}p_{34}p_{35}=1,\nonumber\\
\text{Tetra 4}:&&p_{34}^{-1}p_{24}^{-1}p_{14}^{-1}p_{45}=1,\nonumber\\
\text{Tetra 5}:&&p_{25}^{-1}p_{35}^{-1}p_{45}^{-1}p_{15}^{-1}=1,\nonumber\\
\text{``crossing''}:&&p_{13}^{(1)}=p_{24}p_{13}^{(2)}p_{24}^{-1}.\label{pi1M4}
\ee
where $\text{Tetra}\ a$ $(a=1,\cdots,5)$ label the 5 tetrahedra forming the boundary of 4-simplex. Each tetrahedron corresponds to a vertex of the $\G_5$ graph. By comparing Eq.\Ref{pi1M3} and Eq.\Ref{pi1M4}, it is obvious that
\be
\pi_1\lt(S^3\setminus\G_5\rt)\simeq\pi_1\Big(\mathrm{sk}(\mathrm{Simplex})\Big).
\ee
Given the identification of the tetrahedra of 4-simplex and the vertices of $\G_5$, and certain orientation compatibility, the isomorphism between $\pi_1\lt(S^3\setminus\G_5\rt)$ and $\pi_1\Big(\mathrm{sk}(\mathrm{Simplex})\Big)$ is unique \cite{HHKR}. We consider Lorentzian 4-simplex geometries and choose the structure group $G$ to be $\PSlc\simeq \mathrm{SO}^+(3,1)$. Then the spin connections on 4-simplex are related to the flat connections on $S^3\setminus\G_5$ by the commuting triangle \Ref{Smap} and Eq.\Ref{FlatSpin}. We can also consider the structure group $G=\Slc$ then Eq.\Ref{FlatSpin} is modified by $\pm$ sign.

However $\o_{\mathrm{Spin}}$ obtained here, as a representation of $\pi_1(\mathrm{sk}(M_d))$, only gives a set of holonomies along the 1-skeleton. So $\o_{\mathrm{Spin}}$ doesn't contain enough information about the geometry on $M_d$ unless there is additional input. Now we assume that the 4-simplex is embedded in the constant curvature spacetime, such that all the triangles are flatly embedded surfaces (vanishing extrinsic curvature). Different 4-simplex geometries give different edge lengths, while the interior of 4-simplex is of constant curvature geometry \footnote{Gluing constant curvature simplices give a large simplicial geometry which is locally constant curvature. But the simplicial geometry can approximate arbitrary smooth geometry when the simplicial complex is sufficiently refined.}. For a 2-surface $f$ flat embedded in a constant curvature spacetime, the holonomy of spin connection along $\partial f$ relates the area $\mathbf{a}_f$ of the 2-surface and the bivector $\ce_f$ at the holonomy base point $O$:
\be
h_{\partial f}(\o_{\mathrm{Spin}})=\exp\lt[\frac{\kappa}{3}\mathbf{a}_f\ce_f(O)\rt]\in\Slc.\label{spinhol}
\ee
Here both $\kappa$ and $\mathbf{a}_f$ are dimensionful and defined with respect to a certain length unit. But the product $\kappa\mathbf{a}_f$ is dimensionless and independent of unit. The bivector is expressed in terms of the area element $\epsilon_{\mu\nu}$ and tetrad $e^{\mu I}$ by $\ce^{IJ}=\epsilon_{\mu\nu}e^{\mu I} e^{\nu J} \in \slc$. The proof of this result is straight-forward (see \cite{HHKR} for details). From the holonomies of spin connection along the closed paths in $\pi_1\Big(\mathrm{sk}(\mathrm{Simplex})\Big)$, we can read the areas $\mathbf{a}_{ab}$ and the normal bivectors $\ce_{ab}$ of the triangles $\ff_{ab}$ \footnote{Note that there is a subtlety in identifying unambiguously the area and bivector of each triangle. But the subtlety can be resolved by requiring the tetrahedron convexity \cite{HHKR,curvedMink}.}. Then it turns out that the area and bivector data fix completely the constant curvature 4-simplex geometry, including the sign of constant curvature $\sgn(\kappa)$. Here we state the result and refer the reader to the proof in \cite{HHKR,3dblockHHKR} 

\begin{Theorem}\label{3d/4d}

There is a class of framed flat connection in $\cm_{\mathrm{flat}}(S^3\setminus\G_5,\Slc)$, in which each flat connection determines uniquely a nondegenerate, convex, geometrical 4-simplex with constant curvature $\kappa$ in Lorentzian signature \footnote{$\sgn(\kappa)$ is also determined by the flat connection. The magnitude of $\kappa$ depends on the length unit.}. The tetrahedra of the resulting 4-simplex are all space-like.

\end{Theorem}

If the class of framed flat connection is projected to $\cm_{\mathrm{flat}}(S^3\setminus\G_5,\PSlc)$, the correspondence with constant curvature 4-simplex is bijective. There are two ways to describe the class of framed flat connections satisfying the above correspondence with 4-simplex geometry: 

\begin{itemize}

\item We impose the following boundary condition on $\partial(S^3\setminus\G_5)=\Sig_{g=6}$: As it is manifest in the ideal triangulation, the closed surface $\Sig_{g=6}$ can be decomposed into five 4-holed spheres $\cs_{a=1,\cdots,5}$. We require that the $\Slc$ flat connection reduces to SU(2) flat connection, when it is restricted in each $\cs_a$ \cite{HHKR,3dblockHHKR} \footnote{It doesn't mean that the flat connections satisfying the boundary condition are SU(2) on entire $\Sig_{g=6}$, because different $\cs_a$ associate with different SU(2) subgroups in $\Slc$.}. The framed flat connection satisfying the boundary condition satisfies the correspondence to a constant curvature 4-simplex in Theorem \ref{3d/4d}. 

\item The framed flat connections on $S^3\setminus\G_5$ belong to a Lagrangian submanifold $\cl_{S^3\setminus\G_5}$ embedded in the phase space $\cp_{\partial(S^3\setminus\G_5)}$. The flat connections satisfying the correspondence in Theorem \ref{3d/4d} live in 2 branches $\a_{\rm 4d},\tilde{\a}_{\rm 4d}$ of $\cl_{S^3\setminus\G_5}$. The 2 branches $\a_{\rm 4d},\tilde{\a}_{\rm 4d}$ are related by a 4-simplex parity \cite{HHKR,3dblockHHKR}. Namely, given a flat connection $A$ on $\a_{\rm 4d}$, there is a unique flat connection $\tilde{A}$ on $\tilde{\a}_{\rm 4d}$, such that (1) $A,\tilde{A}$ determines the same 4-simplex geometry, but with opposite 4-simplex orientations; (2) $A,\tilde{A}$ induce the same SU(2) flat connections on all $\cs_a$. The pair of flat connection $A,\tilde{A}$ are referred to as a \emph{parity pair}, being an analog of the situation in \cite{semiclassical,semiclassicalEu,HZ,HZ1,hanPI}. 

\end{itemize}

A similar result can be proved at the level of the pairing $(S^2\setminus\{\mathrm{4pts}\},\mathrm{Tetra})$ from Eq.\Ref{fundtetra} (see \cite{curvedMink} for a proof, see also \cite{HHKR} for a sketch):  Any framed SU(2) flat connection in $\cm_{\mathrm{flat}}(S^2\setminus\{\mathrm{4pts}\},\mathrm{SU(2)})$ determines a uniquely a nondegenerate convex geometrical tetrahedron with constant curvature $\kappa$. As a difference from Theorem \ref{3d/4d}, the flat connections corresponding to tetrahedron geometry are dense in $\cm_{\mathrm{flat}}(S^2\setminus\{\mathrm{4pts}\},\mathrm{SU(2)})$. Again if the structure group is SO(3) instead of SU(2), the correspondence is bijective.

Actually the correspondence between SU(2) flat connection on 4-holed sphere and constant curvature tetrahedron is a preliminary step toward Theorem \ref{3d/4d}. It is because each vertex of $\G_5$ leads to a 4-holed sphere, which corresponds to one of the 5 tetrahedra on the boundary of 4-simplex (comparing Eq.\Ref{pi1M3} and Eq.\Ref{pi1M4}). The tetrahedron geometries reconstructed from SU(2) flat connections form the boundary geometry of 4-simplex. It is also the reason of boundary condition mentioned above. 

Recall the definition of framed flat connection on $M_3$ and $\partial M_3$, here for $S^3\setminus\G_5$ we denotes by $s_{ab}$ the framing flag on the annulus connecting the 4-holed spheres $\cs_a$ and $\cs_b$. We continue $s_{ab}$ to the 4-holed spheres by parallel transportation, although the continuation may introduce branch cuts by the nontrivial monodromies of cusps. Let's fix a point $\fp_a$ on $\cs_a$, and parallel transport $s_{ab}$ to $\fp_a$ and denotes the value by $s_{ab}(\fp_a)$. Firstly $s_{ab}(\fp_a)$ is an eigenvector of the holonomy $H_{ab}$ meridian to annulus cusps $\ell_{ab}$ based at $\fp_{a}$. Secondly $s_{ab}(\fp_a)$ is multivalued since a parallel transportation around $\ell_{ab}$ results in $s_{ab}(\fp_a)\mapsto \l_{ab} s_{ab}(\fp_a)$, where $\l_{ab}$ is the eigenvalue of $H_{ab}$. Because the flat connections we considered become SU(2) flat connections on each $\cs_a$, the holonomy $H_{ab}$ belongs to an SU(2) subgroup of $\Slc$, so it makes sense to endow $\C^2$ a Hermitian inner product $\langle\ ,\ \rangle$. We normalize the vector $s_{ab}(\fp_a)$ by the Hermitian inner product, and denote by 
\be
\xi_{ab}=\frac{s_{ab}(\fp_a)}{||s_{ab}(\fp_a)||}.\label{normalization}
\ee
$\xi_{ab}$ is defined up to a phase since the eigenvalue $\l_{ab}\in\mathrm{U(1)}$. The set of four $H_{ab}$'s ($b\neq a$) are sharing the same base point $\fp_a$ on $\cs_a$, so they satisfy the relation $H_{ab_4}H_{ab_3}H_{ab_2}H_{ab_1}=1$ by the representation of fundamental group. By the correspondence between SU(2) flat connection on 4-holed sphere and constant curvature tetrahedron, each $H_{ab}$ relates the triangle area $\mathbf{a}_{ab}$ and normal vector $\hat{\fn}_{ab}$ of the tetrahedron face by 
\be
\pm H_{ab}=\exp\lt[-\frac{i\kappa}{6}\mathbf{a}_{ab}\hat{\fn}_{ab}\cdot\vec{\sig}\rt]\quad\text{where}\quad \hat{\fn}_{ab}=\lag\xi_{ab},\vec{\sig}\xi_{ab}\rag\label{pmH}
\ee
where $\vec{\sig}$ denotes the Pauli matrices. The relation may be understood by considering Eq.\Ref{spinhol} and letting $e^\mu_0$ be the time-like normal of the 2-surface. $\pm$ comes from the 2-fold covering of SU(2) over SO(3) \footnote{Given an $\Slc$ flat connection on $S^3\setminus\G_5$ satisfying Theorem \ref{3d/4d}, one should first project down to $\PSlc$ which is bijective to constant curvature 4-simplices, then lift it back to $\Slc$. $+$ or $-$ is uniquely determined by asking the lift to be the same as the original $\Slc$ flat connection.}. Here we see that the framing flags of flat connection relate to the normal vectors of the tetrahedron faces \footnote{The dihedral angles between tetrahedron faces are given by (FIG.\ref{4holedsphere} for example) $\cos\phi_{ij}=\hat{\fn}_i\cdot\hat{\fn}_j$ for $(i,j)\neq (2,4)$ and $\cos\phi_{24}=H_1^{-1}\hat{\fn}_2\cdot\hat{\fn}_4$, where $\hat{\fn}_i$ is the normal of the triangle $\ff_{i}$ that doesn't connect to the vertex $i$. $\phi_{ij}$ is the dihedral angle between $\ff_{i}$ and $\Delta_j$. $H_i$ is the representative of $p_i$ in SU(2). The constant curvature $\kappa$ relates to the Gram matrix $\mathrm{Gram}_{ij}=-\cos\phi_{ij}$ by $\sgn(\kappa)=\sgn(\det \mathrm{Gram})$.}.

Let's come back to the moduli space of massive supersymmetric vacua $\cl_{\mathrm{SUSY}}(T_{S^3\setminus\G_5})$, which is isomorphic to the (open) moduli space of framed $\Slc$ flat connections $\cl_{S^3\setminus\G_5}(\mathbf{t})$ defined by the ideal triangulation $\mathbf{t}$ in Section \ref{3d/3dduality}. The flat connections in $\cl_{S^3\setminus\G_5}(\mathbf{t})$ give each ideal tetrahedron in $\mathbf{t}$ non-degenerate cross-ratios $z,z',z''\in \C\setminus\{0,1\}$. Let's check the cross-ratios in each ideal tetrahedron for the flat connections satisfying Theorem \ref{3d/4d}. The ideal triangulation $\mathbf{t}$ in FIG.\ref{5oct} has the following feature: Each ideal tetrahedron touches 4 edges (corresponding to 4 annuli cusps) in the graph $\G_5$ at the tetrahedron vertices. 3 of the 4 edges are connected to the same vertex of $\G_5$. In addition to these 3 edges, the ideal tetrahedron touches another edge of $\G_5$, which doesn't connect to the same vertex. Without loss of generality, take the ideal tetrahedron in Oct(1), which contributes an ideal triangle to $\cs_{2}$ (the tetrahedron with parameter $X_1$). The tetrahedron touches $\ell_{23},\ell_{24},\ell_{25}$ connecting to the vertex 2, and touches $\ell_{35}$ which doesn't connect to the vertex 2. The tetrahedron parameters $e^X,e^{X'},e^{X''}$ are the cross-ratios of the framing flags $s_{23},\ s_{24},\ s_{25}$, and $s_{35}$ when they are parallel transported to the same point. We choose to parallel transport the framing flags to $\fp_2$ on $\cs_2$ and denotes
\be
f_1=\xi_{25},\quad f_2=\xi_{23},\quad f_3=\xi_{24},\quad f_4=G_{23}\xi_{35}\propto G_{25}\xi_{53}
\ee
where $G_{ab}$ denotes a holonomy of flat connection traveling from $\fp_b$ on $\cs_b$ to $\fp_a$ on $\cs_a$. The cross-ratios are given by 
\be
e^X=\frac{(f_4\wedge f_2)(f_1\wedge f_3)}{(f_4\wedge f_1)(f_2\wedge f_3)},\quad 
e^{X'}=\frac{(f_2\wedge f_3)(f_1\wedge f_4)}{(f_2\wedge f_1)(f_3\wedge f_4)},\quad
e^{X''}=\frac{(f_1\wedge f_2)(f_3\wedge f_4)}{(f_1\wedge f_3)(f_2\wedge f_4)}
\ee
where $f\wedge f'\equiv f^1 f'^2-f^2 f'^1$, and all the cross-ratios are $\Slc$ invariant and invariant under the individual scaling of $f_1,\dots,f_4$. It is straight-forward to check that 
\be
e^Xe^{X'}e^{X''}=-1, \quad e^{X''}+e^{-X}=1,\quad e^{X}+e^{-X'}=1,\quad e^{X'}+e^{-X''}=1
\ee
being cyclic invariant under $X\mapsto X'\mapsto X''\mapsto X$. Any of the cross-ratios become $0,\ 1$ or $\infty$ if and only if one of the cross-ratio vanishes, i.e. there are two vectors $f_j$ and $f_k$ collinear. Firstly any pair in $\{f_1,f_2,f_3\}$ being collinear would imply a pair of $\xi$'s collinear at $\fp_2$. Secondly $f_4\propto f_1$ or $f_4\propto f_2$ would imply $\xi_{35}\propto\xi_{32}$ at $\fp_3\in\cs_3$ or $\xi_{53}\propto\xi_{52}$ at $\fp_5\in\cs_5$. Therefore $\cl_{S^3\setminus\G_5}(\mathbf{t})$ doesn't contain the flat connections which lead to collinear $\xi$'s on the 4-holed spheres\footnote{That $\xi$'s are non-collinear excludes some degenerate tetrahedron geometries, from the expression of dihedral angle.}. Thirdly $\cl_{S^3\setminus\G_5}(\mathbf{t})$ doesn't have collinear $f_4$ and $ f_3$. The holonomy $G_{ab}$ has been computed for the flat connections satisfying Theorem \ref{3d/4d} \cite{HHKR,3dblockHHKR,HHKRshort}, and relates to the hyper-dihedral (boost) angle $\Theta_{ab}$ by \footnote{$\Theta_{ab}$ is the boost angle between two tetrahedra sharing $\ff_{ab}$ on the boundary of constant curvature 4-simplex.} 
\footnote{$s_{ab}$ is a flat section over the annulus implies $G_{ab}s_{ab}(\fp_b)=s_{ab}(\fp_a)$. In terms of $\xi_{ab}$, we have $G_{ab}\xi_{ba}=\frac{||s_{ab}(\fp_a)||}{||s_{ab}(\fp_b)||}\xi_{ab}$ by Eq.\Ref{normalization}. } 
\be
G_{ab}=M_{ab}\left(
\begin{array}{cc}
 e^{-\half\sgn(V_4)\Theta_{ab} } & 0     \\
 0  &  e^{\half\sgn(V_4)\Theta_{ab} }
\end{array}
\right)M_{ba}^{-1},\quad \text{where}\quad
M_{ab}=\left(
\begin{array}{cc}
 \xi^1_{ab} &  -\bar{\xi}^2_{ab}    \\
 \xi^2_{ab}  &  \bar{\xi}^1_{ab}
\end{array}
\right).\label{Gab}
\ee
Between parity pair $A,\tilde{A}$, the sign of 4-volume $\sgn(V_4)$ flips sign, while the angle $\Theta_{ab}$ does't flip sign. That $f_4$ and $f_3$ are not collinear implies \footnote{If any cross-ratio is assumed to be degenerate $z,z',z''\in \{0,1,\infty\}$ in any of the 4 ideal tetrahedron in Oct(1), it would only imply a pair of collinear $\xi$'s at certain $\fp_a\in\cs_a$, or imply Eq.\Ref{3524}, and no more. Eq.\Ref{3524} means that $\cl_{S^3\setminus\G_5}(\mathbf{t})$ doesn't contain the flat connection which makes $s_{35}$ and $s_{24}$ collinear when parallel transport to the same point. This condition is the same in all the 4 ideal tetrahedron in Oct(1). There are 4 more conditions similar to Eq.\Ref{3524} from other 4 ideal octahedra.}
\be
\left(
\begin{array}{cc}
 e^{-\half\sgn(V_4)\Theta_{23} } & 0     \\
 0  &  e^{\half\sgn(V_4)\Theta_{23}}
\end{array}
\right)M_{32}^{-1} \xi_{35}\not\propto  M^{-1}_{23}\xi_{24}\label{3524}
\ee
It means that for certain $\xi$'s, special values of $\Theta_{ab}$ might not be included in $\cl_{S^3\setminus\G_5}(\mathbf{t})$. See Appendix \ref{collinear} for some additional geometrical meanings of the condition Eq.\Ref{3524}. $\cl_{S^3\setminus\G_5}(\mathbf{t})$ as an open subset includes generic flat connections that satisfies the correspondence in Theorem \ref{3d/4d}. Some special flat connections, which are not captured by $\cl_{S^3\setminus\G_5}(\mathbf{t})$, might still satisfy Theorem \ref{3d/4d}. But they form some lower dimensional subspaces, and can be included by the closure of $\cl_{S^3\setminus\G_5}(\mathbf{t})$. As a result, the moduli space of massive supersymmetric vacua $\cl_{\mathrm{SUSY}}(T_{S^3\setminus\G_5})$, when we take the closure, includes all the nondegnerate, convex 4-simplex of constant curvature.

\begin{figure}[h]
\begin{center}
\includegraphics[width=12cm]{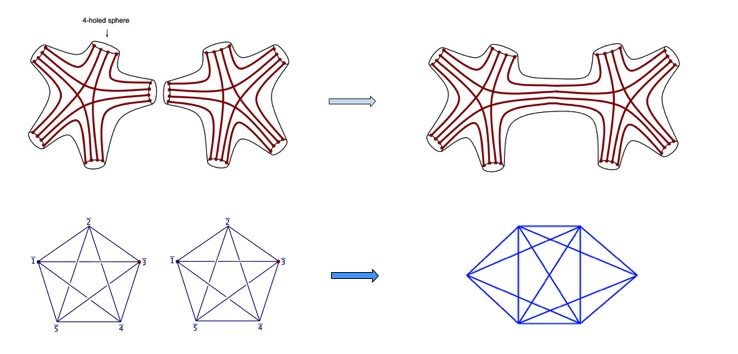}
\caption{$\sm_3$ are obtained by gluing a number of $S^3\setminus\G_5$, and relates to a simplicial manifold with the same number of 4-simplices. }
\label{m3andm4}
\end{center}
\end{figure}

The result can be generalized to $\sm_3$ obtained by gluing a number of $S^3\setminus\G_5$, which relates to a simplicial manifold with the same number of 4-simplices (FIG.\ref{m3andm4}). When two copies of $S^3\setminus\G_5$ are glued through a pair of 4-holed spheres $\cs,\cs'$, the fundamental group $\pi_1(\sm_3)$ of the resulting $\sm_3$ are given by two copies of $\pi_1(S^3\setminus\G_5)$ modulo the identification of the generators on $\cs$ and $\cs'$ ($\pi_1(\cs)\simeq\pi_1(\cs')$ with the isomorphism denoted by $\ci$). The resulting $\pi_1(\sm_3)$ is isomorphic to the fundamental group of a 1-skeleton $\pi_1(\mathrm{sk}(\sm_4))$ from the 4d polyhedron $\sm_4$ obtained by gluing a pair of 4-simplices. However here the 1-skeleton $\mathrm{sk}(\sm_4)$ includes the edges of the interface (the tetrahedron shared by the pair of 4-simplices), as it is drawn in FIG.\ref{m3andm4}. It is the key point to make general simplicial geometries on $\sm_4$ which make curvatures in gluing 4-simplices.

Given two (framed) flat connections $A,A'$ as homomorphisms (representations) $\pi_1(S^3\setminus\G_5)\to \Slc$ modulo conjugation, they can be glued and give a flat connection $\sa$ on $\sm_3$ if they induce the same representation to $\pi_1(\cs)$ and $\pi_1(\cs')$ (i.e. $A=A'\circ \ci$). Let's consider $A,A'$ satisfying Theorem \ref{3d/4d} and corresponding to 2 nondegenerate, convex, constant curvature 4-simplices $\Fs,\Fs'$. When $A,A'$ glue and give a flat connection $\sa$ on $\sm_3$, they induce the same SU(2) representation to $\pi_1(\cs)$ and $\pi_1(\cs')$. The SU(2) representation (modulo conjugation) reconstructs a unique geometrical tetrahedron of constant curvature. The constant curvature tetrahedron belongs to both 4-simplices $\Fs,\Fs'$ for $A,A'$ satisfying Theorem \ref{3d/4d}. Therefore the flat connection $\sa$ on $\sm_3$ determines a 4-dimensional geometrical polyhedron obtained by gluing constant curvature 4-simplices $\Fs,\Fs'$. The procedure can be continued to arbitrary $\sm_3=\cup_{i=1}^N (S^3\setminus\G_5)$. It identifies a class of framed flat connections on $\sm_3$, which determine all 4-dimensional geometrical polyhedron $(\sm_4,g)$ obtained by gluing $N$ nondegenerate, convex, constant curvature 4-simplices \footnote{All 4d polyhedron geometries are covered because at the level of a single 4-simplex, all nondegenerate, convex, constant curvature 4-simplex geometry are covered by the closure of $\cl_{S^3\setminus\G_5}(\mathbf{t})$.  }.

There are two remarks for the geometrical polyhedron $(\sm_4,g)$ determined by the pair $(\sm_3,\sa)$:

\begin{itemize}

\item The simplicial manifold $\sm_4$ can have all possible discrete geometries\footnote{The discrete geometries are of the Regge type \cite{regge}. The only difference is that here the geometries are formed by gluing constant curvature 4-simplices rather than flat 4-simplices.}. The only restriction is that locally inside each 4-simplex, the geometry is of constant curvature with fixed $\kappa$. But the 4-simplices can have e.g. different shapes and edge-lengths. The curvature of the resulting geometrical polyhedron $(\sm_4,g)$ is generic, which is described by the deficit angles $\eps({\ff})$ located at the triangles $\ff$.

\item $(\sm_4,g)$ may not have a global orientation, i.e. different 4-simplices may obtain different orientations, because of the existence of parity pairs $A,\tilde{A}$. The parity pairs $A,\tilde{A}$ induce the same SU(2) flat connection on all $\cs_a$. There is freedom to choose $A$ or $\tilde{A}$ on each individual 4-simplex. All $2^N$ choices give $2^N$ flat connections on $\sm_3$ which determines the same geometry on $\sm_3$, but with different assignment of 4-simplex orientations. Within the $2^N$ choices, there are a unique pair of flat connections, denoted by $(\sa,\tilde{\sa})$, correspond to a geometry on $\sm_3$ with uniform orientations in all 4-simplices. $(\sa,\tilde{\sa})$ may be referred to as \emph{global parity pair}, since they determine two opposite global orientations of $\sm$. 

\end{itemize}

The ideal triangulation $\mathbf{t}$ of $\sm_3$ is easily obtained by gluing the ideal triangulation of $S^3\setminus\G_5$. The flat connections $\sa$ corresponding to 4-geometry $(\sm_4,g)$ are obtained by gluing the flat connections $A$ satisfying Theorem \ref{3d/4d} on $S^3\setminus\G_5$. Generically $A$ satisfying Theorem \ref{3d/4d} induce nondegenerate cross-ratios $z,z',z''\in \C\setminus\{0,1\}$ in all ideal tetrahedra in $S^3\setminus\G_5$. Then almost all $\sa$ having 4-geometry correspondence induce nondegenerate cross-ratios $z,z',z''\in \C\setminus\{0,1\}$ in all ideal tetrahedra in $\sm_3$, thus belong to $\cl_{\sm_3}(\mathbf{t})$. The exceptional $\sa$'s can be included by taking the closure of $\cl_{\sm_3}(\mathbf{t})$. As a result, the moduli space of massive supersymmetric vacua $\cl_{\mathrm{SUSY}}(T_{\sm_3})$, when we take the closure, includes all the simplical geometries $(\sm_4,g)$, which is made by gluing nondegnerate, convex, constant curvature 4-simplices.

\subsection{Complex Fenchel-Nielsen Coordinate and Geometrical Quantities}\label{FNandSG}

In the last subsection, we have established the correspondence between a class of supersymmetric vacua in $\cl_{\mathrm{SUSY}}(T_{\sm_3})$ and the 4d simplicial geometries on $\sm_4$ with constant curvature 4-simplices. The correspondence allows us to make a dictionary between the parameters $x_i,y_i$ in $\cl_{\mathrm{SUSY}}(T_{\sm_3})$ and geometrical quantities of $(\sm_4,g)$. 

In the supersymmetric gauge theory $T_{\sm_3}$, the parameter $x_i=\exp(m_i)$ where $m_i$ is the 3d real mass $m_i^{\mathrm{3d}}$ complexified by the Wilson-line along the $S^1$ fiber over $D^2$. By the construction of $T_{\sm_3}$ in Section \ref{3d/3dduality}, $x_i$ map to the position coordinates (which are also labelled by $x_i=\exp(m_i)$) in the phase space $\cp_{\partial\sm_3}$, because of the isomorphism $\cl_{\mathrm{SUSY}}(T_{\sm_3})\simeq \cl_{\sm_3}(\mathbf{t})$. Similarly the effective background FI parameters $y_i$ map to the momentum coordinates $y_i=\exp(p_i)$ in $\cp_{\partial\sm_3}$. 

In general, the boundary components of $\sm_3$ can be classified into (1) the geodesic boundary components being a set of 4-holed spheres $\cs$, (2) the cusp annuli $\ell$ connecting to a pair of boundary 4-holed spheres, and (3) the torus cusps $T^2$ which doesn't connect to the geodesic boundary. The internal torus cusp happens e.g. when we glue a pair of $S^3\setminus\G_5$ through 2 pairs of 4-holed spheres, or when we glue 3 copies of $S^3\setminus\G_5$, and each pair of $S^3\setminus\G_5$ share a 4-holed sphere. In each of these cases, $\sm_3$ is the complement of a graph in a 3-manifold with nontrivial cycles. The corresponding 4d simplicial manifold $\sm_4$ has a set of internal triangles $\ff_{T^2}$ which are not contained by a tetrahedron on the boundary $\partial\sm_4$. Here a tetrahedron (labelled by $\ft_\cs$) on $\partial\sm_4$ relates to a 4-holed sphere $\cs$ on $\partial\sm_3$. 

Because of the classification of $\partial\sm_3$, the position coordinates $x_i$ contains 3 different types of coordinates $\vec{x}=(\l^2_{T^2},\l_{\ell},{\sf m}_\cs)$. Both $\l_{T^2}=\exp(\L_{T^2})$ and $\l_\ell=\exp(\L_\ell)$ are the eigenvalues of meridian monodromy to the cusps. $\l_\ell$ is the complex FN length variable constructed in Section \ref{coordinates} in the case of $S^3\setminus\G_5$. ${\sf m}_\cs=\exp(M_\cs)$ is the position coordinate associates to each 4-holed sphere $\cs\subset\partial\sm_3$. In Section \ref{coordinates}, we have constructed the phase space coordinates ${\sf m}_a=\exp(M_a),{\sf p}_a=\exp(P_a)$ for each 4-holed sphere $\cs_a$.

$\l_{T^2},\l_{\ell}$ are eigenvalues of the holonomies $\o_{\mathrm{flat}}(\fl_{T^2}),\o_{\mathrm{flat}}(\fl_{\ell})$ along the cycles $\fl_{T^2},\fl_\ell$ meridian to the cusps. By the isomorphism $\pi_1(\sm_3)\simeq \pi_1(\mathrm{sk}(\sm_4))$, $\fl_{T_2},\fl_\ell$ map to the closed paths $p_{T^2},p_{\ell}$ around the triangles $\ff_{T^2}, \ff_{\ell}$. $\ff_{T^2}$ is an internal triangle which are not contained by a tetrahedron on $\partial\sm_4$. $\ff_{\ell}$ is a boundary triangle shared by a pair of tetrahedra $\ft_{\cs_1},\ft_{\cs_2}$ on $\partial\sm_4$, where $\cs_1,\cs_2$ are connected by the annulus $\ell$ in $\sm_3$. The isomorphism $\pi_1(\sm_3)\simeq \pi_1(\mathrm{sk}(\sm_4))$ relates the spin connection $\o_{\mathrm{Spin}}$ on $\sm_4$ to the flat connection $o_{\mathrm{flat}}$ on $\sm_3$ by the commuting triangle \Ref{Smap} and Eq.\Ref{FlatSpin}. So we have $\pm\o_{\mathrm{flat}}(\fl_{T^2})=\o_{\mathrm{Spin}}(p_{T^2})$ and $\pm\o_{\mathrm{flat}}(\fl_{\ell})=\o_{\mathrm{Spin}}(p_{\ell})$ ($\pm$ appears because we consider $\Slc$ flat connection instead of $\PSlc$). The holonomy of spin connection is given by Eq.\Ref{spinhol}, then up to conjugations $\o_{\mathrm{flat}}(\fl_{T^2}),\o_{\mathrm{flat}}(\fl_{\ell})$ can be expressed by Eq.\Ref{pmH}. It is manifest that the eigenvalues $\l_{T^2},\l_{\ell}$ (the complexified twisted mass parameters in $T_{\sm_3}$) are relates to the triangle areas:
\be
\l_{T^2}^2=\exp\lt[-\frac{i\kappa}{3}\mathbf{a}(\ff_{T^2})\rt],\quad\quad 
\l_{\ell}=\exp\lt[-\frac{i\kappa}{6}\mathbf{a}(\ff_{\ell})+\pi i\fs_{\ell}\rt]
\ee 
where $\fs_\ell\in\{0,1\}$ parametrizes the lifts from $\PSlc$ to $\Slc$.

The momentum coordinates in $\cp_{\partial\sm_3}$, being the effective FI parameters $y_i$ of $T_{\sm_3}$, also contain 3 different types of coordinates $\vec{y}=(\t_{T^2},\t_{\ell},{\sf p}_\cs)$. Here $\t_{T^2}$ is the eigenvalue of the longitude holonomy on torus cusp. The longitude holonomy is nontrivial because the longitude cycle of a torus cusp is not contractible in $\sm_3$. $\t_{\ell}$ is the complex FN twist coordinate along the annulus $\ell$. $\t_\ell$ has been constructed in Section \ref{coordinates} in the case of $S^3\setminus\G_5$. ${\sf p}_\cs=\exp(P_\cs)$ is the momentum coordinate associated to the 4-holed sphere $\cs$.

For the supersymmetric vacua corresponding to 4d simplicial geometry, the effective FI parameters $\t_{T^2},\t_{\ell}$ relate to the deficit angles $\eps(\ff_{T^2})$ and dihedral angles $\Theta(\ff_{\ell})$ in $(\sm_4,g)$. Let's first consider the longitude eigenvalue $\t_{T^2}$. By Eq.\Ref{Gab} in $S^3\setminus\G_5$ translated into the suitable notation for $\sm_3$, we obtain
\be
G(\cs_2,\cs_1)\,\xi_\ell(\cs_1)=e^{-\half\sgn(V_4)\Theta_{\cs_2,\cs_1}(\ell)}\,\xi_{\ell}(\cs_2)
\ee
where $\xi_{\ell}(\cs)$ is the framing flag $s_\ell$ parallel transport to a point $\fp_\cs\in\cs$, followed by a normalization by Eq.\Ref{normalization}. $G(\cs_2,\cs_1)$ is a holonomy of flat connection  along any contour from $\fp_{\cs_1}$ to $\fp_{\cs_2}$ on the annulus $\ell$. $\Theta_{\cs_2,\cs_1}(\ell)\equiv \Theta_\Fs(\ff_\ell)$ is the hyper-dihedral boost angle between the two tetrahedra $\ft_{\cs_1},\ft_{\cs_2}$ sharing the triangle $\ff_\ell$ in a 4-simplex $\Fs$. When two copies $S^3\setminus \G_5$ are glued such that $\cs_1$ becomes internal, the annulus $\ell$ is extended in the resulting $\sm_3$ and connecting $\cs_2, \cs_0\subset\partial\sm_3$. We denote by $G(\cs_2,\cs_0)=G(\cs_2,\cs_1)G(\cs_1,S_0)$, then we have
\be
G(\cs_2,\cs_0)\,\xi_\ell(\cs_0)&=&e^{-\half\sgn(V_4)\Theta_{\cs_1,\cs_0}(\ell)}\,G(\cs_2,\cs_1)\xi_{\ell}(\cs_1)\nonumber\\
&=&e^{-\half\sgn(V_4)\lt[\Theta_{\cs_2,\cs_1}(\ell)+\Theta_{\cs_1,\cs_0}(\ell)\rt]}\,\xi_{\ell}(\cs_2)
\ee
Here we always consider the flat connections that correspond to globally oriented $(\sm_4,g)$, in which $\sgn(V_4)$ is a constant. We can continue to have more $S^3\setminus\G_5$ glued, and in general 
\be
G(\cs_n,\cs_0)\,\xi_\ell(\cs_0)&=&e^{-\half\sgn(V_4)\sum_{i=1}^n\Theta_{\cs_i,\cs_{i-1}}(\ell)}\,\xi_{\ell}(\cs_n)\label{Gxiexi}
\ee
When $\ell=T^2$ is a torus cusp, $\cs_0$ is identified with $\cs_n$ so that $\xi_\ell(\cs_0)=\xi_{\ell}(\cs_n)$ up to a phase $e^{\frac{i}{2}\theta(\ff_{T^2})}$. Then $G(S_n,S_0)$ is the holonomy along a longitude cycle, whose eigenvalue relates the deficit angle $\eps(\ff_{T^2})$ by 
\be
\t_{T^2}=e^{-\half\sgn(V_4)\,\eps(\ff_{T^2})-\frac{i}{2}\theta(\ff_{T^2})},\ \ \ \ \text{where}\ \ \ \ \eps(\ff_{T^2})=\sum_{i=1}^n\Theta_{\cs_i,\cs_{i-1}}(T^2)=\sum_{\Fs,\ \ff_{T^2}\subset \Fs}\Theta_\Fs(\ff_{T^2}).\label{tauT2}
\ee
See e.g. \cite{foxon} for the definition of deficit angle $\eps(\ff_{T^2})$ in Lorentzian signature. The angle $\theta(\ff_{T^2})$ also depends on the choice of longitude cycle since the definition of $\t_{T^2}$ depends on the choice. It turns out that there is a longitude cycle whose holonomy gives 
\be
\theta(\ff_{T^2})=\eta(\ff_{T^2})\pi,\ \ \ \ \text{where}\ \ \ \ \eta(\ff_{T^2})\in \{0,1,-1\}
\ee
This result is explained in Appendix \ref{0andpi}. We keep this longitude cycle as a part of the definition for the coordinate $\t_{T^2}$. Importantly a simplicial spacetime $(\sm_4,g)$ being globally time-oriented implies $\theta(\ff_{T^2})=0$.

\begin{figure}[h]
\begin{center}
\includegraphics[width=7cm]{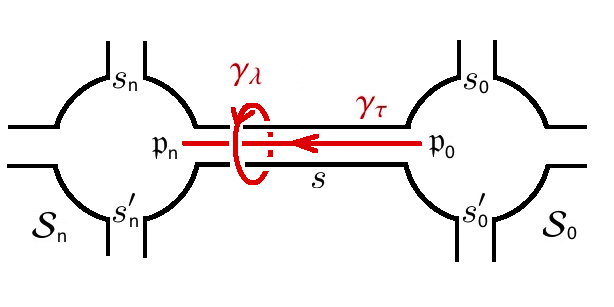}
\caption{The contour $\g_\t$ used to define the complex FN twist $\t_\ell$, and the meridian cycle $\g_\l$ used to define the complex FN length $\l_\ell$.}
\label{xandtau}
\end{center}
\end{figure}

Let's consider the annulus cusps $\ell$ and the complex FN twist coordinate $\t_\ell$. Given $\ell$ connecting a pair of 4-holed spheres $\cs_0,\cs_n$, the FN twist $\t_\ell$ is defined in the following way: Let $s$ be the framing flag for $\ell$, and $s_{0,n},s_{0,n}'$ be the framing flags for a pair of other cusps connecting $\cs_{0,n}$. Then the complex FN twist is defined by (see e.g. \cite{DGV})
\be
\t_\ell=-\frac{( s_{0}\wedge s_{0}')}{( s_{0}\wedge s)( s_{0}'\wedge s )}\frac{( s_{n}\wedge s)(s_{n}'\wedge s)}{( s_{n}\wedge s_{n}')}.
\ee 
where $s\wedge s'$ are evaluated at a common point after parallel transportation. Without loss of generality, we evaluate the first ratio with factors ${( s_{0}\wedge s_{0}')},( s_{0}\wedge s),( s_{0}'\wedge s )$ at a point $\fp_0 \in\cs_0$, and evaluate the second ratio with factors ${( s_{n}\wedge s),(s_{n}'\wedge s)},{( s_{n}\wedge s_{n}')}$ at a point $\fp_n\in\cs_n$. The evaluation involves both $s(\fp_0)$ and $s(\fp_n)$ at two ends of $\ell$, while the parallel transportation between $s(\fp_0)$ and $s(\fp_n)$ depends on a choice of contour $\g_\t$ connecting $\fp_0,\fp_n$ (FIG.\ref{xandtau}). Different $\g_\t$ may transform $s(\fp_n)\to\l_\ell s(\fp_n)$ but keep $s(\fp_0)$ invariant. Moreover by definition, $\t_\ell$ also depend on the choice of two other auxiliary cusps for each of $\cs_0,\cs_n$. The choices of $\g_\t$ and the auxiliary cusps are part of the definition for $\t_\ell$. It turns out that the choices in defining $\t_\ell$ doesn't affect our result in Section \ref{HBQG}.

Recall $\xi_\ell(\cs_0)= s(\fp_0)/||s(\fp_0)||$ and $\xi_\ell(\cs_n)= s(\fp_n)/||s(\fp_n)||$, and denotes $\xi(\cs_{0,n})= s_{0,n}(\fp_{0,n})/||s_{0,n}(\fp_{0,n})||$ and $\xi'(\cs_{0,n})= s'_{0,n}(\fp_{0,n})/||s'_{0,n}(\fp_{0,n})||$. We consider $G(\cs_n,\cs_0)$ to be the parallel transportation along $\g_\t$, and compute
\be
\t_\ell&=&-\frac{\Big( \xi(\cs_{0})\wedge \xi'(\cs_{0})\Big)}{\Big( \xi(\cs_{0})\wedge \xi_\ell(\cs_0) \Big)\Big( \xi'(\cs_{0})\wedge \xi_\ell(\cs_0) \Big)} 
\frac{\Big( \xi(\cs_{n})\wedge G(\cs_n,\cs_0)\,\xi_\ell(\cs_0)\Big)\Big(\xi'(\cs_{n})\wedge  G(\cs_n,\cs_0)\,\xi_\ell(\cs_0)\Big)}{\Big( \xi(\cs_{n})\wedge \xi'(\cs_{0,n})\Big)}\nonumber\\
&=&e^{-\sgn(V_4)\,\Theta(\ff_\ell)}\chi_{\ell}(\xi)\label{tauell}
\ee
where $\chi_{\ell}(\xi)$ is a short-hand notation for
\be
\chi_{\ell}(\xi)=-\frac{\Big( \xi(\cs_{0})\wedge \xi'(\cs_{0})\Big)}{\Big( \xi(\cs_{0})\wedge \xi_\ell(\cs_0) \Big)\Big( \xi'(\cs_{0})\wedge \xi_\ell(\cs_0) \Big)} \frac{\Big( \xi(\cs_{n})\wedge \xi_{\ell}(\cs_n)\Big)\Big(\xi'(\cs_{n})\wedge  \xi_{\ell}(\cs_n)\Big)}{\Big( \xi(\cs_{n})\wedge \xi'(\cs_{n})\Big)}.
\ee
$\Theta(\ff_\ell)$ is the hyper-dihedral angle at the boundary triangle $\ff_\ell$. The hyper-dihedral angle is between two boundary tetrahedra $\ft_{\cs_0}$ and $\ft_{\cs_n}$ sharing $\ff_\ell$. Here $\chi_{\ell}(\xi)$ depends on the phase difference between $\xi_\ell(\cs_0),\xi_\ell(\cs_n)$, thus depends choice of $\g_\t$. But $\Theta(\ff_\ell)$ is unambiguous (independent of the choices in defining $\t_\ell$) for the flat connections satisfying Theorem \ref{3d/4d} \footnote{The essential reason is that for the flat connections satisfying Theorem \ref{3d/4d}, they reduce to SU(2) flat connections on each 4-holed sphere.}.

To summarize, for the supersymmetric vacua in $\cl_{\mathrm{SUSY}}(T_{\sm_3})$ corresponding to the simplicial geometries on $\sm_4$, the complex twisted mass parameters are given by $\vec{x}=(\l^2_{T^2},\l_{\ell},{\sf m}_\cs)$, in which $\l^2_{T^2},\l_{\ell}$ relate to the areas $\mathbf{a}(\ff_{T^2}),\mathbf{a}(\ff_{\ell})$ of internal and boundary triangles $\ff_{T^2},\ff_{\ell}$. The effective FI parameters are given by $\vec{y}=(\t_{T^2},\t_{\ell},{\sf p}_\cs)$, in which $\t_{T^2}$ relates to the deficit angle $\eps(\ff_{T^2})$ at an internal triangle $\ff_{T^2}$, and $\t_{\ell}$ relates to the hyper-dihedral angle $\Theta(\ff_{\ell})$ at a boundary triangle $\ff_{\ell}$. The pair ${\sf m}_\cs,{\sf p}_\cs$ parametrize the shapes of tetrahedron $\ft_\cs$ at the boundary $\partial \sm_4$ \cite{curvedMink}.

The symplectic structure $\O_{\mathrm{SUSY}}$ in Eq.\Ref{Osusy} is written as
\be
\O_{\mathrm{SUSY}}=\sum_{T^2}\frac{\rmd \t_{T^2}}{\t_{T^2}}\wedge \frac{\rmd \l^2_{T^2}}{\l^2_{T^2}}+\sum_{\ell}\frac{\rmd \t_{\ell}}{\t_\ell}\wedge\frac{\rmd \l_\ell}{\l_\ell}+\sum_{\cs}\frac{\rmd {\sf p}_\cs}{{\sf p}_\cs}\wedge\frac{\rmd {\sf m}_\cs}{{\sf m}_\cs}
\ee
Being the boundary components in $\partial\sm_3$, $T^2$ denotes a torus cusp, $\ell$ denotes an annulus cusp, and $\cs$ denotes a 4-holed sphere. By the identification between $(\vec{x},\vec{y})$ and the symplectic coordinates in $\cp_{\partial \sm_3}$, $\O_{\mathrm{SUSY}}$ coincides with the Atiyah-Bott-Goldman symplectic form $\O=\int_{\partial \sm_3}\tr\lt(\delta \Fa\wedge\delta \Fa\rt)$.

There is a description of global parity pair $(\sa,\tilde{\sa})$ in terms of the coordinates $\vec{x},\vec{y}$. As it is mentioned before, $(\sa,\tilde{\sa})$, as a pair of $\Slc$ flat connections on $\sm_3$, induce the same SU(2) flat connection on each 4-holed sphere, including $\cs\subset\partial\sm_3$ and the interface for gluing copies of $S^3\setminus\G_5$. Therefore $(\sa,\tilde{\sa})$ have the same $\vec{x}=(\l^2_{T^2},\l_{\ell},{\sf m}_\cs)$ and the same ${\sf p}_\cs$. But they have different $\t_{T^2}$ and $\t_{\ell}$, because of Eqs.\Ref{tauT2} and \Ref{tauell} where $\sgn(V_4)$ flips sign between $\sa,\tilde{\sa}$.

We pick a ``boundary data'' $\vec{x}=(\l^2_{T^2},\l_{\ell},{\sf m}_\cs)$ and ${\sf p}_\cs$ where (1) $\l^2_{T^2}\in\mathrm{U(1)}$, and (2) ${\sf m}_\cs, {\sf p}_\cs$ parametrize SU(2) flat connections on $\cs\subset\partial\sm_3$ with conjugacy class $\l_{\ell}\in\mathrm{U(1)}$ at holes. The boundary data of this type, although include both position and momentum variables ${\sf m}_\cs,{\sf p}_\cs$, is natural from the point of view of geometries on $\sm_4$. Indeed correspondingly on $\sm_4$, $\l^2_{T^2},\l_{\ell}$ give areas to all internal and boundary triangles, and ${\sf m}_\cs, {\sf p}_\cs$ determines the shapes of boundary tetrahedra with the given areas. There are finitely many (locally constant curvature) simplicial geometries on $\sm_4$ satisfying the data $(\l^2_{T^2},\l_{\ell},{\sf m}_\cs,{\sf p}_\cs)$, which corresponds to finitely many supersymmetric vacua in $\cl_{\mathrm{SUSY}}(T_{\sm_3})\simeq \cl_{\sm_3}(\mathbf{t})$. Varying $\vec{x}=(\l^2_{T^2},\l_{\ell},{\sf m}_\cs)$ and ${\sf p}_\cs$ gives finitely many branches of supersymmetric vacua in $\cl_{\mathrm{SUSY}}(T_{\sm_3})$, which correspond to simplicial geometries on $\sm_4$. We denote these branches by $\a_{\mathrm{4d}}$. In general the collection of $\a_{\mathrm{4d}}$ is a subset of all branches in  $\cl_{\mathrm{SUSY}}(T_{\sm_3})$, because ${\sf p}_\cs$ is also specified in addition to $\vec{x}$. The boundary data ${\sf p}_\cs={\sf p}_\cs^{(\a_{\rm 4d})}(\l_T^2,\l_\ell,{\sf m}_\cs)$ is a solution of $\mathbf{A}_i(x,y)=0$, and is of the same value by all $\a_{\rm 4d}$ determined by $(\l_T^2,\l_\ell,{\sf m}_\cs,{\sf p}_\cs)$.

The data $(\l^2_{T^2},\l_{\ell},{\sf m}_\cs,{\sf p}_\cs)$ are constrained for the geometries on $\sm_4$. Firstly we have seen that ${\sf p}_\cs$ has to satisfy $\mathbf{A}_i(x,y)=0$. In addition, $\l^2_{T^2},\l_{\ell}\in\mathrm{U(1)}$, which give triangle areas in $\sm_4$, in general can not be arbitrary for simplicial geometries. The allowed values of triangle areas are usually called \emph{Regge-like} areas \cite{CFsemiclassical,HZ}.

Here we only consider the simplicial geometries with a global orientation, i.e. $\sgn(V_4)$ is a constant on $\sm_4$. Thus each $\a_{\mathrm{4d}}$ is paired by $\tilde{\a}_{\mathrm{4d}}$, because the global parity pair $(\sa,\tilde{\sa})$ share the same data $\vec{x}=(\l^2_{T^2},\l_{\ell},{\sf m}_\cs)$ and ${\sf p}_\cs$. $(\sa,\tilde{\sa})$ determine the same geometry on $\sm_4$ but give opposite orientations.

\section{Holomorphic Block and 4-dimensional Quantum Geometry}\label{HBQG}

Recall that given the complex twisted masses $\vec{x}$, there are a finite number of supersymmetric massive ground states 
\be
|\a\rangle\quad \leftrightarrow\quad \vec{s}^{(\a)}(\vec{x})\quad \rightarrow\quad \vec{y}^{(\a)}(\vec{x}).
\ee 
Varying $\vec{x}$, $\a$ then labels the branches of supersymmetric vacua in $\cl_{\mathrm{SUSY}}(T_{\sm_3})$. Let's pick the branches $\{\a_{\rm 4d}\}\subset \{\a\}$ which correspond to simplicial geometries on $\sm_4$. We propose that the holomorphic block $B^{\a}_{\sm_3}(\vec{x},q)$ from the theory $T_{\sm_3}$ with $\a=\a_{\rm 4d}$ is a quantum state for 4-dimensional simplicial geometry on $\sm_4$. 

The reason of the proposal is simple: $B^{\a}_{\sm_3}(\vec{x},q)$ satisfies the line-operator Ward identity Eq.\Ref{AB=0}, thus is a state from the quantization of Lagrangian subamnifold $\cl_{\mathrm{SUSY}}(T_{\sm_3})$ at the branch $\a$. The branch $\a=\a_{\rm 4d}$ of $\cl_{\mathrm{SUSY}}(T_{\sm_3})$ contains the supersymmetric vacua which correspond to simplicial geometries on $\sm_4$. These supersymmetric vacua in the branch $\a_{\rm 4d}$ are parametrized by $\vec{x}=(\l^2_{T^2},\l_{\ell},{\sf m}_\cs)$ with some restrictions, i.e. (1) $\l^2_{T^2},\l_{\ell}\in\mathrm{U(1)}$, and (2) ${\sf m}_\cs, {\sf p}^{(\a_{\rm 4d})}_\cs$ give SU(2) flat connections on $\cs\subset\partial\sm_3$ with conjugacy class $\l_{\ell}$ at holes.

Being a state for 4d simplicial geometry, $B^{\a}_{\sm_3}(\vec{x},q)$ with $\a=\a_{\rm 4d}$ should encode certain dynamics of 4d geometry on $\sm_4$. To understand the dynamics, we consider the perturbative behavior of holomorphic block as $\hbar\to 0$ discussed in Section \ref{BPS}.  Recall Eq.\Ref{Bvth}, as well as apply $\a=\a_{\rm 4d}$ and the Liouville 1-form
\be
\vth=\sum_{T^2}\ln{ \t_{T^2}}\frac{\rmd \l^2_{T^2}}{\l^2_{T^2}}+\sum_{\ell}\ln{ \t_{\ell}}\frac{\rmd \l_\ell}{\l_\ell}+\sum_{\cs}\ln{{\sf p}_\cs}\frac{\rmd {\sf m}_\cs}{{\sf m}_\cs}
\ee
with $\rmd\vth=\O_{\mathrm{SUSY}}$. When $\vth$ is restricted in submanifold in the branch $\a_{\rm 4d}$ in which the supersymmetric vacua correspond to simplicial geometries on $\sm_4$, it can be expressed in terms of the geometrical quantities in 4-dimensions:
\be
\vth&=&\frac{i\kappa}{6}\sum_{\ff\ \text{internal}}\Big[\sgn(V_4)\,\eps(\ff)+{i}\pi\eta(\ff)\Big]\,\rmd \mathbf{a}(\ff)+\nonumber\\
&&+\frac{i\kappa}{6}\sum_{\ff\ \text{boundary}}\Big[\sgn(V_4)\,\Theta(\ff)-\ln\chi_{\ell}(\xi)\Big]\,\rmd \mathbf{a}(\ff)
+\sum_{\cs}\ln{{\sf p}_\cs}\frac{\rmd {\sf m}_\cs}{{\sf m}_\cs}
\ee 
where $\eps(\ff)$ is the deficit angle at an internal triangle, and $\Theta(\ff)$ is the hyper-dihedral angle at a boundary triangle:
\be
\eps(\ff)=\sum_{\Fs,\ \ff\subset \Fs}\Theta_\Fs(\ff),&& \ \ \text{for internal}\ \ \ff\nonumber\\
\Theta(\ff)=\sum_{\Fs,\ \ff\subset \Fs}\Theta_\Fs(\ff), && \ \ \text{for boundary}\ \ \ff
\ee 
where $\Theta_\Fs(\ff)$ is the hyper-dihedral angle at $\ff$ within a 4-simplex $\Fs$. We use Schl\"afli identity \cite{eva}
\be
\sum_{\ff\subset \Fs}\mathbf{a}(\ff)\,\rmd \Theta_\Fs(\ff)=\kappa\,\rmd V_4(\Fs)
\ee
to write the terms involving $\eps(\ff),\Theta(\ff)$ as an total differential:
\be
\vth&=&\frac{i\kappa}{6}\sgn(V_4)\,\rmd\lt[\sum_{\ff\ \text{internal}}\mathbf{a}(\ff)\,\eps(\ff)-\kappa\sum_{\Fs} V_4(\Fs)+\sum_{\ff\ \text{boundary}}\mathbf{a}(\ff)\,\Theta(\ff)\rt]-\frac{\kappa\pi}{6}\sum_{\ff\ \text{internal}}\eta(\ff)\,\rmd\mathbf{a}(\ff)-\nonumber\\
&&-\frac{i\kappa}{6}\sum_{\ff\ \text{boundary}}\ln\chi_{\ell}(\xi)\,\rmd \mathbf{a}(\ff)
+\sum_{\cs}\ln{{\sf p}_\cs}\frac{\rmd {\sf m}_\cs}{{\sf m}_\cs}
\ee 
The contour integral in Eq.\Ref{Bvth} gives ($\mathrm{C}^{(\a_{\rm 4d})}$ is an integration constant)
\be
\int_{\Fc\subset\cl_{\mathrm{SUSY}}}^{(x_i,y_i^{(\a_{\rm 4d})})} \vth&=&\frac{i\kappa}{6}\sgn(V_4)\lt[\sum_{\ff\ \text{internal}}\mathbf{a}(\ff)\,\eps(\ff)-\kappa\sum_{\Fs} V_4(\Fs)+\sum_{\ff\ \text{boundary}}\mathbf{a}(\ff)\,\Theta(\ff)\rt]-\frac{\kappa\pi}{6}\sum_{\ff\ \text{internal}}\eta(\ff)\,\mathbf{a}(\ff)-\nonumber\\
&&-\frac{i\kappa}{6}\sum_{\ff\ \text{boundary}}\int_{\Fc\subset\cl_{\mathrm{SUSY}}}^{(x_i,y_i^{(\a_{\rm 4d})})} \ln\chi_{\ell}(\xi)\,\rmd \mathbf{a}'(\ff)
+\sum_{\cs}\int_{\Fc\subset\cl_{\mathrm{SUSY}}}^{(x_i,y_i^{(\a_{\rm 4d})})} \ln{{\sf p}'_\cs}\frac{\rmd {\sf m}'_\cs}{{\sf m}'_\cs}+\mathrm{C}^{(\a_{\rm 4d})}.\label{intvth}
\ee
The first term is precisely Einstein-Regge action $\mathbf{S}_{\rm Regge}$ on $(\sm_4,g)$ with a cosmological constant term ($\kappa$ is the cosmological constant)
\be
\mathbf{S}_{\rm Regge}(\sm_4,g)=\sum_{\ff\ \text{internal}}\mathbf{a}(\ff)\,\eps(\ff)-\kappa\sum_{\Fs} V_4(\Fs)+\sum_{\ff\ \text{boundary}}\mathbf{a}(\ff)\,\Theta(\ff). 
\ee
Einstein-Regge action is a discretization of Einstein-Hilbert action of gravity $\mathbf{S}_{\rm EH}=\half\int_{\sm_4}(R-2\kappa)+\int_{\partial \sm_4} K$ using constant curvature 4-simplices \cite{regge,BD,BD1,foxon,Hartle1981,Sorkin1975}. 

The second term in Eq.\Ref{intvth} contains the index $\eta(\ff)$. When the spacetime $(\sm_4,g)$ is globally time-oriented, $\eta(\ff)=0$ so that the second term vanishes.

The last two integrals in Eq.\Ref{intvth} are two boundary terms, since they only involve the boundary triangles $\ff\subset\partial\sm_4$ and $\cs\in\partial\sm_3$ which corresponds to $\ft_\cs\subset\partial\sm_4$. When $\sm_4$ doesn't have a boundary, $\sm_3$ doesn't have 4-holed sphere and annulus cusps in the boundary. Then all the boundary terms in Eq.\Ref{intvth} disappear.

Let's fix the boundary data $(\l_T^2,\l_\ell,{\sf m}_\cs,{\sf p}_\cs)$ which determines the triangle areas in $\sm_4$ and the shapes of boundary tetrahedra $\ft_\cs\in\partial\sm_4$. As it has been mentioned, the data $(\l_T^2,\l_\ell,{\sf m}_\cs,{\sf p}_\cs)$ determine finitely many branches $\a_{\rm 4d}$ in $\cl_{\rm SUSY}(T_{\sm_3})$, thus give finitely many holomorphic blocks $B^{\a_{\rm 4d}}_{\sm_3}(\vec{x},q)$ where the argument $\vec{x}=(\l_T^2,\l_\ell,{\sf m}_\cs)$ is given by the boundary data. It turns out that the last two integrals in Eq.\Ref{intvth} have the same result (up to integration constant) in all holomorphic blocks $B^{\a_{\rm 4d}}_{\sm_3}(\vec{x},q)$. 

Indeed let's consider a variation of the boundary data, which corresponds to a continuous variation of geometry on $\sm_4$. We have an 1-parameter family $(\l_T^2(r),\l_\ell(r),{\sf m}_\cs(r),{\sf p}_\cs(r))$, which reduces to the original data at $r=0$. The 1-parameter family gives a curve $c^{\a_{\rm 4d}}$ on each branch $\a_{\rm 4d}$, which is an extension of the contour $\Fc$ in Eq.\Ref{intvth}. However the curves $c^{\a_{\rm 4d}}$ coincide when they are projected to the $(\l_\ell,{\sf m}_\cs,{\sf p}_\cs)$-subspace \footnote{In $\cl_{\sm_3}(\mathbf{t})$, $(x_i,y_i^{(\a_{\rm 4d})})$ are the flat connections on $\sm_3$ whose boundary values are consistent with $(\l_T^2,\l_\ell,{\sf m}_\cs,{\sf p}_\cs)$ for all $\a_{\rm 4d}$. So they reduces to the same set of flat connections on 4-holed spheres $\cs\subset\partial\sm_3$. Therefore ${\sf p}_\cs={\sf p}_\cs^{(\a_{\rm 4d})}(\l_T^2,\l_\ell,{\sf m}_\cs)$, as solutions to $\mathbf{A}_i(x,y)=0$, are the same for all $\a_{\rm 4d}$.}. Thus the following variation of the integral is independent of $\a_{\rm 4d}$ 
\be
&&-\frac{i\kappa}{6}\sum_{\ff\ \text{boundary}}\delta \int_{\Fc\subset\cl_{\mathrm{SUSY}}}^{(x_i,y_i^{(\a_{\rm 4d})})} \ln\chi_{\ell}(\xi)\,\rmd \mathbf{a}'(\ff)
+\sum_{\cs}\delta \int_{\Fc\subset\cl_{\mathrm{SUSY}}}^{(x_i,y_i^{(\a_{\rm 4d})})} \ln{{\sf p}'_\cs}\frac{\rmd {\sf m}'_\cs}{{\sf m}'_\cs}\nonumber\\
&=&-\frac{i\kappa}{6}\sum_{\ff\ \text{boundary}}\int_{c^{\a_{\rm 4d}}} \ln\chi_{\ell}(\xi)\,\rmd \mathbf{a}'(\ff)
+\sum_{\cs}\int_{c^{\a_{\rm 4d}}} \ln{{\sf p}'_\cs}\frac{\rmd {\sf m}'_\cs}{{\sf m}'_\cs},\label{intvar}
\ee
because both integrands only depend on $(\l_\ell,{\sf m}_\cs,{\sf p}_\cs)$. Note that $\chi_{\ell}(\xi)$ only depends on the flat connection on 4-holed spheres parametrized by $(\l_\ell,{\sf m}_\cs,{\sf p}_\cs)$ (It also depends on some global choices in defining the coordinates, e.g. the framing data, as well as the choice of $\g_\t$ in defining $\t_\ell$). Integrating the variation, the two integrals have the same result (up to integration constant) for all $\a_{\rm 4d}$. We denote by 
\be
\hbar \Xi_B=-\frac{i\kappa}{6}\sum_{\ff\ \text{boundary}}\int_{\Fc\subset\cl_{\mathrm{SUSY}}}^{(x_i,y_i^{(\a_{\rm 4d})})} \ln\chi_{\ell}(\xi)\,\rmd \mathbf{a}'(\ff)
+\sum_{\cs}\int_{\Fc\subset\cl_{\mathrm{SUSY}}}^{(x_i,y_i^{(\a_{\rm 4d})})} \ln{{\sf p}'_\cs}\frac{\rmd {\sf m}'_\cs}{{\sf m}'_\cs}
\ee
From the point of view in 4-dimensional geometry on $\sm_4$, $\Xi_B$ is a constant boundary term independent of bulk variations. If we make a variation of data $(\l_T^2(r),\l_\ell,{\sf m}_\cs,{\sf p}_\cs)$ with constant $\l_\ell,{\sf m}_\cs,{\sf p}_\cs$, the variations of integrals in Eq.\Ref{intvar} vanish. 

As a result, we obtain the semiclassical asymptotic behavior of holomorphic block
\be
B^{\a_{\rm 4d}}_{\sm_3}(\vec{x},q)&=&\exp\lt\{\frac{i\kappa\,\sgn(V_4)}{6\hbar}\lt[\sum_{\ff\ \text{internal}}\mathbf{a}(\ff)\,\eps(\ff)-\kappa\sum_{\Fs} V_4(\Fs)+\sum_{\ff\ \text{boundary}}\mathbf{a}(\ff)\,\Theta(\ff)\rt]+o\lt(\ln\hbar\rt)\rt\}\times\nonumber\\
&&\times\ \  \exp\lt[\Xi_B+\mathrm{C}^{(\a_{\rm 4d})}-\frac{\kappa\pi}{6\hbar}\sum_{\ff\ \text{internal}}\eta(\ff)\,\mathbf{a}(\ff)\rt]\label{asympB}
\ee
For $(\sm_4,g)$ being globally oriented ($\sgn(V_4)$ is constant) and time-oriented ($\eta(\ff)=0$), the semiclassical dynamics encoded in the corresponding holomorphic block is Einstein-Regge gravity in 4-dimensions. The gravitational coupling $1/\ell_P^2$ in front of Einstein-Regge action in Eq.\Ref{asympB} is
\be
\ell_P^2=\frac{6\hbar}{\kappa}
\ee
However from the supersymmetric gauge theory perspective $\hbar=2\pi i\b\eps\in i\R$ on $D^2\times_q S^1$. In order that $\ell_P^2\in\R$, $\hbar$ in $B^{\a_{\rm 4d}}_{\sm_3}(\vec{x},q)$ has to be real (or $q\in\R$). i.e. Eq.\Ref{asympB} is the asymptotic behavior of the block integral Eq.\Ref{blockintegral} as $q=e^{\hbar}\in\R$ ($\cj_\a$ should be also computed accordingly). The asymptotic behavior Eq.\Ref{asympB} suggests that the holomorphic block $B^{\a_{\rm 4d}}_{\sm_3}(\vec{x},q)$ is a semiclassical wave functions of 4-dimensional simplicial gravity on $\sm_4$, whose dominant contribution comes from the geometry on $\sm_4$ determined by $(\vec{x},\vec{y}^{(\a)})$.





\section*{Acknowledgements}

I gratefully acknowledge Roland van der Veen for various enlightening discussions, and in particular, teaching me how to make the ideal triangulations for graph complement 3-manifolds. I also gratefully acknowledge Xiaoning Wu for the invitation to visiting Chinese Academy of Mathematics in Beijing and for his hospitality. I would like to thank Hal Haggard, Wojciech Kami\'nski, Aldo Riello, Yuji Tachikawa, Junbao Wu, Junya Yagi, Gang Yang, Hongbao Zhang for useful discussions. I also thank an anonymous referee for his helpful comments on an earlier version of the paper. The research is supported by the funding received from Alexander von Humboldt Foundation.

\appendix

\section{Ellipsoid Partition Function of $T_{S^3\setminus\G_5}$}\label{ellipsoidpartition}

As a way to understand the idea behind the above manipulation of 3d supersymmetric gauge theories according to the symplectic transformations, we consider the partition function $Z_{S^3_b}(T_{M_3})$ of the theory $T_{M_3}$ on a 3d ellipsoid $S_b^3$ (here we follow the discussion in \cite{DGG11} and apply the construction to $M_3=S^3\setminus\G_5$). The ellipsoid $S_b^3$ is a deformation of the ordinary 3-sphere defined by
\be
b^2|z_1|^2+b^{-2}|z_2|^2=1,\ \ \ \ z_1,z_2\in\C
\ee
where $b$ is the squashing parameter. $S_b^3$ preserves only a $\mathrm{U(1)}\times \mathrm{U(1)}$ symmetry.  

The lagrangians in the last subsection are written for the supersymmetric field theories on a flat background. When we put the theories on a curve space, the couplings are turned on between the conserved currents and the background (nondynamical) supergravity multiplet \cite{SUSYcurve}. In particular, the conserved current of the unbroken U(1) R-symmetry is coupled with a background U(1) gauge field $A^R$. Given that the chiral superfield $\Phi_Z$ has a R-charge $R_{Z}$, the fermion in the chiral multiplet has R-charge $R_Z-1$. In the same way as we mentioned before, the fermion would generate an anomaly to break the R-symmetry. In order to preserve the R-symmetry, some additional Chern-Simons terms relating to $A^R$ has to be added to cancel the anomaly, in the same way as the cancellation of gauge anomaly mentioned before. For example, for $T_\Delta$ containing a single chiral multiplet, the additional Chern-Simons coupling has to be added to Eq.\Ref{chiral} so that the resulting Chern-Simons level matrix $k_{ij}$ is given by \cite{3dblock}
\be
\begin{array}{ccc}
 \ \ \ \,\big|& F & R \\
\hline
F  \ \big|& -\half & \half (1-R_Z)\\
R \ \,\big|& \half (1-R_Z) & -\half (1-R_Z)^2
\end{array}
\ee
Here $F$ stands for the gauge field coupled with the flavor current. The anomalies generated by the fermions shift the Chern-Simons level matrix according to \cite{3dSUSY}
\be
(k_{ij})_{\mathrm{eff}}=k_{ij}+\half \sum_{\mathrm{fermions}} (q_f)_i(q_f)_j\sgn(m_f)
\ee
where $q_f$'s are the flavor charges or R-charges carried by the fermions, and $m_f$'s are the fermions masses. We set the background gauge field $V_Z=m_{Z}^{\mathrm{3d}}\bar{\theta}^\a\theta_\a$ which turns on a 3d real mass $m_{Z}^{\mathrm{3d}}$ (VEV of scalar field $\sig_Z^{\mathrm{3d}}$ in the vector multiplet). The resulting Chern-Simons levels after the shift are all integers.

The ellipsoid partition function of $T_\Delta$ can be computed exactly by the SUSY localization technique (see e.g.\cite{Hama11,Kapustin09}). The resulting partition function $Z_{S^3_b}(T_{\Delta})$ is a function of complexfied mass parameter $\mu_Z=m_Z+\frac{i}{2}(b+b^{-1})R_Z$ and the squashing parameter $b$ 
\be
Z_{S^3_b}\lt(T_{\Delta};\mu_Z\rt)=\exp\lt[\frac{i\pi}{2}\lt(\mu_Z-\frac{i}{2}(b+b^{-1})\rt)^2\rt]s_b\lt(\frac{i}{2}(b+b^{-1})-\mu_Z\rt) 
\ee
where $s_b(X)$ is a variant of the noncompact quantum dilogarithm function \cite{Faddeev:1995nb} commonly used in Liouville theory
\be
s_b(X)=e^{-\frac{i\pi}{2}X^2} \prod_{r=1}^\infty\frac{1+e^{2\pi bX+2\pi i b^2(r-\half)}}{1+e^{2\pi b^{-1}X-2\pi i b^{-2}(r-\half)}} 
\ee
We define some notations:
\be
&Z\equiv 2\pi b\mu_Z, \ \ \ \tilde{Z}\equiv 2\pi b^{-1}\mu_Z,\ \ \ \hbar\equiv 2\pi i b^2,\ \ \ \tilde{\hbar}\equiv 2\pi i b^{-2}&\nonumber\\  
&z=\exp Z,\ \ \ \tilde{z}=\exp\tilde{Z},\ \ \ q=\exp\hbar,\ \ \ \tilde{q}=\exp\tilde{\hbar}&\label{Zandmu}
\ee
where $\tilde{\hbar}=-\frac{4\pi^2}{\hbar}$ and $\tilde{Z}=\frac{2\pi i}{\hbar} Z$ relate to the S-duality (modular) transformation in \cite{Dimofte2011}. The partition function $Z_{S^3_b}(T_{\Delta})$ can be analytic continued to an entire cut plane $\hbar\in\C\setminus\{i\R_{<0}\}$.
\be
Z_{S^3_b}\lt(T_{\Delta};\mu_Z\rt)
\equiv\Psi_\hbar(Z)
=\lt\{\begin{array}{lr}
\prod_{r=0}^\infty\dfrac{1-q^{r+1}z^{-1}}{1-\tilde{q}^{-r}\tilde{z}^{-1}} & |q|<1\\
& \\
\prod_{r=0}^\infty\dfrac{1-\tilde{q}^{r+1}\tilde{z}^{-1}}{1-q^{-r}z^{-1}} & |q|>1
\end{array}\rt..\label{tetraduality}
\ee

In \cite{Dimofte2011}, it has been shown that $\Psi_\hbar(Z)$ is the partition function of analytic continued SL(2) Chern-Simons theory on a single ideal tetrahedron, where $Z$ is the position coordinate of the phase space $\cp_{\partial\Delta}$. The 3d-3d correspondence (in DGG-construction) is illustrated by Eq.\Ref{tetraduality} as the simplest example: a single $\cn=2$ chiral multiplet with the background Chern-Simons coupling corresponds to SL(2) Chern-Simons theory on a single ideal tetrahedron \cite{DGG11}. 

The T-type and S-type symplectic transformations considered in the last subsection can be translated into the transformations of ellipsoid partition functions. The T-type transformation adds a background Chern-Simons term to the Lagrangian. After the SUSY localization of the path integral, the Chern-Simons term of level $k$ contributes $e^{-i\pi k\mu^2}$. Therefore Eqs.\Ref{TAB}
\be
T:\ \ \ \ Z_{S^3_b}\lt(\vec{\mu}\rt)\mapsto e^{-i\pi\, \vec{\mu}^T(\mathbf{A}\mathbf{B}^T)\vec{\mu}}Z_{S^3_b}\lt(\vec{\mu}\rt)\label{TTTT}
\ee
The S-type transformation Eq.\Ref{Strans} can be obtained in a similar way. Since the gauge fields becomes dynamical in S-type transformation, the corresponding complexfied mass $\vec{\mu}$ are integrated:
\be
S:\ \ \ \ Z_{S^3_b}\lt(\vec{\mu}\rt)\mapsto \int\rmd^{15}\vec{\mu}\ e^{2\pi i\, \vec{\mu}^T\vec{\mu}'}Z_{S^3_b}\lt(\vec{\mu}\rt)\label{SSSS}
\ee
which is a Fourier transformation. The GL-type transformation is simply a field redefinition in the partition function.  

It is straight-forward to compute the ellipsoid partition function (up to an overall constant) for the theory $T_{S^3\setminus\G_5}$ defined in the last subsection
\be
Z_{S^3_b}\lt(T_{S^3\setminus\G_5};\vec{\mu}_{\xi}\rt)=\int\rmd^{15}\vec{\sig}\ Z_\times\lt(\vec{\sig}\rt)
e^{\frac{1}{2\hbar}\vec{\sig}^T\mathbf{B}^{-1}\mathbf{A}\vec{\sig}-\frac{1}{\hbar}\vec{\sig}^T\mathbf{B}^{-1}\vec{\xi}+\frac{1}{2\hbar}\vec{\xi}^T\mathbf{D}\mathbf{B}^{-1}\vec{\xi}}\label{Z1}
\ee
where $\vec{\xi}$ relates $\vec{\mu}_\xi$ by $\vec{\xi}=2\pi b\vec{\mu}_{\xi}$. Here $Z_{\times}$ is a product of ellipsoid partition functions of $T_{\mathrm{Oct}}$:
\be
Z_{\times}\lt(\vec{\sig}\rt)=\prod_{\a=1}^5 Z_{S^3_b}\lt(T_{\mathrm{Oct}};\mu_{X_\a},\mu_{Y_\a},\mu_{Z_\a}\rt)=\prod_{\a=1}^5\Psi_{\hbar}\lt(X_\a\rt)\Psi_{\hbar}\lt(Y_\a\rt)\Psi_{\hbar}\lt(Z_\a\rt)\Psi_{\hbar}\lt(W_\a\rt)
\ee
where $W_\a$ is expressed in terms of $X_\a,Y_\a,Z_\a$ because of the constraint imposed by the superpotential $\cw=\Phi_X\Phi_Y\Phi_Z\Phi_W$, i.e.
\be
\mu_{X_\a}+\mu_{Y_\a}+\mu_{Z_\a}+\mu_{W_\a}=i(b+b^{-1})\ \ \ \Leftrightarrow\ \ \ C_\a=X_\a+{Y_\a}+{Z_\a}+W_\a=2\pi i+\hbar.
\ee
Here $\vec{\sig}=(X_\a,Y_\a,Z_\a)_{\a=1}^5$ and $X_\a=2\pi b\mu_{X_\a}$ and similar for $Y_\a, Z_\a$. The constraint $C_\a=2\pi i+\hbar$ from the field theory on $S^3_b$ contains a quantum correction comparing to the classical constraint in Section \ref{coordinates}. This quantum correction indeed matches the quantization of SL(2) Chern-Simons theory \cite{Dimofte2011}.

Recall that there is a classical affine shift parametrized by $\vec{\nu}_1,\vec{\nu}_2\in(\Z/2)^{15}$ in the definition Eq.\Ref{ABCD} of the symplectic coordinates. The classical affine shift by $i \pi\vec{\nu}_1$ in position coordinates corresponds in field theory the shifts of the R-charge, which gives
\be
\vec{\xi}=2\pi b \vec{m}^{\mathrm{3d}}+\lt(i\pi+\frac{\hbar}{2}\rt)\vec{R}_{\xi}&\ \ \ \ \mapsto\ \ \ \ 
&\vec{m}=2\pi b \vec{m}^{\mathrm{3d}}+\lt(i\pi+\frac{\hbar}{2}\rt)\vec{R}_{m},\ \ \ \ \text{with}\ \ \ \ \vec{R}_{m}=\vec{R}_{\xi}+\vec{\nu}_1.
\ee  
In $T_{S^3\setminus\G_5}$ defined in Eq.\Ref{GCtheory}, the R-charges $\vec{R}_{m}$ complexifies the real masses $\vec{m}^{\mathrm{3d}}=\langle\vec{\sig}^{\mathrm{3d}}\rangle$ from the background vector multiplets $\vec{V}$, which are all coupled to the topological U(1) currents. Thus $\vec{R}_{m}$ should be understood as the R-charges of the chiral fields in certain dual description, which carry the topological (monopole) charges. 

It is interesting to see that the resulting $\vec{m}$ has the quantum corrections comparing to the classical expressions in Section \ref{coordinates}. The quantum correction simply replaces each $i\pi$ in the affine shifts by $i\pi+\hbar/2$. Again this correction given by the field theory indeed matches the quantization of SL(2) Chern-Simons theory, and is consistent with the quantum correction for the constraint $C_\a=2\pi i+\hbar$ \cite{Dimofte2011}.

The same quantum corrections have to be implement to the momentum coordinates $\vec{p}$ in Eq.\Ref{ABCD}. Then the affine shifts by $i \pi\vec{\nu}_2$ act on the field theory by adding mixed Chern-Simons terms between the background R-symmetry and flavor symmetry gauge fields \cite{3dblock}. It adds a term $\frac{1}{\hbar}\lt(i \pi+\frac{\hbar}{2}\rt)\vec{\nu}_2\cdot\vec{m}$ on the exponential in Eq.\Ref{Z1}.

We obtain the final ellipsoid partition function of $T_{S^3\setminus\G_5}$:
\be
&&Z_{S^3_b}\lt(T_{S^3\setminus\G_5};\vec{\mu}_{m}\rt)\equiv Z_{S^3\setminus\G_5}\lt(\vec{m}\rt)\nonumber\\
&&\quad\quad =\int\rmd^{15}\vec{\sig}\ Z_\times\lt(\vec{\sig}\rt)
e^{\frac{1}{2\hbar}\vec{\sig}^T\mathbf{B}^{-1}\mathbf{A}\vec{\sig}-\frac{1}{\hbar}\vec{\sig}^T\mathbf{B}^{-1}\vec{m}+\frac{1}{2\hbar}\vec{m}^T\mathbf{D}\mathbf{B}^{-1}\vec{m}+\frac{1}{\hbar}\lt(i\pi+\frac{\hbar}{2}\rt)\vec{\sig}^T\mathbf{B}^{-1}\vec{\nu}_1-\frac{1}{\hbar}\lt(i\pi+\frac{\hbar}{2}\rt)\vec{m}^T\mathbf{D}\mathbf{B}^{-1}\vec{\nu}_1+\frac{1}{\hbar}\lt(i \pi+\frac{\hbar}{2}\rt)\vec{\nu}_2\cdot\vec{m}},
\ee
with $\vec{m}=2\pi b\mu_m$. The resulting ellipsoid partition function, understood as a wave function of position coordinates $\vec{m}\equiv(\L_{ab},M_a)$, is a partition function of the analytic continued SL(2) Chern-Simons theory on the graph complement 3-manifold $S^3\setminus\G_5$. The resulting finite-dimensional integral coincides with the ``state-integral model'' of Chern-Simons theory on $S^3\setminus\G_5$, by applying the construction in \cite{Dimofte2011} to our graph complement 3-manifold. The analysis here provides an example involving the graph complement 3-manifold illustrating the general argument in \cite{DGG11,CJ,LY,LTYZ} of 3d-3d correspondence.

The ellipsoid partition function $Z_{S_b^3}(T_{\sm_3})$ can be computed by apply the symplectic transformations to a product of two ellipsoid partition functions for $T_{S^3\setminus\G^5}$ (see Eqs.\Ref{TTTT} and \Ref{SSSS} for S-type and T-type, while GL-type is simply a field redefinition). The resulting $Z_{S_b^3}(T_{\sm_3})$ is a state-integral model for $\sm_3=S^3\setminus\G_5\cup S^3\setminus\G_5$
\be
&&Z_{\sm_3}\lt(\L_{1a},\L_{ab},\L_{ab}',M_a,M_a'\rt)= \frac{1}{(2\pi i \hbar)^{1/2}}\times\nonumber\\
&\times&\int \rmd \check{\sig}\ Z_{S^3\setminus\G_5}\Big(\L_{1a},\L_{ab},\lt(i\pi+{\hbar}/{2}\rt)\zeta_1-\check{\sig},M_a\Big)\,Z_{S^3\setminus\G_5}\lt(-\L_{1a},\L_{ab}',\check{\sig},M_a'\rt)\,e^{\frac{1}{\hbar}\lt(i\pi+\frac{\hbar}{2}\rt)\zeta_2 \check{\sig}}\label{ZM3}
\ee
where $a,b\neq 1$, and we have written $Z_{S^3\setminus\G_5}(\vec{m})$ as $Z_{S^3\setminus\G_5}(\L_{1a},\L_{ab},M_1,M_a)$.

\section{Review of Holomorphic Block}\label{BPS}

From Eq.\Ref{tetraduality}, one may have noticed that the ellipsoid partition function $Z_{S^3_b}(T_\Delta)$ of a single chiral multiplet admits a ``holomorphic factorization'', i.e. it can be factorized into two ``holomorphic blocks'' each of which is either holomorphic in $q,z$ or holomorphic in $\tilde{q},\tilde{z}$:
\be
Z_{S^3_b}\lt(T_{\Delta}\rt)=B_\Delta(z,q)B_\Delta(\tilde{z},\tilde{q}),\quad\text{where}\quad
B_\Delta(z,q):=(qz^{-1};q)_{\infty}=
\lt\{\begin{array}{lr}
\prod_{r=0}^\infty(1-q^{r+1}z^{-1}) & |q|<1\\
& \\
\prod_{r=0}^\infty(1-q^{-r}z^{-1})^{-1} & |q|>1
\end{array}\rt.
\ee 
Here $(qz^{-1};q)_{\infty}$ is a meromorphic function of $z\in\C$ and $q\in\C\setminus\{|q|=1\}$, and has no analytic continuation between the 2 regimes $|q|<1$ and $|q|>1$. A similar factorization also happens to the supersymmetric spherical index (partition function on $S^2\times_q S^1$) studied in \cite{3dindice}, with the same holomorphic block function $B_\Delta(z,q)$. Furthermore in \cite{Pasquetti:2011fj,3dblock}, it is argued that the holomorphic factorization happens for generic 3d $\cn=2$ supersymmetric gauge theories, which include all the gauge theories labelled by 3-manifold defined in \cite{DGG11}. Therefore the holomorphic blocks can be defined for the supersymmetric gauge theories defined in the last section. In this section, we review briefly the construction and properties of holomorphic blocks from 3d $\cn=2$ supersymmetric gauge theories. The discussion in this subsection mainly follows \cite{3dblock} (see also \cite{YK}).

The holomorphic block of a 3d $\cn=2$ theory $T_{M_3}$ can be understood as the partition function of $T_{M_3}$ on a curved background $D^2\times_q S^1$ with a torus boundary. $D^2\times_q S^1$ is a solid torus with metric 
\be
\rmd s^2=\rmd r^2+f(r)^2\lt(\rmd\varphi+\eps\beta\rmd\theta\rt)^2+\b^2\rmd\theta^2
\ee
where the coordinates $r\in[0,\infty)$ and $r,\varphi$ are periodic with period $2\pi$. $f(r)\sim r$ as $r\to0$ and $f(r)\sim\rho$ as $r\to\infty$ where $\rho$ is the length of the cigar $D^2$. $\times_q$ means that around the $\theta$ circle, the holomorphic variable $z=r e^{i\varphi}$ are identified by $(z,0)\sim (q^{-1}z,2\pi)$, where 
\be
q= e^\hbar,\ \ \ \ \hbar={2\pi i\b\eps}.
\ee
It is also convenient to consider the geometry of $D^2\times_q S^1$ as a cigar $D^2$ with $S^1$ fibers.

A topological twist has to be made in order to preserve $\cn=2$ SUSY on $D^2\times_q S^1$ \cite{wittenTQFT,phaseN=2,TATfusion}. For $\cn=2$ supersymmetric gauge theories on $D^2\times_q S^1$, the topological twist can be made if the U(1) R-symmetry is preserved. It turns out that there are 2 possible ways to implement the topological twist, which are called the topological twist and anti-topological twist. The resulting partition functions are shown to be equivalent to the supersymmetic BPS indices with Q-exact Hamiltonian $H_\pm$ \cite{3dblock}
\be
Z^{(\a)}_{\mathrm{BPS}}\lt(x_+,q\rt)&=&\mathrm{Tr}_{\ch(D^2;\a)}(-1)^R e^{-2\pi\b H_+}q^{-J_3-\frac{R}{2}} x_+^{-e}\quad\quad\text{(topological twist)} \nonumber\\
Z^{(\a)}_{\overline{\mathrm{BPS}}}\lt(x_-,q\rt)&=&\mathrm{Tr}_{\ch(D^2;\a)}(-1)^R e^{-2\pi\b H_-}q^{-J_3+\frac{R}{2}} x_-^{e}\quad\quad\text{(anti-topological twist)}\label{BPSantiBPS}
\ee
Here $\a$ labels a massive supersymmetric ground state on which the Hilbert space $\ch(D^2;\a)$ is generated by supercharges. $J_3$ is the generator of the rotation isometry of $D^2$, and $R$ is the generator of R-symmetry. The complex fugacities $x_\pm=\exp X_\pm$ where $X_\pm$ reads
\be
X_\pm=2\pi\b m^{\mathrm{3d}}\mp i\oint_{S^1_\b|_{r=0}}A^f.
\label{Xpm}
\ee
Here $m^{\mathrm{3d}}=(m^{\mathrm{3d}}_1,\cdots,m^{\mathrm{3d}}_N)$ are the 3d real mass deformations. $e=(e_1,\cdots,e_N)$ are the generators (charges) of the abelian flavor symmetries, coupled with external gauge fields $A^f=(A_1^f,\cdots,A^f_N)$. $X_\pm$ is understood as a 2-dimensional twisted masses when the 3d theory is dimensional reduced to 2d along $S^1$.

Holomorphic block is defined by combining $Z^{(\a)}_{\mathrm{BPS}}$ and $Z^{(\a)}_{\overline{\mathrm{BPS}}}$, and analytic continuing $q=e^{2\pi i\b\eps}$ to either outside or inside of the unit circle
\be
B^\a\lt(x,q\rt)=\lt\{\begin{array}{cc}
Z^{(\a)}_{\mathrm{BPS}}\lt(x,q\rt),\  &\  |q|<1\\
\\
Z^{(\a)}_{\overline{\mathrm{BPS}}}\lt(x,q\rt),\ &\ |q|>1
\end{array}\rt.
\ee
The resulting $B^\a(x,q)$ are a set of meromorphic functions of $x\in\C$ and $q\in\C\setminus\{|q|=1\}$, with no analytic continuation from $|q|<1$ to $q>1$. The perturbative expansion of $B^\a(x,q)$ in $\hbar=\ln q$ match on the inside and outside of the unit circle $|q|=1$ order by order.

It is argued in \cite{3dblock,Pasquetti:2011fj} that the ellipsoid partition function of generic 3d $\cn=2$ theories can be expressed as a sum over product of holomorphic blocks:
\be
Z_{S^3_b}=\sum_{\a}B^\a(x,q)B^\a(\tilde{x},\tilde{q}).\label{factor}
\ee
Here the pair of holomorphic blocks are identical, but with different entries of $(x,q)$ and $(\tilde{x},\tilde{q})$. Explicitly $(x,q)$ and $(\tilde{x},\tilde{q})$ are related by $q=e^\hbar,\tilde{q}=e^{-\frac{4\pi}{\hbar}}$ and $x=e^X, \tilde{x}=e^{\frac{2\pi i}{\hbar}X}$. The same type of factorization as Eq.\Ref{factor} happens for the spherical index (partition function on $S^2\times_q S^1$) in \cite{3dindice}, with the same set of holomorphic blocks $B^\a$ for a given theory. The only difference between the spherical index and $Z_{S^3_b}$ is that $(x,q)$ and $(\tilde{x},\tilde{q})$ have different relations. In the spherical index, $x=q^{\frac{m}{2}}\zeta,\tilde{x}=q^{\frac{m}{2}}\zeta^{-1}$ and $\tilde{q}=q^{-1}$. Two different relations between $(x,q)$ and $(\tilde{x},\tilde{q})$ come from the two different ways of gluing a pair of $D^2\times_q S^1$, in order to produce respectively $S^3_b$ and $S^2\times_q S^1$. There is an important note: for a given 3d $\cn=2$ theory, in order that both the ellipsoid partition function and the spherical index can be factorized into the same set of holomorphic blocks $B^\a$, we have to require that all fields in the theory are of integer R-charge assignment (see \cite{3dblock} for details).

For a 3d $\cn=2$ supersymmetric gauge theory with abelian gauge group $\mathrm{U(1)}^r$ and flavor symmetry group $\mathrm{U(1)}^N$ (including all $T_{M_3}$ labelled by 3-manifolds $M_3$), the holomorphic block $B^\a(x,q)$ of a massive supersymmetric vacuum $\a$ is expressed as a $r$-dimensional integral \cite{3dblock,YK}
\be
B^\a(x,q)=\int_{\cj_\a}\prod_{a=1}^r\frac{\rmd s_a}{2\pi i s_a} \mathrm{CS}[k,\nu;x,s,q]\prod_{\Phi }B_\Delta\Big(z_{\Phi}(x,s;R_{\Phi}),q\Big)\label{blockintegral}
\ee
Given the 3d theory on $D^2\times_q S^1$, its Kaluza-Klein reduction on $S^1$ results in a $\cn=(2,2)$ theory on $D^2$. The variables $s_a$ and $x_i$ relates the complex (twisted) scalars $\sig_a$ and complex (twisted) masses $m_i$ in the $\cn=(2,2)$ theory by
\be
 s_a=e^{\sig_a},\ \ \ x_i=e^{m_i}
\ee
The periodicities of $\sig_a$ and $m_i$ are consequences of the gauge invariance in 3d theory.  

In Eq.\Ref{blockintegral} $B_\Delta$ is the holomorphic block of a chiral multiplet. The factor $\mathrm{CS}[k,\nu;x,s,q]$, being the contribution from Chern-Simons terms, is a finite product of Jacobi theta functions ($b_t,n_t\in\Z$ and $a_t$ is a $(N+r)$-dim vector of integers)
\be
\mathrm{CS}[k,\nu;x,s,q]=\prod_{t}\theta\lt((-q^{\half})^{b_t}(x,s)^{a_t};q\rt)^{n_t},\quad\text{with}\quad \sum_{t}n_t a_t(a_t)^T=-k,\ \ \ \sum_t n_t b_t a_t =-\nu.\label{atbtnt}
\ee
$k=(k_{ab},k_{ai},k_{ij})$ are the integer Chern-Simons levels mixing gauge-gauge, gauge-flavor, and flavor-flavor symmetries. $\nu=(\nu_{R a},\nu_{Ri})$ is a $(r+N)$ dimensional vector being the integer Chern-Simons levels mixing R-symmetry and gauge/flavor symmetries. $(x,s)^{a_t}$ stands for the monomial $x_1^{a_t^1}\cdots x_N^{a_t^N}s_1^{a_t^{N+1}}\cdots s_r^{a_t^{N+t}}$ where $a_t^\mu$ is the $\mu$-th component of $a_t$. A general way to construct the exponents $a_t,b_t,n_t$ from $k,\nu$ is given in Appendix \ref{abn}.  The above factorizations of Chern-Simons levels $k,\nu$ relate to the anomaly cancellation between 3d $\cn=2$ theory and boundary $\cn=(0,2)$ theory, where $a_t,b_t$ relate to the charges of the boundary Chiral and Fermi multiplets \cite{GGP13,YK}.

The integration cycle $\cj_\a$ in Eq.\Ref{blockintegral} is uniquely associates to a critical point $\{s^{(\a)}_a(x_1,\cdots,x_N)\}_{a=1}^r$ of the integral. The asymptotic boundary condition of $D^2\times_q S^1$ determines the massive supersymmetric ground state $|\a\rangle$, which corresponds to a critical point of the holomorphic block integral. We can write the integrand of Eq.\Ref{blockintegral} as $\exp \frac{1}{\hbar} \widetilde{\cw}$ (we view $\widetilde{\cw}$ as a perturbative expansion in $\hbar$). $\widetilde{\cw}$ is the twisted superpotential of $\cn=(2,2)$ theory from the Kaluza-Klein reduction of the 3d theory on $S^1$. The massive supersymmetric ground states $|\a\rangle$ are idenified as solutions of \footnote{The equation may correspond to the Bethe ansatz of certain integrable system \cite{NekrasovWitten,NS2009}.}
\be
\frac{\partial\widetilde{\cw}}{\partial\sig_a}=2\pi i n_a\ \ \ \ \text{or}\ \ \ \ \exp\lt(s_a\frac{\partial\widetilde{\cw}}{\partial s_a}\rt)=1.\label{expvac}
\ee
A ground state $|\a\rangle$ uniquely corresponds to a solution $s_a^{(\a)}$ to the above equations\footnote{Here we assume there are a finite number of distinct solutions, i.e. the vacua are all massive. }: 
\be
|\a\rangle\quad \leftrightarrow\quad \{s_a^{(\a)}(x_1,\cdots,x_N)\}_{a=1}^r.
\ee
We consider the flow equation on $\R_+$ generated by $\widetilde{\cw}$:
\be
\frac{\rmd \sig_a}{\rmd t}+g_{a\bar{b}}\frac{\partial \mathrm{Im}\cw}{\partial\bar{\sig}_b}=0\label{floweqn}
\ee 
with the boundary condition $s_a\sim s_a^{(\a)}$ as $t\to\infty$. The flow equation results in a space of solutions $s^{(\a)}_a(t)$ satisfying the boundary condition. The solutions $s^{(\a)}_a(0)$ at the other boundary $t=0$ span a cycle $\cj_\a$ in the space $\cm$ of $s_a$, which is known as the Lefschetz Thimble associated with the critical point $\sig_a^{(\a)}$ \cite{newQM,analcs}. The real dimension $\dim_\R\cj_\a=i_\a$ is the Morse index, which is half of $\dim_\R\cm$ for a nondegernate critical point. The integral along $\cj_\a$ is independent of the target space metric $g_{a\bar{b}}$.

The holomorphic block integral can be studied by the stationary phase analysis. The leading order contribution to the holomorphic block is the effective twisted superpotential $\widetilde{\cw}_{\mathrm{eff}}(m_i)$ \cite{CSSduality,DGG11}:
\be
B^\a(x,q)=\exp\lt[\frac{1}{\hbar}\widetilde{\cw}_{\mathrm{eff}}^{(\a)}(m_i)+o\lt(\ln\hbar\rt)\rt],\quad\text{where}\quad \widetilde{\cw}_{\mathrm{eff}}^{(\a)}(m_i)=\widetilde{\cw}\lt(s^{(\a)}_a(x_i),m_i,\hbar=0\rt)
\ee
We now define a new variable $y_i$ (effective background FI parameter preserving supersymmetry) from the twisted superpotential by
\be
y_i=\exp\lt(x_i\frac{\partial\widetilde{\cw}}{\partial x_i}\rt). \label{yi}
\ee
Eliminating $s_a$ by combining Eqs.\Ref{expvac} and \Ref{yi} results in a set of $N$ algebraic equations $\mathbf{A}_i(x,y)=0$ ($i=1,\cdots,N$) of $2N$ complex parameters $x_i,y_i$ (see \cite{DGLZ,DG12} for more discussions). A solution $s^{(\a)}_a({x}_i)$ of Eqs.\Ref{expvac} corresponds to a unique solution to $\mathbf{A}_i(x,y)=0$
\be
y_i^{(\a)}(x_i)=\exp\lt(x_i\frac{\partial\widetilde{\cw}}{\partial x_i}\rt)\Big|_{s_a=s^{(\a)}_a({x}_i)}=\exp\lt(x_i\frac{\partial\widetilde{\cw}_{\mathrm{eff}}^{(\a)}}{\partial x_i}\rt).
\ee
$\mathbf{A}_i(x,y)=0$ defines locally a holomorphic Lagrangian submanifold $\cl_{\mathrm{SUSY}}$ in the space of $x_i,y_i$ (locally $\C^{2N}$), with respect to the holomorphic symplectic structure 
\be
\O_{\mathrm{SUSY}}=\sum_{i=1}^N\frac{\rmd y_i}{y_i}\wedge\frac{\rmd x_i}{x_i}\label{Osusy}
\ee
As an important result in the 3d-3d correspondence, the parameters $x_i,y_i$ comes from the theory $T_{M_3}$ correspond to a system of holomorphic coordinates in the moduli space of framed $\Slc$ flat connections $\cp_{\partial M_3}$ on $\partial M_3$, where $\O_{\mathrm{SUSY}}$ coincides with the Atiyah-Bott-Goldman symplectic form. As a holomorphic Lagrangian submanifold, it turns out that there is an isomorphism bewteen the supersymmetric parameter space $\cl_{\mathrm{SUSY}}(T_{M_3})$ from $T_{M_3}$ and the moduli space of framed $\Slc$ flat connections $\cl_{M_3}$:
\be
\cl_{\mathrm{SUSY}}(T_{M_3})\simeq \cl_{M_3}\subset\cm_{\mathrm{flat}}\lt(M_3,\Slc\rt) 
\ee
when the ideal triangulation of $M_3$ used in the DGG-construction is fine enough \cite{Dimofte2011,DGV}. It is not a global isomorphism between $\cl_{\mathrm{SUSY}}(T_{M_3})$ and the full moduli space of flat connections $\cm_{\mathrm{flat}}\lt(M_3,\Slc\rt)$, because the reducible flat connections on $M_3$ are not captured by $\cl_{\mathrm{SUSY}}(T_{M_3})$ from DGG-construction. However there is a recent construction in \cite{33revisit} by putting an addition U(1) flavor symmetry in $T_{M_3}$, which realizes the global isomorphism between $\cl_{\mathrm{SUSY}}$ and $\cm_{\mathrm{flat}}\lt(M_3,\Slc\rt)$ in many examples. 

The algebraic equations $\mathbf{A}_i(x,y)=0$ from Eq.\Ref{yi} characterize $\cl_{M_3}$ as a holomorphic Lagrangian submanifold embedded in $\cp_{\partial M_3}$. In the case that $M_3$ is the complement of a knot in $S^3$, the boundary of $M_3$ is a torus, and the number of the flavor symmetries in $T_{M_3}$ from DGG construction is $N=1$. Then there is a single algebraic equation $\mathbf{A}(x,y)=0$ defining $\cl_{\mathrm{SUSY}}$. The polynomial $\mathbf{A}(x,y)$ is often referred to as ``A-polynomial'' of the knot\footnote{$\mathbf{A}(x,y)$ from DGG-construction is sometimes referred to as the irreducible A-polynomial, since it doesn't capture the reducible flat connections \cite{Dimofte2011,DGG11}.}, which is known as a classical knot invariant \cite{Apolynomial,knot,Gukov3dgravity}.

Once we find the Lagrangian submanifold $\cl_{\mathrm{SUSY}}$, the effective twisted superpotential $\widetilde{\cw}_{\mathrm{eff}}$ can be written as an 1-dimensional integral along a contour $\Fc$ in a cover space $\hat{\cl}_{\mathrm{SUSY}}$, where the logarithmic variables are single valued, so that 
\be
B^\a(x,q)=\exp\lt[\frac{1}{\hbar}\int_{\Fc\subset\hat{\cl}_{\mathrm{SUSY}}}^{(x_i,y_i^{(\a)})} \vth+o\lt(\ln\hbar\rt)\rt],\quad\text{where}\quad \vth=\sum_{i=1}^N\ln y_i\,{\rmd \ln x_i}.\label{Bvth}
\ee
We find that the integrand $\vth$ is the Liouville 1-form from $\O_{\mathrm{SUSY}}$, i.e. $\rmd\vth=\O_{\mathrm{SUSY}}$. The end point of the contour integral is at a solution $(x_i,y^{(\a)}_i)$ of the equations $\mathbf{A}_i(x,y)=0$. $\vth$ is an exact 1-form on $\hat{\cl}_{\mathrm{SUSY}}$ so that the integral is independent of the choice of contour $\Fc$.

The perturbative expression Eq.\ref{Bvth} suggests that the holomorphic block $B^\a(x,q)$ is a WKB solution of a set of operator equations 
\be
\hat{\mathbf{A}}_i(\hat{x},\hat{y},q)\, B^\a(x,q) =0, \ \ \ i=1,\cdots,N\label{AB=0}
\ee
The operators $\hat{\mathbf{A}}_i(\hat{x},\hat{y},q)$ quantize the classical $\mathbf{A}_i(x,y)$, with $\hat{x},\hat{y}$ satisfying the commutation relation $\hat{x}_i\hat{y}_i=q \hat{y}_i\hat{x}_i$ and $\hat{x}_i\hat{y}_j=\hat{y}_j\hat{x}_i$ ($i\neq j$). Indeed the operator equations are a set of line-operator ward identities satisfied (nonperturbatively) by the holomorphic block from supersymmetric gauge theory \cite{3dblock}. From the 3d supersymmetric gauge theory $T_{M_3}$, the resulting operator equations $\hat{\mathbf{A}}_i(\hat{x},\hat{y},q)\,\psi(x,q) =0$ quantizes the moduli space $\cm_{\mathrm{flat}}\lt(M_3,\Slc\rt)$. In the case of $M_3$ being the complement of a knot in $S^3$, the resulting operator equation quantizes the A-polynomial equation, and motivates the so called ``AJ-conjecture'' for the colored Jones polynomial \cite{Garoufalidis04,Gelca02,FGL02} \footnote{In general, $\hat{\mathbf{A}}_i(\hat{x},\hat{y},q)\,\psi(x,q) =0$ ($i=1,\cdots,N$) give a set of $q$-difference equations. For $M_3$ being the complement of a knot in $S^3$, the single quantum A-polynomial equation $\hat{\mathbf{A}}(\hat{x},\hat{y},q)\,\psi(x,q) =0$ gives a recursion relation of the colored Jones polynomial of the knot, which is known as AJ conjecture. The conjecture has also been generalized to the graph complement 3-manifold \cite{satoshi}}.

\section{Exponents in the Product of Theta Functions inside Block Integral}\label{abn}

In this appendix, we discuss the exponents $a_t,b_t,n_t$ in the product of theta functions $\mathrm{CS}[k,\nu;x,s,q]=\prod_{t}\theta\lt((-q^{\half})^{b_t}(x,s)^{a_t};q\rt)^{n_t}$ entering the block integral Eq.\Ref{blockintegral}. The guideline of relating $a_t,b_t,n_t$ and $k,\nu$ is to look at the perturbative behavior as $\hbar\to 0$
\be
\prod_{t}\theta\lt((-q^{\half})^{b_t}x^{a_t};q\rt)^{n_t}\sim \exp\lt[\frac{1}{2\hbar}\sum_{i,j}k_{ij}X_iX_j+\frac{1}{\hbar}\nu_iX_i \lt(i\pi+\frac{\hbar}{2}\rt)\rt]
\ee
by keeping in mind that
\be
\theta(x;q)\sim \exp\lt(-\frac{1}{2\hbar} X^2\rt).
\ee
Here we don't distinguish $s$ from gauge fields and $x$ from external fields, and denote them only by $x$, as well as denote $X_i=\ln x_i$. It is convenient to include both $k$ and $\nu$ into a single matrix $\ck$ for Chern-Simons levels for all gauge-gauge, gauge-flavor, gauge-R, flavor-R, and R-R couplings
\be
\begin{array}{cccc}
     &\Big|& X  & m_R  \\
     \hline
X    &\Big|& k  &  \nu \\
m_R  &\Big|& \nu^T   & 0   
\end{array}
\ee 
where $m_R=i\pi+\hbar/2$ can be viewed effectively as the complex mass from R-symmetry and $e^{m_R}=-q^{\half}$. $(-q^{\half})^{b_t}x^{a_t}$ is a monomial of all the complex masses. We consider the following equality between quadratic forms with integer coefficients
\be
\sum_{\a,\b=1}^{N+r+1}\ck_{\a\b}Z^\a Z^\b=\sum_{\a<\b}\ck_{\a\b}\lt(Z^\a +Z^\b\rt)^2+\sum_{\a=1}^{N+r+1}(\ck_{\a\a}-\sum_{\b\neq\a}\ck_{\a\b})Z^\a Z^\a
\ee
where $Z^{\a}=X^\a$ ($\a=1,\cdots,N+r$) and $Z^{N+r+1}=m_R$. We define $t$ as a set of labels $\a$ and $(\a,\b)$, $n_t\in\Z$ as the coefficients of the quadratic form, and $\fa_t=(a_t,b_t)$ as a $(N+r+1)$-dimensional vector with $\fa_t^\mu=a^\mu\in\Z\ (\nu=1,\cdots,N+r)$ and $\fa_t^{N+r+1}=b_t\in\Z$: 
\be
t=\lt\{\,\{\a\}_{\a=1}^{N+r+1},\{(\a,\b)\}_{\a<\b}^{N+r+1}\rt\},\quad\quad\quad&&\nonumber\\
n_\a=-\ck_{\a\a}+\sum_{\b\neq\a}\ck_{\a\b},\quad\quad n_{(\a,\b)}=-\ck_{\a\b},\ \ &&\nonumber\\
\fa_\a^\mu=\delta_{\a}^\mu,\quad\quad\quad\quad\quad \fa_{(\a,\b)}^\mu=\delta^\mu_\a+\delta^\mu_\b.&&\label{exponentrule}
\ee
So we have
\be
\sum_{t}n_t\fa_t^\mu\fa_t^\nu=-\ck_{\mu\nu}
\ee
where is identical to the relation between $a_t,b_t,n_t$ and $k,\nu$ in Eq.\Ref{atbtnt}. 

The above gives us a general way to obtain the integer exponents $a_t,b_t,n_t$ from $k,\nu$. However $a_t,b_t,n_t$ satisfying Eq.\Ref{atbtnt} is not unique \cite{3dblock}. The holomorphic block $B^\a(x,q)$ has the ambiguity corresponding to multiplying the block integral by a factor 
\be
c(x,q)=\prod_t\theta\lt((-q^{\half})^{b_t'}x^{a_t'};q\rt)^{n_t'},\quad \text{with}\quad  \sum_{t}n_t' a_t'(a_t')^T=0,\ \ \ \sum_t n'_t b_t' a_t' =0.
\ee
The ambiguity doesn't affect the perturbative behavior of $B^\a(x,q)$, but makes the nonperturbative contributions ambiguous. The physical meaning of $c(x,q)$ is not clear so far, and deserves further studies.

As an example we apply the above procedure to obtain $a_t,b_t,n_t$ of the theory $T_{S^3\setminus\G_5}$. The total $31\times 31$ Chern-Simons level matrix $\ck$ for $T_{S^3\setminus\G_5}$ is given by 
\be
\begin{array}{ccccc}
     &\Big|& \sig & m & m_R  \\
     \hline
\sig  &\Big|& -{\bf B}^{-1}{\bf A}  & {\bf B}^{-1} & -{\bf B}^{-1}\vec{\nu}_1 \\
m    &\Big|& ({\bf B}^{-1})^T  & -{\bf D}{\bf B}^{-1}  & {\bf D}{\bf B}^{-1}\vec{\nu}_1-\vec{\nu}_2\\
m_R  &\Big|& -\vec{\nu}_1^T({\bf B}^{-1})^T\ \   & \vec{\nu}_1^T({\bf D}{\bf B}^{-1})^T-\vec{\nu}_2^T\ \  & 0   
\end{array}
\ee 
with $\ck_{\a\b}\in\Z$. The exponents $a_t,b_t$ has been given in Eq.\Ref{exponentrule}. We obtain $n_\a\ (\a=1,\cdots,31)$: 
\be
(4, -1, -1, 2, 1, 0, 5, 4, 1, 8, 4, 3, 2, 5, 3, -1, 5, -3, 5, 0, 3, -1, -1, 0, 1, 2, -3, -1, -2, 1, 0).\label{nalpha}
\ee
and $n_{(\a,\b)}$ ($\a,\b=1,\cdots,31$, $n_{(\a,\b)}=n_{(\b,\a)}$, and the diagonal entries are set to zero):
\be
\left(
\begin{array}{ccccccccccccccccccccccccccccccc}
 0 & 0 & 0 & 0 & 0 & 0 & 0 & 0 & 0 & 0 & 0 & 0 & -1 & 0 & 0 & 0 & 0 & 0 & 0 & 0 & 1 & -1 & 0 & 0 & 0 & 0 & -1 & 0
   & 0 & 0 & -1 \\
 0 & 0 & 0 & 0 & 0 & 0 & -1 & 0 & 0 & -1 & 0 & 0 & -1 & 0 & 0 & 1 & 0 & 0 & 0 & 1 & 1 & 1 & 0 & 0 & 0 & 0 & 0 & 0
   & 0 & 0 & 1 \\
 0 & 0 & 0 & -1 & 0 & 0 & 0 & 0 & 0 & 0 & 1 & 0 & -1 & -1 & -1 & 0 & 1 & 0 & 0 & 1 & 0 & 0 & 1 & 1 & 0 & 0 & 0 & 0
   & 0 & 0 & 1 \\
 0 & 0 & -1 & 0 & 1 & 1 & -1 & -1 & 0 & 0 & 0 & 1 & 0 & 1 & 1 & 1 & -1 & 0 & -1 & 0 & 1 & 0 & -1 & 0 & 1 & 0 & 0 &
   0 & 0 & 0 & -2 \\
 0 & 0 & 0 & 1 & 0 & 1 & 0 & 0 & 1 & 0 & 0 & 1 & 0 & 0 & 1 & 1 & -1 & -1 & -1 & 0 & 0 & 0 & 0 & 0 & 0 & 0 & 0 & 0
   & 0 & 0 & -3 \\
 0 & 0 & 0 & 1 & 1 & 0 & 0 & 0 & 0 & 0 & 0 & 1 & 0 & 0 & 0 & 0 & 0 & 1 & -1 & 0 & 0 & 0 & 0 & 0 & 0 & -1 & 0 & 0 &
   0 & 0 & -1 \\
 0 & -1 & 0 & -1 & 0 & 0 & 0 & 1 & 1 & 1 & 1 & 0 & 0 & -1 & -1 & -1 & 1 & 0 & 0 & 0 & -1 & -1 & 0 & 1 & -1 & 0 & 0
   & 0 & -1 & 0 & 0 \\
 0 & 0 & 0 & -1 & 0 & 0 & 1 & 0 & 1 & 0 & 0 & -1 & 0 & -1 & -1 & -1 & 1 & 0 & 1 & 0 & -1 & 0 & 0 & 1 & -1 & 0 & 0
   & 0 & -1 & 1 & 0 \\
 0 & 0 & 0 & 0 & 1 & 0 & 1 & 1 & 0 & 0 & 0 & 0 & 0 & -1 & -1 & -1 & 1 & 1 & 0 & 0 & -1 & 0 & 0 & 0 & 0 & -1 & 0 &
   0 & -1 & 0 & 1 \\
 0 & -1 & 0 & 0 & 0 & 0 & 1 & 0 & 0 & 0 & 1 & 1 & 0 & 0 & 0 & -1 & 0 & 0 & -1 & -1 & 0 & -1 & 0 & 0 & 0 & 0 & -1 &
   0 & 0 & -1 & -2 \\
 0 & 0 & 1 & 0 & 0 & 0 & 1 & 0 & 0 & 1 & 0 & 1 & 0 & -1 & -1 & -1 & 1 & 0 & -1 & 1 & -1 & 0 & 1 & 0 & 0 & 0 & 0 &
   -1 & 0 & -1 & -2 \\
 0 & 0 & 0 & 1 & 1 & 1 & 0 & -1 & 0 & 1 & 1 & 0 & 0 & 0 & 0 & 0 & 0 & 0 & -2 & 0 & 0 & 0 & 0 & -1 & 1 & 0 & 0 & 0
   & 0 & -1 & -2 \\
 -1 & -1 & -1 & 0 & 0 & 0 & 0 & 0 & 0 & 0 & 0 & 0 & 0 & 1 & 1 & 0 & -1 & 0 & 0 & 0 & 0 & 0 & -1 & 0 & 0 & 0 & 0 &
   1 & 0 & 0 & 1 \\
 0 & 0 & -1 & 1 & 0 & 0 & -1 & -1 & -1 & 0 & -1 & 0 & 1 & 0 & 2 & 0 & -2 & 0 & 0 & -1 & 2 & 0 & -1 & -1 & 1 & 0 &
   0 & 1 & 1 & 0 & -1 \\
 0 & 0 & -1 & 1 & 1 & 0 & -1 & -1 & -1 & 0 & -1 & 0 & 1 & 2 & 0 & 0 & -2 & 1 & 0 & -1 & 1 & 0 & -1 & -1 & 1 & 0 &
   0 & 1 & 0 & 0 & 0 \\
 0 & 1 & 0 & 1 & 1 & 0 & -1 & -1 & -1 & -1 & -1 & 0 & 0 & 0 & 0 & 0 & 0 & 1 & 0 & 0 & 0 & 1 & 0 & -1 & 1 & -1 & 0
   & 0 & 0 & 0 & 2 \\
 0 & 0 & 1 & -1 & -1 & 0 & 1 & 1 & 1 & 0 & 1 & 0 & -1 & -2 & -2 & 0 & 0 & -1 & 0 & 1 & -1 & 0 & 1 & 1 & -1 & 0 & 0
   & -1 & 0 & 0 & 0 \\
 0 & 0 & 0 & 0 & -1 & 1 & 0 & 0 & 1 & 0 & 0 & 0 & 0 & 0 & 1 & 1 & -1 & 0 & 0 & 0 & 0 & 0 & 0 & 0 & 0 & 1 & 0 & 0 &
   0 & 0 & -2 \\
 0 & 0 & 0 & -1 & -1 & -1 & 0 & 1 & 0 & -1 & -1 & -2 & 0 & 0 & 0 & 0 & 0 & 0 & 0 & 0 & 0 & 0 & 0 & 1 & -1 & 0 & 0
   & 0 & 0 & 1 & 2 \\
 0 & 1 & 1 & 0 & 0 & 0 & 0 & 0 & 0 & -1 & 1 & 0 & 0 & -1 & -1 & 0 & 1 & 0 & 0 & 0 & -1 & 1 & 1 & 0 & 0 & 0 & 1 &
   -1 & 0 & 0 & 0 \\
 1 & 1 & 0 & 1 & 0 & 0 & -1 & -1 & -1 & 0 & -1 & 0 & 0 & 2 & 1 & 0 & -1 & 0 & 0 & -1 & 0 & 0 & 0 & -1 & 1 & 0 & 0
   & 0 & 1 & 0 & -2 \\
 -1 & 1 & 0 & 0 & 0 & 0 & -1 & 0 & 0 & -1 & 0 & 0 & 0 & 0 & 0 & 1 & 0 & 0 & 0 & 1 & 0 & 0 & 0 & 0 & 0 & 0 & 1 & 0
   & 0 & 0 & 2 \\
 0 & 0 & 1 & -1 & 0 & 0 & 0 & 0 & 0 & 0 & 1 & 0 & -1 & -1 & -1 & 0 & 1 & 0 & 0 & 1 & 0 & 0 & 0 & 1 & 0 & 0 & 0 & 0
   & 0 & 0 & 1 \\
 0 & 0 & 1 & 0 & 0 & 0 & 1 & 1 & 0 & 0 & 0 & -1 & 0 & -1 & -1 & -1 & 1 & 0 & 1 & 0 & -1 & 0 & 1 & 0 & -1 & 0 & 0 &
   0 & 0 & 0 & 0 \\
 0 & 0 & 0 & 1 & 0 & 0 & -1 & -1 & 0 & 0 & 0 & 1 & 0 & 1 & 1 & 1 & -1 & 0 & -1 & 0 & 1 & 0 & 0 & -1 & 0 & 0 & 0 &
   0 & 1 & -1 & -1 \\
 0 & 0 & 0 & 0 & 0 & -1 & 0 & 0 & -1 & 0 & 0 & 0 & 0 & 0 & 0 & -1 & 0 & 1 & 0 & 0 & 0 & 0 & 0 & 0 & 0 & 0 & 0 & 0
   & 0 & 0 & 0 \\
 -1 & 0 & 0 & 0 & 0 & 0 & 0 & 0 & 0 & -1 & 0 & 0 & 0 & 0 & 0 & 0 & 0 & 0 & 0 & 1 & 0 & 1 & 0 & 0 & 0 & 0 & 0 & 0 &
   0 & 0 & 3 \\
 0 & 0 & 0 & 0 & 0 & 0 & 0 & 0 & 0 & 0 & -1 & 0 & 1 & 1 & 1 & 0 & -1 & 0 & 0 & -1 & 0 & 0 & 0 & 0 & 0 & 0 & 0 & 0
   & 0 & 0 & 1 \\
 0 & 0 & 0 & 0 & 0 & 0 & -1 & -1 & -1 & 0 & 0 & 0 & 0 & 1 & 0 & 0 & 0 & 0 & 0 & 0 & 1 & 0 & 0 & 0 & 1 & 0 & 0 & 0
   & 0 & 0 & 2 \\
 0 & 0 & 0 & 0 & 0 & 0 & 0 & 1 & 0 & -1 & -1 & -1 & 0 & 0 & 0 & 0 & 0 & 0 & 1 & 0 & 0 & 0 & 0 & 0 & -1 & 0 & 0 & 0
   & 0 & 0 & 2 \\
 -1 & 1 & 1 & -2 & -3 & -1 & 0 & 0 & 1 & -2 & -2 & -2 & 1 & -1 & 0 & 2 & 0 & -2 & 2 & 0 & -2 & 2 & 1 & 0 & -1 & 0
   & 3 & 1 & 2 & 2 & 0 \\
\end{array}
\right).\nonumber\\ \label{nalphabeta}
\ee

\section{$\cl_{S^3\setminus\G_5}(\mathbf{t})$ as a Holomorphic Lagrangian Submanifold in $\cp_{\partial (S^3\setminus\G_5)}$}\label{AppA}

In this appendix, we apply the procedure in \cite{Dimofte2011} to realize the moduli space of framed flat connections $\cl_{S^3\setminus\G_5}(\mathbf{t})$ as a holomorphic Lagrangian submanifold in $\cp_{\partial (S^3\setminus\G_5)}$. Here $\cl_{S^3\setminus\G_5}(\mathbf{t})$ might depend on the 3d ideal triangulation $\mathbf{t}$, because the construction is based on the ideal triangulation specified in Section \ref{idealtriangulation}.

The ideal triangulation in Section \ref{idealtriangulation} decomposes $S^3\setminus\G_5$ into 5 ideal octahedra, each of which is decomposed into 4 ideal tetrahedra. Therefore our starting point is a set of 20 algebraic equations for $\cl_\Delta\hookrightarrow\cp_{\partial\Delta}$. However we express the equations in terms of $(X_\mu,P_{X_\mu}),(Y_\mu,P_{Y_\mu})(Z_\mu,P_{Z_\mu})(C_\mu,\G_\mu)$ where $C_\mu=X_\mu+Y_\mu+Z_\mu+W_\mu$ and $\mu=1,\cdots,5$
\be
e^{-X_\mu}+e^{P_{X_\mu}+\G_\mu}=1,\quad e^{-Y_\mu}+e^{P_{Y_\mu}+\G_\mu}=1,\quad e^{-Z_\mu}+e^{P_{Z_\mu}+\G_\mu}=1,\quad e^{-C_\mu+X_\mu+Y_\mu+Z_\mu}+e^{\G_\mu}=1.
\ee
We can now eliminate $\G_\mu$ and set $C_\mu=2\pi i$ at this stage, because the symplectic transformation from $(X_\mu,Y_\mu,Z_\mu;P_{X_\mu}, P_{Y_\mu},P_{Z_\mu})$ to $(\vec{m},\vec{p})$ doesn't involve $C_\mu,\G_\mu$ (all $C_\mu$ have been set to be $2\pi i$ in the definition of $(\vec{m},\vec{p})$). The elimination lefts us 15 equations
\be
\lt(1-e^{X_\mu+Y_\mu+Z_\mu}\rt)=e^{-P_{X_\mu}}\lt(1-e^{-X_\mu}\rt)=e^{-P_{Y_\mu}}\lt(1-e^{-Y_\mu}\rt)=e^{-P_{Z_\mu}}\lt(1-e^{-Z_\mu}\rt)
\ee
Then what we need to do is simply rewrite the equations in terms of $(\vec{m},\vec{p})$. We may proceed by following the steps in Eq.\Ref{4steps} 
\begin{itemize}

\item Let $\vec{\Fx}=(\mathbf{B}^T)^{-1}\vec{\varphi}$ and $\vec{\Fp}=\mathbf{B}\vec{\chi}$ where $\vec{\varphi}\equiv(X_\a,Y_\a,Z_\a)_{\a=1}^5$ and $\vec{\chi}\equiv(P_{X_\a},P_{Y_\a},P_{Z_\a})_{\a=1}^5$. By changing of variables
\be
\lt(1-e^{\mathbf{B}^T_{X_\mu,\, J}\Fx_J+\mathbf{B}^T_{Y_\mu,\, J}\Fx_J+\mathbf{B}^T_{Z_\mu,\, J}\Fx_J}\rt)=e^{-\mathbf{B}^{-1}_{IJ}{\Fp}_J}\lt(1-e^{-\mathbf{B}^T_{IJ}\Fx_J}\rt)
\ee
where the label $I\in\{X_\mu,Y_\mu,Z_\mu\}$

\item Let $\vec{m}-i\pi\vec{\nu}_1=\mathbf{A}\mathbf{B}^T\vec{\Fx}+\vec{\Fp}\equiv\vec{\xi}$, where $\vec{m}$ are the position coordinates in Eq.\Ref{ABCD}. By changing of variables, 
\be
\lt(1-e^{\mathbf{B}^T_{X_\mu,\, J}\Fx_J+\mathbf{B}^T_{Y_\mu,\, J}\Fx_J+\mathbf{B}^T_{Z_\mu,\, J}\Fx_J}\rt)e^{\mathbf{B}^{-1}_{IJ}(\vec{m}-i\pi\vec{\nu}_1)_J}=e^{(\mathbf{B}^{-1}\mathbf{A}\mathbf{B}^T)_{IK}\Fx_K}\lt(1-e^{-\mathbf{B}^T_{IJ}\Fx_J}\rt)\label{11}
\ee
We find the resulting equations are identical to Eqs.\Ref{sIxI}, if we identify $\sig_{X,\,\mu}=\mathbf{B}^T_{X_\mu,\, J}\Fx_J=X_\mu$ and similar for $Y_\mu$ and $Z_\mu$.

\item The momentum variables $\vec{p}$ in Eq.\Ref{ABCD} are given from Eq.\Ref{4steps} by 
\be
\vec{p}=\mathbf{D}\mathbf{B}^{-1}\vec{\xi}-\vec{\Fx}+i\pi\vec{\nu}_2=\mathbf{D}\mathbf{B}^{-1}\lt(\vec{m}-i\pi\vec{\nu}_1\rt)-(\mathbf{B}^{-1})^{T}\vec{\sig}+i\pi\vec{\nu}_2\label{22}
\ee
which are identical to Eqs.\Ref{sIinyI} after exponentiation. Finally we may combine Eqs.\Ref{11} and \Ref{22} and eliminate $\vec{\Fx}$, which is precisely how we deal with Eqs.\Ref{sIxI} and \Ref{sIinyI} from the twisted superpotential. The finally resulting algebraic equations $\mathbf{A}_I(x,y)=0$ ($I=1,\cdots,15$) realizes $\cl_{S^3\setminus\G_5}(\mathbf{t})$ as a holomorphic Lagrangian submanifold in $\cp_{\partial (S^3\setminus\G_5)}$. The identification of the equations indicates the isomorphism between $\cl_{\mathrm{SUSY}}(T_{S^3\setminus\G_5})$ and $\cl_{S^3\setminus\G_5}(\mathbf{t})$.

\end{itemize}

\noindent
The finally resulting $\mathbf{A}_I(x,y)=0$ ($I=1,\cdots,15$) may be written as
\be
&&\lt(1-e^{\mathbf{B}^T_{X_\mu,\, J}\lt[(\mathbf{D}\mathbf{B}^{-1})_{JK}\lt(\vec{m}-i\pi\vec{\nu}_1\rt)_K-{p}_J+i\pi({\nu}_2)_J\rt]+\mathbf{B}^T_{Y_\mu,\, J}\lt[(\mathbf{D}\mathbf{B}^{-1})_{JK}\lt(\vec{m}-i\pi\vec{\nu}_1\rt)_K-{p}_J+i\pi({\nu}_2)_J\rt]+\mathbf{B}^T_{Z_\mu,\, J}\lt[(\mathbf{D}\mathbf{B}^{-1})_{JK}\lt(\vec{m}-i\pi\vec{\nu}_1\rt)_K-{p}_J+i\pi({\nu}_2)_J\rt]}\rt)\nonumber\\
&=&e^{-\mathbf{B}^{-1}_{IJ}(\vec{m}-i\pi\vec{\nu}_1)_J}e^{(\mathbf{B}^{-1}\mathbf{A}\mathbf{B}^T)_{IK}\lt[(\mathbf{D}\mathbf{B}^{-1})_{KL}\lt(\vec{m}-i\pi\vec{\nu}_1\rt)_L-{p}_K+i\pi({\nu}_2)_K\rt]}\lt(1-e^{-\mathbf{B}^T_{IJ}\lt[(\mathbf{D}\mathbf{B}^{-1})_{JK}\lt(\vec{m}-i\pi\vec{\nu}_1\rt)_K-{p}_J+i\pi({\nu}_2)_J\rt]}\rt).
\ee

\section{Geometrical Meaning of Eq.\Ref{3524}}\label{collinear}

The condition Eq.\Ref{3524} means that the parallel transportations of $\xi_{35}$ and $\xi_{24}$ to a common point shouldn't be collinear:
\be
G_{23}\xi_{35}\not\propto\xi_{24}
\ee
Let's find the consequence if we assume that they would be collinear with the ratio $k\in\C$. Since $\xi_{ab}=M_{ab}(1,0)^T$, we would have
\be
\left(
\begin{array}{c}
1  \\
0      
\end{array}
\right)=kM_{24}^{-1}G_{23}M_{35}\left(
\begin{array}{c}
1  \\
0      
\end{array}
\right)\quad\Rightarrow\quad M_{24}^{-1}G_{23}M_{35}=\left(
\begin{array}{cc}
k^{-1} & \mu  \\
0  & k    
\end{array}
\right)
\ee 
We define the vector $J\xi_{ab}$ by writing $M_{ab}=(\xi_{ab},J\xi_{ab})$. Consequently 
\be
G_{23}\xi_{35}=k^{-1}\xi_{24}, \quad\text{and}\quad G_{23}J\xi_{35}=\mu\xi_{24}+kJ\xi_{24}.
\ee
We view $\xi_{ab}$ and $J\xi_{ab}$ to be Weyl spinors, and construct null vectors $\xi_{ab}\xi_{ab}{}^\dagger$ and $(J\xi_{ab})(J\xi_{ab})^\dagger$ in Minkowski space $(\R^4,\eta)$. Then $G_{23}$ acts on the bivector would give 
\be
G_{23}\act (J\xi_{35})(J\xi_{35})^\dagger\wedge \xi_{35}\xi_{35}{}^\dagger=k^{-1}\bar{k}^{-1}\Big(\mu\xi_{24}+kJ\xi_{24}\Big)\lt(\bar{\mu}\xi_{24}{}^\dagger+\bar{k}(J\xi_{24})^\dagger\rt)\wedge \xi_{24}\xi_{24}{}^\dagger\nonumber\\
=\lt(\mu k^{-1}\xi_{24}(J\xi_{24})^\dagger+\bar{\mu} \bar{k}^{-1}(J\xi_{24})\xi_{24}{}^\dagger\rt)\wedge \xi_{24}\xi_{24}{}^\dagger
+(J\xi_{24})(J\xi_{24})^\dagger\wedge \xi_{24}\xi_{24}{}^\dagger\label{GJJxixi}
\ee
The bivector $G_{23}\act (J\xi_{35})(J\xi_{35})^\dagger\wedge \xi_{35}\xi_{35}{}^\dagger$ and $(J\xi_{24})(J\xi_{24})^\dagger\wedge \xi_{24}\xi_{24}{}^\dagger$ would have the null direction $\xi_{24}\xi_{24}{}^\dagger$ in common. For the flat connection satisfying Theorem \ref{3d/4d}, the resulting bivector in Eq.\Ref{GJJxixi} is interpreted as $\star\ce_{35}$ (the normal bivector of $\ff_{35}$, see Eq.\Ref{spinhol}) parallel transported by spin connection to a certain point (say, vertex $\bar{1}$) on the 4-simplex, while $(J\xi_{24})(J\xi_{24})^\dagger\wedge \xi_{24}\xi_{24}{}^\dagger$ is $\star\ce_{24}$ parallel transported to the same point. Then as a consequence,
\be
\eps_{IJKL}\ce_{35}^{IJ}(\bar{1})\ce_{24}^{KL}(\bar{1})=0.\label{cece}
\ee 
By the definition of $\ce$, Eq.\Ref{cece} implies that the 4 edges meeting at vertex $\bar{1}$ of 4-simplex have the tangent vectors lying in a 3d subspace of the tangent space at $\bar{1}$, i.e. the 4-simplex geometry is degenerate. Therefore a nondegenerate 4-simplex geometry always satisfies Eq.\Ref{3524}.

Eq.\Ref{3524} is a condition derived from the ideal octahedron Oct(1). There are 4 more similar conditions derived from other 4 ideal octahedra:
\be
\mathrm{Oct(2):}&& G_{14}\xi_{45}\not\propto\xi_{13}\quad \Leftarrow\quad \eps_{IJKL}\ce_{13}^{IJ}(\bar{1})\ce_{45}^{KL}(\bar{1})\neq0\nonumber\\
\mathrm{Oct(3):}&& G_{14}\xi_{45}\not\propto\xi_{12}\quad \Leftarrow\quad \eps_{IJKL}\ce_{12}^{IJ}(\bar{1})\ce_{45}^{KL}(\bar{1})\neq0\nonumber\\
\mathrm{Oct(4):}&& G_{13}\xi_{35}\not\propto\xi_{12}\quad \Leftarrow\quad \eps_{IJKL}\ce_{12}^{IJ}(\bar{1})\ce_{35}^{KL}(\bar{1})\neq0\nonumber\\
\mathrm{Oct(5):}&& G_{13}\xi_{34}\not\propto\xi_{12}\quad \Leftarrow\quad \eps_{IJKL}\ce_{12}^{IJ}(\bar{1})\ce_{34}^{KL}(\bar{1})\neq0
\ee   
However $\eps_{IJKL}\ce_{ab}^{IJ}(\bar{1})\ce_{cd}^{KL}(\bar{1})=0$ doesn't always implies the degeneracy of a constant curvature 4-simplex, because there are some parallel transportations involved in $\ce_{ab}(\bar{1})$. The situation is quite similar to the case of constant curvature tetrahedron, that a pair of $\xi$'s are collinear doesn't always imply the degeneracy of the tetrahedron, because there is a holonomy $H$ involved in a dihedral angle $\phi_{ij}$.

Nevertheless, the configurations violating any of the above conditions are special, and they can be viewed as ``almost degenerate geometries''. If the 4-simplex was flat, $\eps_{IJKL}\ce_{ab}^{IJ}(\bar{1})\ce_{cd}^{KL}(\bar{1})=0$ would always mean to be degnerate. For constant curvature 4-simplex, the deviation of these configurations from degeneracy is of order $\kappa L^2$ where $L$ is the length scale of the 4-simplex. If we assume $\kappa L^2$ is small and expand the volume of constant curvature 4-simplex
\be
V_4=V_4^{(\mathrm{flat})}+o(L^6\kappa)
\ee
$\eps_{IJKL}\ce_{ab}^{IJ}(\bar{1})\ce_{cd}^{KL}(\bar{1})=0$ implies the flat 4-simplex volume $V_4^{(\mathrm{flat})}=0$, and the volume of constant curvature 4-simplex $V_4\sim o(L^6\kappa)$.

\section{Determining $\theta(\ff_{T^2})$}\label{0andpi}

{The discussion in this appendix relies on the material in \cite{HHKR} of proving the correspondence between flat connection on $S^3\setminus\G_5$ and constant curvature 4-simplex.}

First we consider a single 4-simplex and triangle $\ff_{ab}$ (which doesn't connect to the vertices $\bar{a},\bar{b}$). At a triangle vertex, we denote by $b_1(\ff_{ab}),b_2(\ff_{ab})$ the tangent vectors along the two edges connecting to the vertex. $\ff_{ab}$ is a face of $\mathrm{Tetra}_a$ (the tetrahedron doesn't connect to the vertex $\bar{a}$). At the base point of the tetrahedron where all paths in FIG.\ref{4holedsphere} start and end, we denote the tangent vectors by $b_1,b_2$. For example at vertex 4 in FIG.\ref{4holedsphere}, 3 tangent vector of the adjacent edges can be determined by 
\be
b_i\cdot\hat{\fn}_j=\delta_{ij}\lt[\epsilon_{\mu\nu\rho\sig}\fu^\mu \hat{\fn}_1^\nu\hat{\fn}_2^\rho\hat{\fn}_3^\sig\rt],\quad\text{or}\quad b_k^\mu=\half\epsilon_{\ \ k}^{ij}\epsilon^\mu_{\ \nu\rho}\hat{\fn}_i^\nu\hat{\fn}_j^\rho\label{bn1}
\ee
where $\fu^\mu$ is the future-pointing time-like normal of $\fu^I=(1,0,0,0)^T$, and $\epsilon_{\mu\nu\rho\sig}\fu^\mu=\epsilon_{\nu\rho\sig}$. The tangent vectors of nonadjacent edges at vertex 4 is determined similarly, which relates the true tangent vectors located on the edges by a parallel transportation. 

\begin{Lemma}\label{Xbb}

There is a factorization $G_{ab}=X_a(a,b)^{-1}X_b(a,b)$, such that 
\be
\hat{X}_a(a,b)\, (b_1,b_2)=e^{i \eta_{ab}\pi} \lt(b_1(\ff_{ab}),b_2(\ff_{ab})\rt)\quad where\quad \eta_{ab}\in\{0,1\}.
\ee
$\hat{X}_a(a,b)\in\mathrm{SO}^+(3,1)$ denotes the vector representation of ${X}_a(a,b)$.
\end{Lemma}

\noindent
\textbf{Proof:} In the factorization, $\hat{X}_a(a,b)$ is a parallel transportation from the local frame adapted to the $a$-th tetrahedron to a generic frame located at a vertex of $\ff_{ab}$. The time-like vector 
\be
N_a(\ff_{ab})=\hat{X}_a(a,b) (1,0,0,0)^T
\ee 
is a future-pointing time-like normal vector to $\ff_{ab}$ at the triangle vertex. This can be checked by considering the inner product with $\ce_{ab}$ at the triangle vertex. A space-like normal of $\ff_{ab}$ is given by 
\be
\hat{\fn}_{3}(\ff_{ab})\equiv \hat{\fn}_{ab}(\ff_{ab})=\hat{X}_a(a,b) \hat{\fn}_{ab}.
\ee 
Obviously $N_a(\ff_{ab})$ and $\hat{\fn}_{ab}(\ff_{ab})$ are orthogonal since $(1,0,0,0)^T\cdot\hat{\fn}_{ab}=0$. At the vertex and consider $\mathrm{Tetra}_a$, there are another 2 triangles $\ff_{ac},\ff_{ae}$ adjacent to the vertex, whose space-like normals are 
\be
\hat{\fn}_{1}(\ff_{ab})\equiv \hat{\fn}_{ac}(\ff_{ab})=\hat{X}_a(a,b) \hat{\fn}_{ac},\quad\quad\hat{\fn}_{2}(\ff_{ab})\equiv \hat{\fn}_{ae}(\ff_{ab})=\hat{X}_a(a,b) \hat{\fn}_{ae}.
\ee
Their time-like normals are the same as $N_a(\ff_{ab})$. Let $b_1(\ff_{ab}),b_2(\ff_{ab}),b_3(\ff_{ab})$ be the tangent vectors at the vertex along the 3 adjacent edges. They relate $\hat{\fn}_{1}(\ff_{ab}),\hat{\fn}_{2}(\ff_{ab}),\hat{\fn}_{3}(\ff_{ab})$ by
\be
b_i(\ff_{ab})\cdot\hat{\fn}_j(\ff_{ab})=\delta_{ij}\lt[\epsilon_{\mu\nu\rho\sig}\,U_a(\ff_{ab})^\mu\, \hat{\fn}_{1}(\ff_{ab})^\nu\,\hat{\fn}_{2}(\ff_{ab})^\rho\,\hat{\fn}_{3}(\ff_{ab})^\sig\rt],\label{bn2}
\ee
where $U_a(\ff_{ab})$ is the geometrical normal vector of $\mathrm{Tetra}_a$ embedded in the boundary of geometrical 4-simplex. $U_a(\ff_{ab})$ may be past-pointing for some tetrahedra, so in general, it relates to the future-pointing normal $N_a(\ff_{ab})$ by 
\be
U_a(\ff_{ab})=\pm N_a(\ff_{ab}).\label{UpmN}
\ee
Comparing Eqs.\Ref{bn1} and \Ref{bn2}, we obtain that \footnote{If minus sign appears in Eq.\Ref{UpmN}, in principle the space-like normal $\hat{\fn}_{ab}$ has to also flip sign, i.e. $\hat{\fn}_{ab}(\ff_{ab})=-\hat{X}_a(a,b) \hat{\fn}_{ab}$, because the geometrical bivector $\star\ce_{ab}(\ff_{ab})=\hat{\fn}_{ab}(\ff_{ab})\wedge U_a(\ff_{ab})=\hat{X}_a(a,b)^I_{\ K}\hat{X}_a(a,b)^{J}_{\ L}(\hat{\fn}_{ab}\wedge \fu)^{KL}$. But the result Eq.\Ref{Xbb=bb} is not affected.}
\be
\hat{X}_a(a,b)\, (b_1,b_2)=e^{i \eta_{ab}\pi} \lt(b_1(\ff_{ab}),b_2(\ff_{ab})\rt)\quad \text{where}\quad \eta_{ab}\in\{0,1\}.\label{Xbb=bb}
\ee
$\eta_{ab}=1$ when $U_a(\ff_{ab})$ is past-pointing.

$\Box$\\

Note that the correspondence between $G_{ab}$ from flat connection on $S^3\setminus\G_5$ and the holonomy of spin connection on 4-simplex uses the isomorphism $\pi_{1}(S^3\setminus\G_5)\simeq \pi_1(\mathrm{sk}(\mathrm{Simplex}))$. Given the path for holonomy on 4-simplex, the contour for $G_{ab}$ is determined up to homotopy. 

Translating the notation from $S^3\setminus\G_5$ to $\sm_3$, $G_{ab}\equiv G(\cs_a,\cs_b)$. By the above result we have for a torus cusp $\ell=T^2$ (${G}_{T^2 }\equiv{G}(\cs_n,\cs_0)$ with $\cs_n=\cs_0$)
\be
\hat{G}_{T^2 }(b_1,b_2)=e^{i \pi\eta(\ff_{T^2})}(b_1,b_2),\quad\quad \eta\lt(\ff_{T^2}\rt)\in\{0,1\}, \label{Gbb}
\ee
where $b_1,b_2$ are the tangent vectors along two edges of $\ff_{T^2}$. Thus $\hat{G}(\cs_n,\cs_0)$ is the composition of a boost and a $\pi$-rotation, which leave the plane spanned by $b_1,b_2$ invariant. On the other hand, by Eq.\Ref{Gxiexi} we find
\be
G_{T^2}=M_{T^2}\left(
\begin{array}{cc}
 e^{-\half\sgn(V_4)\,\eps(\ff_{T^2})-\frac{i}{2}\theta(\ff_{T^2})} & 0     \\
 0  &  e^{\half\sgn(V_4)\,\eps(\ff_{T^2})+\frac{i}{2}\theta(\ff_{T^2})}
\end{array}
\right)M_{T^2}^{-1},\quad \text{where}\quad
M_{T^2}=\left(
\begin{array}{cc}
 \xi^1_{T^2}(\cs) &  -\bar{\xi}^2_{T^2}(\cs)   \\
 \xi^2_{T^2}(\cs)  &  \bar{\xi}^1_{T^2}(\cs)
\end{array}
\right)
\ee
Here $M_{T^2}$ gives a rotation from $xy$-plane in $\R^3$ to the $(b_1,b_2)$-plane. Comparing to Eq.\Ref{Gbb}, we determine that
\be
\theta(\ff_{T^2})=\pm\pi\eta(\ff_{T^2})
\ee 
where $\pm$ is because $\Slc$ is a 2-fold covering over $\mathrm{SO}^+(3,1)$. We have seen in the proof of Lemma \ref{Xbb} that the appearance of the sign $e^{i\pi\eta}$ is because of a flip between future-pointing and past-pointing normals. The parallel transportation of spin connection (along an open path) sometimes gives an element in $\mathrm{SO}^-(3,1)$ as $e^{i\pi\eta}=-1$ \footnote{In a 4-simplex, the spin connection (along the open path) is given by $e^{i \eta_{ab}\pi}\hat{G}_{ab}$.}. For a single 4-simplex it happens at the boundary corners, which are the intersections of 2 causally related spatial slices. However for an internal triangle $\ff_{T^2}$, $\hat{G}_{T^2 }$ relates to a holonomy of spin connection along an internal loop in the (simplicial) spacetime $(\sm_4,g)$. The future-pointing (or past-pointing) time-like normal of the tetrahedron at the stating point should remain future-pointing (or past-pointing) after parallel transporting along a closed loop to the end point, in case $(\sm_4,g)$ is globally time-oriented. Therefore if $(\sm_4,g)$ is globally time-oriented,
\be
\eta(\ff_{T^2})=0 
\ee
and the loop holonomy of spin connection belongs to $\mathrm{SO}^+(3,1)$. When any $\eta(\ff_{T^2})\neq 0$ the corresponding $(\sm_4,g)$ doesn't have a global time-orientation. The classification of globally time-oriented/unoriented $(\sm_4,g)$ by using $\eta(\ff_{T^2})$ actually has been discuss in the work \cite{hanPI}, in the context of simplicial geometry with flat 4-simplices.

\end{document}